\documentclass[11pt,a4paper,dvips]{scrartcl}
\usepackage{cite,mcite}
\usepackage{graphicx}
\usepackage{subfig}
\usepackage{lcd}
\usepackage{lcd_title,ifthen}
\usepackage{mathptmx}
\usepackage{helvet}
\usepackage{amsmath}
\usepackage{color}
\usepackage{units}
\usepackage{url}
\makeatletter
\def\url@myurlfontstyle{%
  \@ifundefined{selectfont}{\def\UrlFont{\sf}}{\def\UrlFont{\small\ttfamily}}}
\makeatother
\long\def\symbolfootnote[#1]#2{\begingroup%
\def\thefootnote{\fnsymbol{footnote}}\footnote[#1]{#2}\endgroup} 
 \lcdnote{2011-003}
\newlength{\capindent}
\newlength{\capwidth}
\newlength{\figwidth}
\newcommand{\icaption}[2][!*!,!]{\hspace*{\capindent}%
  \begin{minipage}{\capwidth}
    \ifthenelse{\equal{#1}{!*!,!}}%
      {\caption{#2}}%
      {\caption[#1]{#2}}
      \vspace*{3mm}
  \end{minipage}}
\begin{document}
\begin{titlepage}
%
\vskip 35mm
%
\mydocversion
%
\title{Determination of Chargino and Neutralino Masses in high-mass SUSY scenarios at CLIC}
%
\author{N.~Alster\affiliated{1} \affiliated{2} and M. Battaglia\affiliated{1} \affiliated{2}}
\affiliations{
\affiliation[1]{University of California at Santa Cruz, Santa Cruz, CA, USA}
\affiliation[2]{CERN, Geneva, Switzerland},\\
}
%
\date{March 11, 2011}
%
\begin{abstract}
\noindent
This note reports the results of a study of the accuracy in the determination of 
chargino and neutralino masses in two high-mass supersymmetric scenarios through 
kinematic endpoints and threshold scans at a multi-TeV $e^+e-$ collider. The effects 
of initial state radiation, beamstrahlung and parton energy resolution are studied
in fully hadronic final states of inclusive SUSY samples. Results obtained at generator 
level are compared to those from fully simulated and reconstructed events for selected 
channels.
\end{abstract}
%
%
\end{titlepage}
%
%
\section{Introduction}

The study of the gaugino sector of Supersymmetry is a complex and important endeavour, 
which appears well suited to a linear collider of sufficient energy and luminosity. 
The main observables of interest are the masses of the $\chi^0$ and $\chi^{\pm}$ 
states and their production cross sections, including those with polarised beams.
$e^+e^-$ collisions offer two independent techniques for determining the mass of 
supersymmetric particles. These are the analysis of the energy spectrum of the SM particle 
produced in association with a lighter supersymmetric state in the two-body decays and the 
study of the pair production cross section near threshold. These techniques have already been 
extensively studied for lower centre-of-mass energies, $\sqrt{s}$, between 0.35 to 
0.5~TeV~\cite{tesla,ild,Aihara:2010zz,Li:2010mq}. 
In this note, we analyse the gaugino pair production and derive the statistical accuracy
on their masses using both techniques and including the effects of initial state radiation 
(ISR), beamstrahlung (BS) and parton energy resolution for multi-TeV $e^+e^-$ collisions. 
We follow the evolution of these accuracies for fully hadronic final states from pure signal 
samples to realistic inclusive SUSY samples and validate the results obtained at generator 
level with analyses performed on 
fully simulated and reconstructed events. The study provides us with requirements on parton 
energy resolution which are complementary to those obtained from other processes, such as 
heavy SUSY Higgs decays, since the kinematics of decays of gaugino pairs with large missing 
energy into pairs of escaping neutralinos does not benefit from the kinematic fits, which are 
instead applicable to processes where the full beam energy is deposited in the detector. 
The estimated mass accuracies can be compared in a next step to those required for the 
reconstruction of the GUT scale SUSY parameters~\cite{Blair:2005ui} and the determination of 
the lightest neutralino contribution to the dark matter relic density in the 
universe~\cite{Baltz:2006fm}. This comparison will provide us with well-motivated quantitative 
requirements on parton energy resolution in SUSY events. 

\section{SUSY Models}

\begin{figure}
\begin{center}
\begin{tabular}{cc}
\subfloat[Model I]{\includegraphics[width=7.25cm]{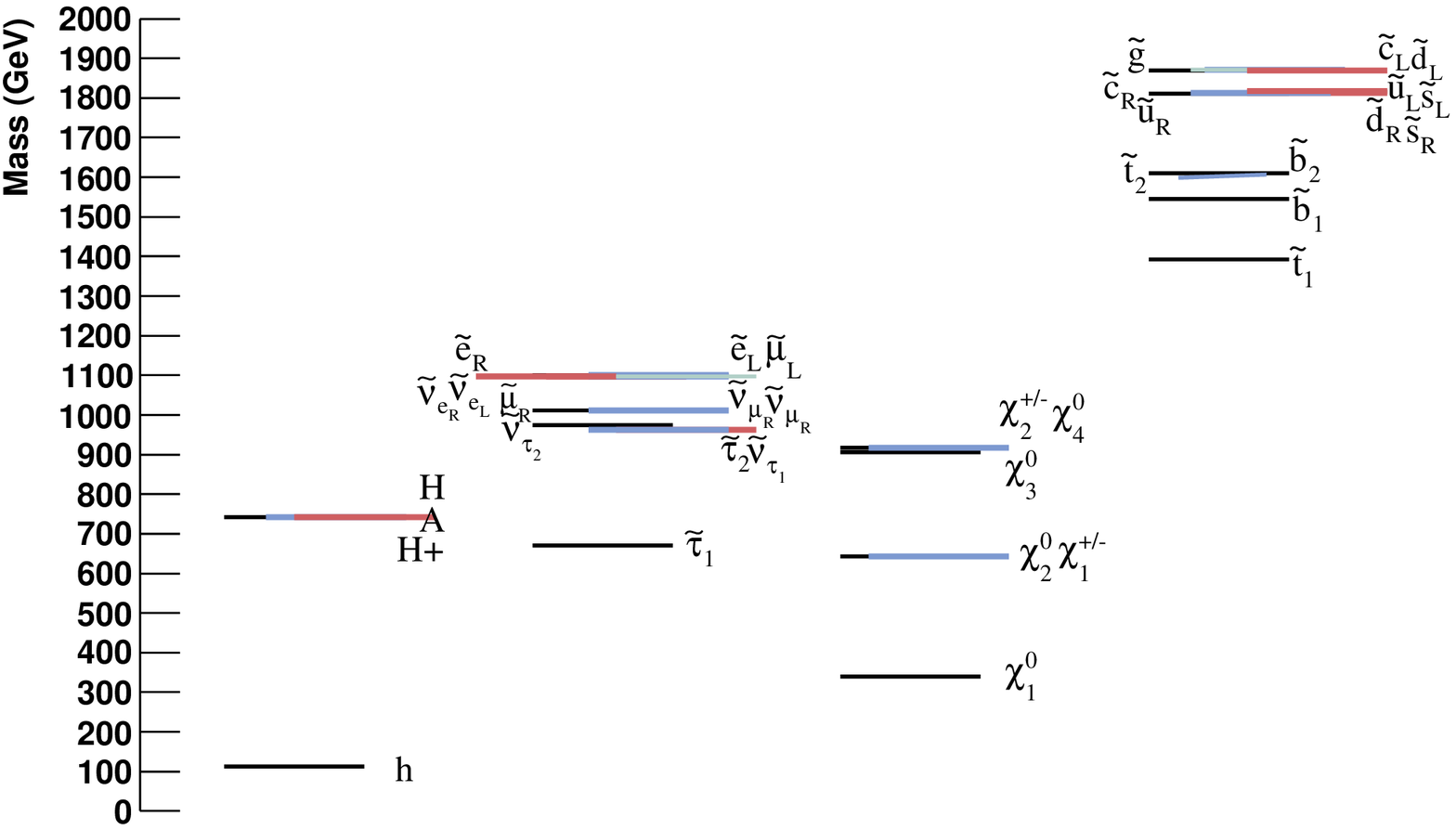}} &
\subfloat[Model II]{\includegraphics[width=7.0cm]{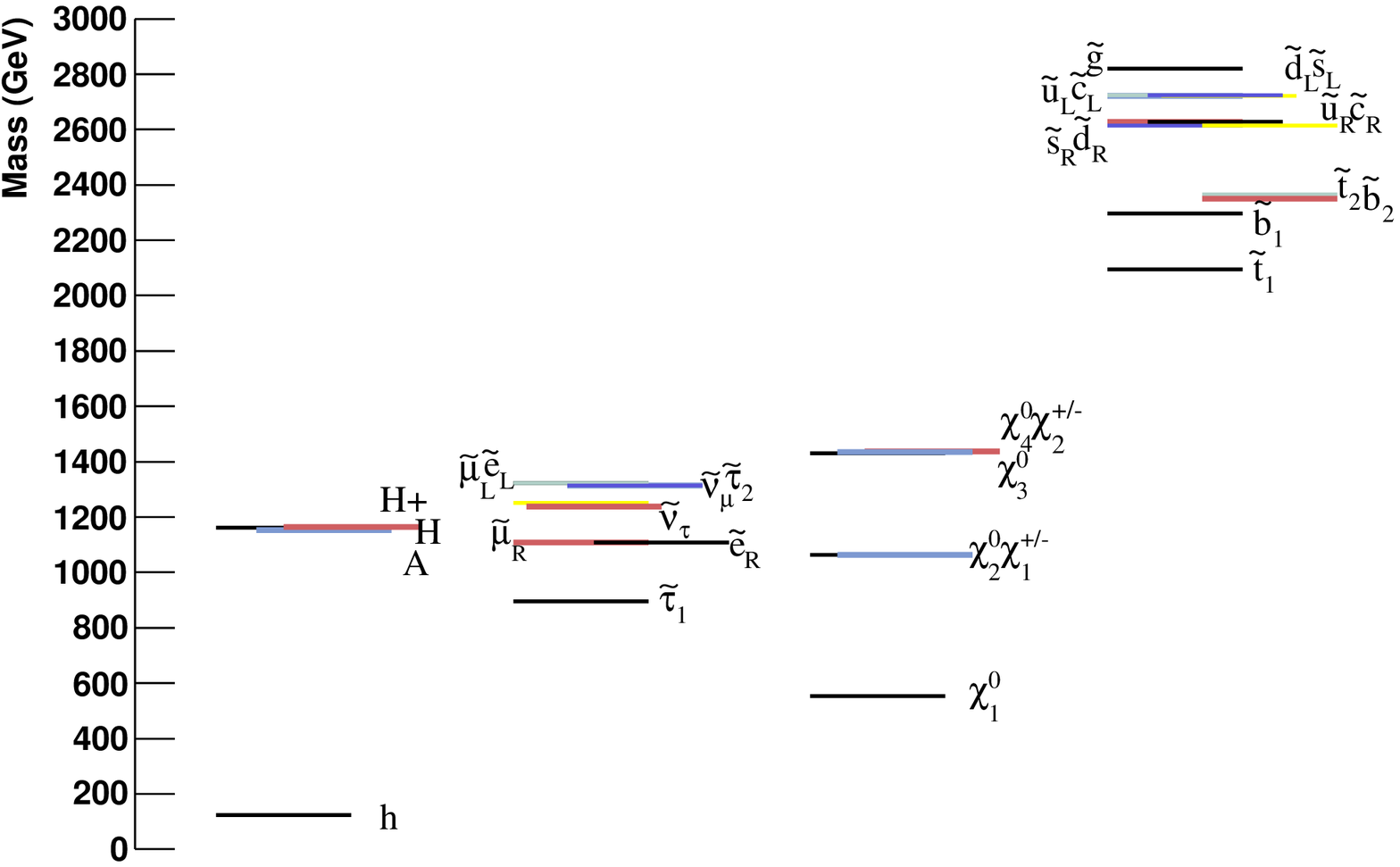}} \\
\end{tabular}
\end{center}
\vspace*{-0.5cm}
\caption{Supersymmetric particle spectra for Model~I and II.}
\label{fig:spectra}
\end{figure}

\begin{figure}
\begin{center}
\begin{tabular}{cc}
\subfloat[Model I]{\includegraphics[width=7.25cm]{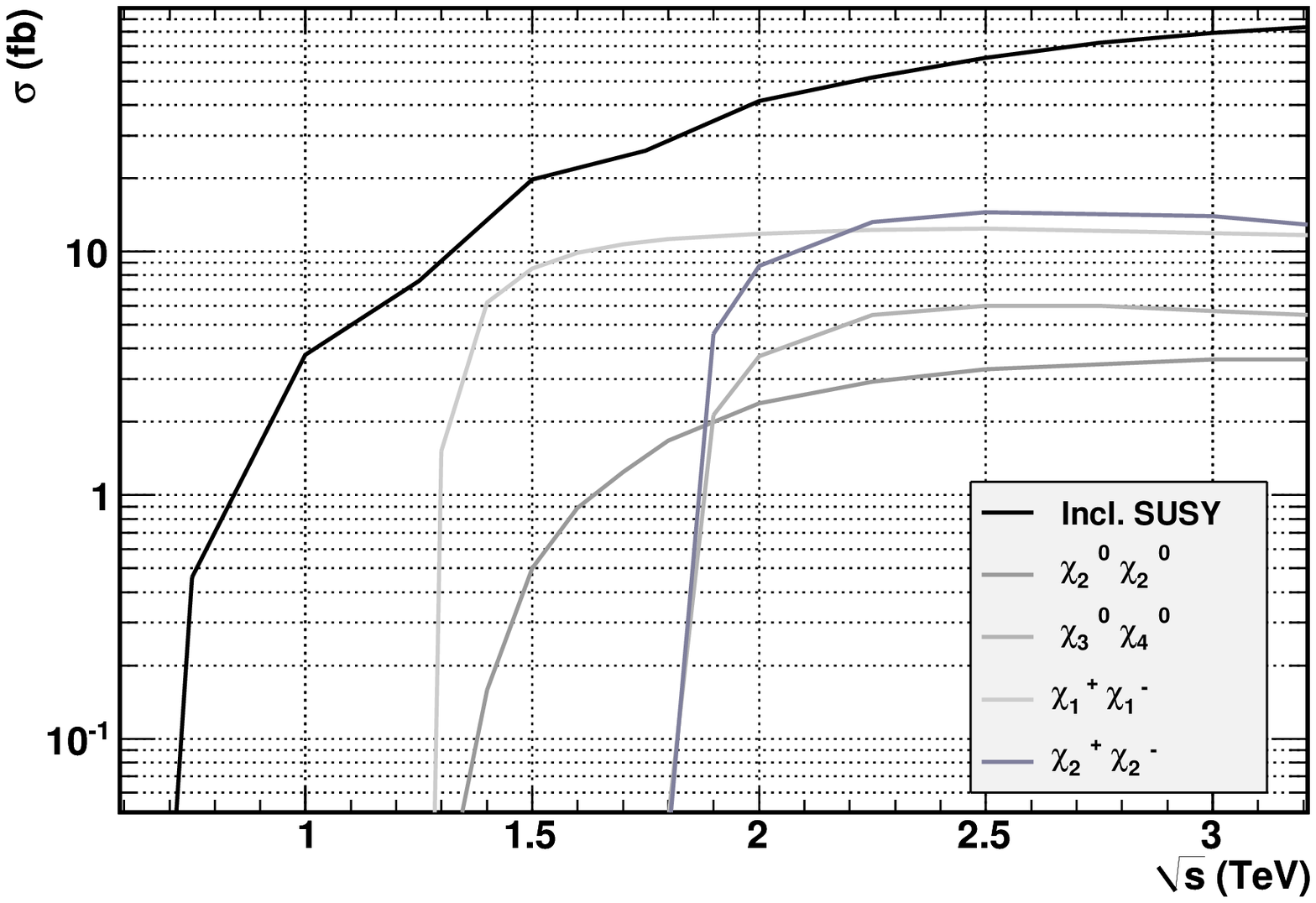}} &
\subfloat[Model II]{\includegraphics[width=7.0cm]{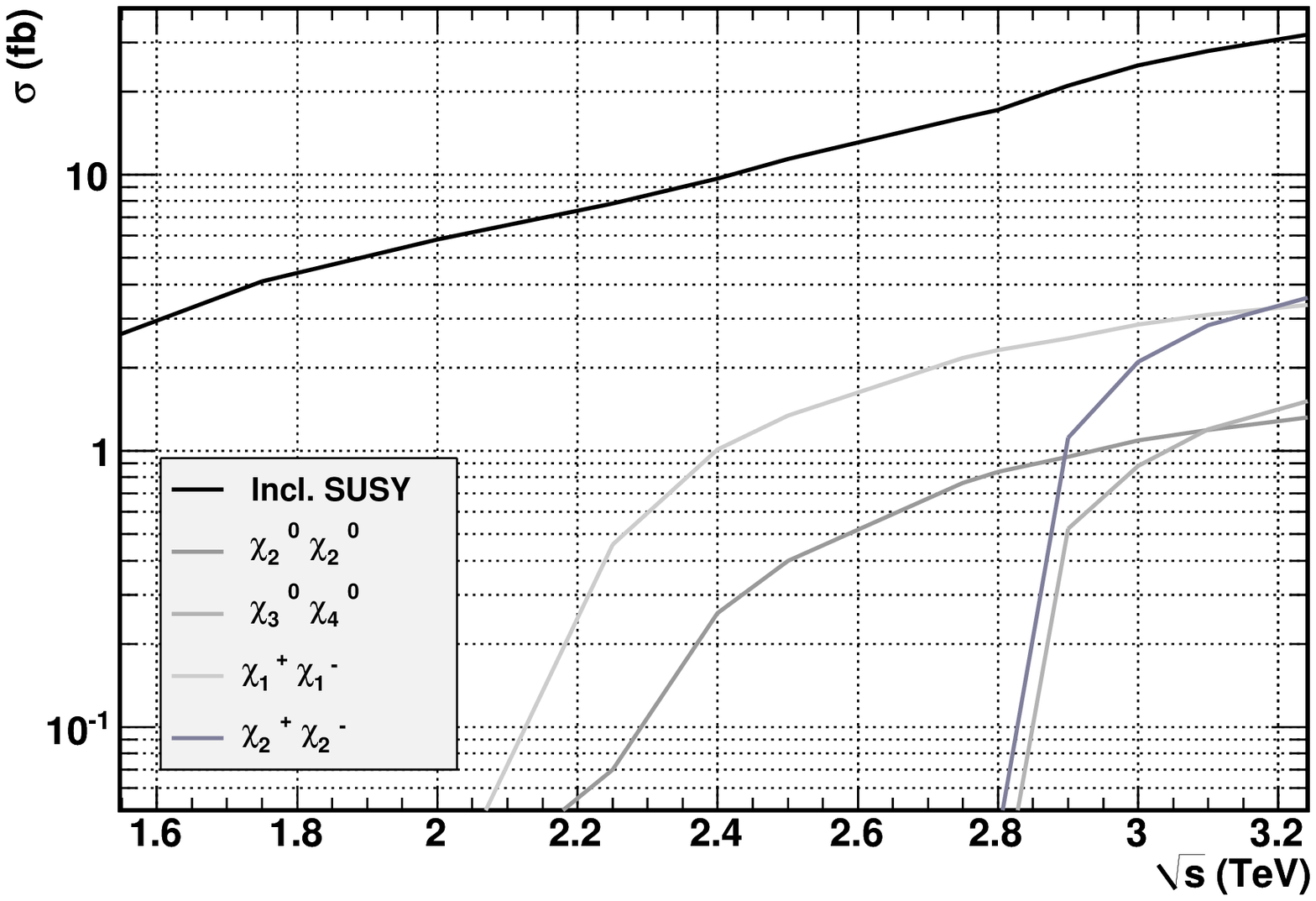}} \\
\end{tabular}
\end{center}
\vspace*{-0.5cm}
\caption{Cross sections for gaugino pair production vs.\ $\sqrt{s}$ in (left) model~I and (right) 
model~II}
\label{fig:xsec}
\end{figure}

This study considers two scenarios in the constrained MSSM (cMSSM) model , which offer 
different experimental challenges. 
Their parameters are given in Table~\ref{tab:modelpar}. The first (model I), adopted as a 
benchmark point for the CLIC CDR studies~\cite{martin}, has the lightest neutralino at 340~GeV 
and the chargino and heavier neutralinos with masses in the range 640 to 917~GeV (see 
Table~\ref{tab:mass} and the left panel of Figure\ref{fig:spectra}). 
At $\sqrt{s}$ = 3~TeV all the gauginos are observables. The relatively low masses 
and the 3~TeV centre-of-mass energy make cross sections sizable but the beamstrahlung 
effects more significant (see Table~\ref{tab:modelpar}).
In the second (model II~\footnote{This is point K' of ref~\cite{Battaglia:2003ab}.}) the lightest 
neutralino has a mass of 554~GeV, while the other neutralinos and the charginos have masses in 
the range from 1064 to 1414~GeV (see Table~\ref{tab:mass} and the right panel of 
Figure\ref{fig:spectra})~\cite{Battaglia:2003ab}. 
At 3~TeV, most gauginos are close to threshold for pair production and cross sections are small.
This minimises the beamstrahlung effects, since the production cross section drops significantly 
when the beams lose energy due to radiation. The cross sections are given in Table~\ref{tab:xsec}
and Figure~\ref{fig:xsec}.
\begin{table}
\caption{Parameters of the two cMSSM models adopted in this study}
\begin{center}
\begin{tabular}{|l|c|c|}
\hline
Parameter        & Model~I & Model~II \\
\hline
$m_0$ (GeV)      & 966     & 1001     \\
$m_{1/2}$ (GeV)  & 800     & 1300     \\
$\tan \beta$     & 51      & 46       \\
$A_0$            & 0.      & 0.       \\
sgn($\mu$)       & +       & -        \\
$m_{top}$ (GeV)  & 173.3   & 175      \\
\hline
\end{tabular}
\label{tab:modelpar}
\end{center}
\end{table}

\begin{table}
\caption{Gaugino mass spectrum in the two cMSSM models adopted in this study}
\begin{center}
\begin{tabular}{|l|c|c|c|c|}
\hline
Particle       & Mass  & Width & Mass  & Width \\
               & (GeV) & (GeV) & (GeV) & (GeV) \\
\hline
$\chi^0_1$     & 340.3 & -    & ~554.3 & -     \\
$\chi^0_2$     & 643.2 & 0.02 & 1064.2 & 0.04  \\
$\chi^0_3$     & 905.5 & 4.55 & 1407.2 & 6.75  \\
$\chi^0_4$     & 916.7 & 4.64 & 1413.8 & 6.85  \\
$\chi^{\pm}_1$ & 643.2 & 0.02 & 1064.3 & 0.04  \\
$\chi^{\pm}_2$ & 916.7 & 4.63 & 1413.7 & 8.08  \\
\hline
\end{tabular}
\label{tab:mass}
\end{center}
\end{table}

\begin{table}
\caption{Cross sections for gaugino pair production in the two cMSSM models adopted in this study}
\begin{center}
\begin{tabular}{|l|c|c|c|c|c|}
\hline
Process                        & no Rad    & ISR    & ISR+BS & Pol +0.8/0.0 & Pol +0.8/-0.6 \\
                               & (fb)      & (fb)   & (fb)   &  (fb)       &   (fb)      \\
\hline
Model I                        &           &        &        &             &             \\ \hline
Inclusive SUSY                 & 103.3     & 97.4   &  79.3  &             &             \\ 
$e^+e^- \to \chi^+_1 \chi^-_1$ & 11.5      & 11.6   &  11.9  &  21.4       &   34.0      \\ 
$e^+e^- \to \chi^0_2 \chi^0_2$ & ~4.2      & ~4.1   &  ~3.6  &  ~6.5       &   10.4      \\ 
$e^+e^- \to \chi^+_2 \chi^-_2$ & 14.5      & 14.4   &  13.8  &  21.1       &   33.3      \\ 
$e^+e^- \to \chi^0_3 \chi^0_4$ & ~6.1      & ~6.1   &  ~6.0  &  ~6.8       &   10.3      \\ \hline
Model II                       &           &        &        &             &             \\ \hline
Inclusive SUSY                 &  45.4     & 39.0   &  25.2  &             &             \\
$e^+e^- \to \chi^+_1 \chi^-_1$ & ~4.9      & ~4.3   &  ~2.9  &  ~5.2       &   ~8.3      \\ 
$e^+e^- \to \chi^0_2 \chi^0_2$ & ~1.9      & ~1.7   &  ~1.1  &  ~2.0       &   ~3.1      \\ 
$e^+e^- \to \chi^+_2 \chi^-_2$ & ~6.2      & ~4.6   &  ~2.1  &  ~3.2       &   ~5.1      \\ 
$e^+e^- \to \chi^0_3 \chi^0_4$ & ~2.6      & ~2.0   &  ~0.9  &  ~1.0       &   ~1.5      \\ 
\hline 
\end{tabular}
\label{tab:xsec}
\end{center}
\end{table}

\begin{figure}
\begin{center}
\begin{tabular}{cc}
\includegraphics[width=5.5cm,clip=]{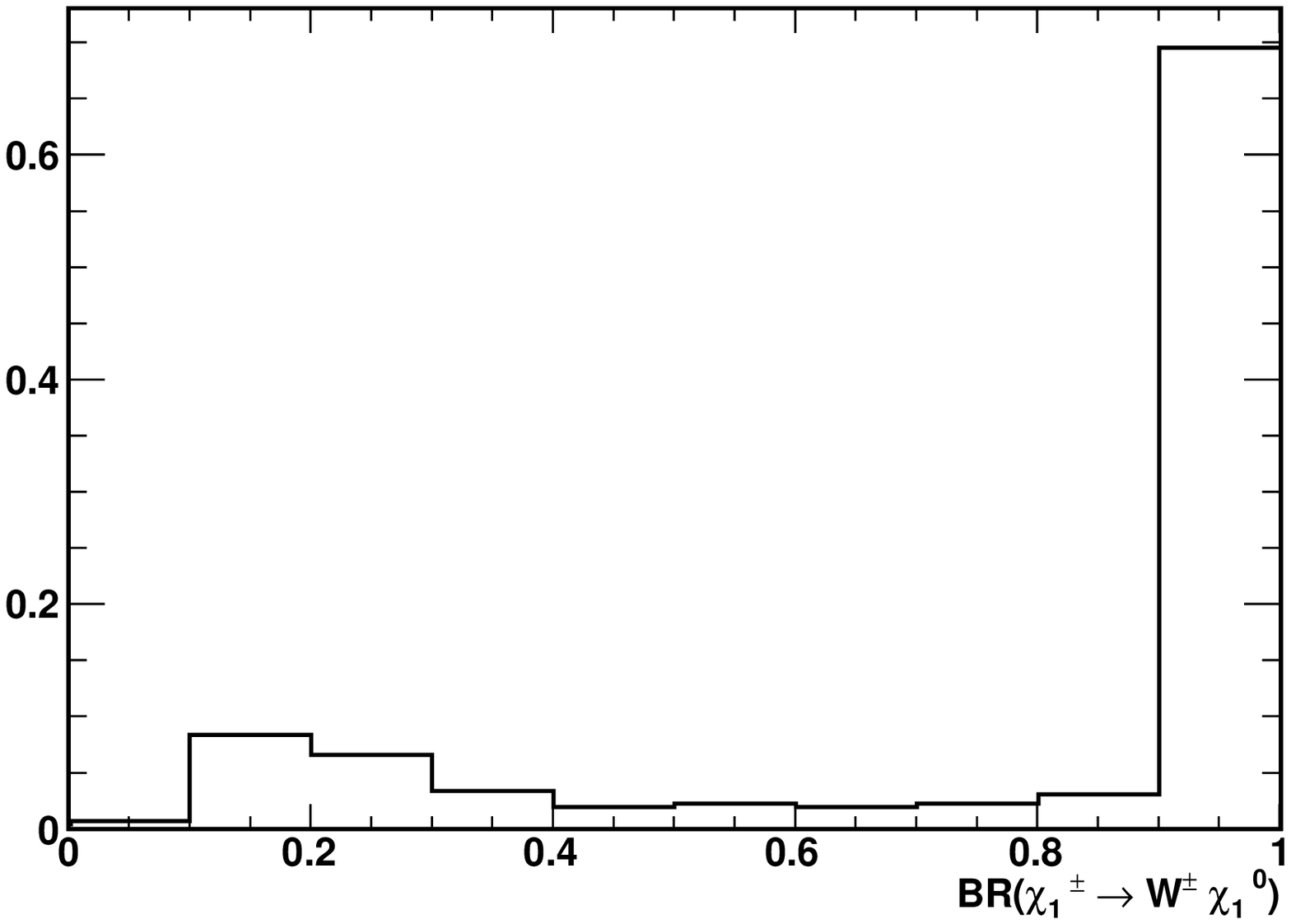} &
\includegraphics[width=5.5cm,clip=]{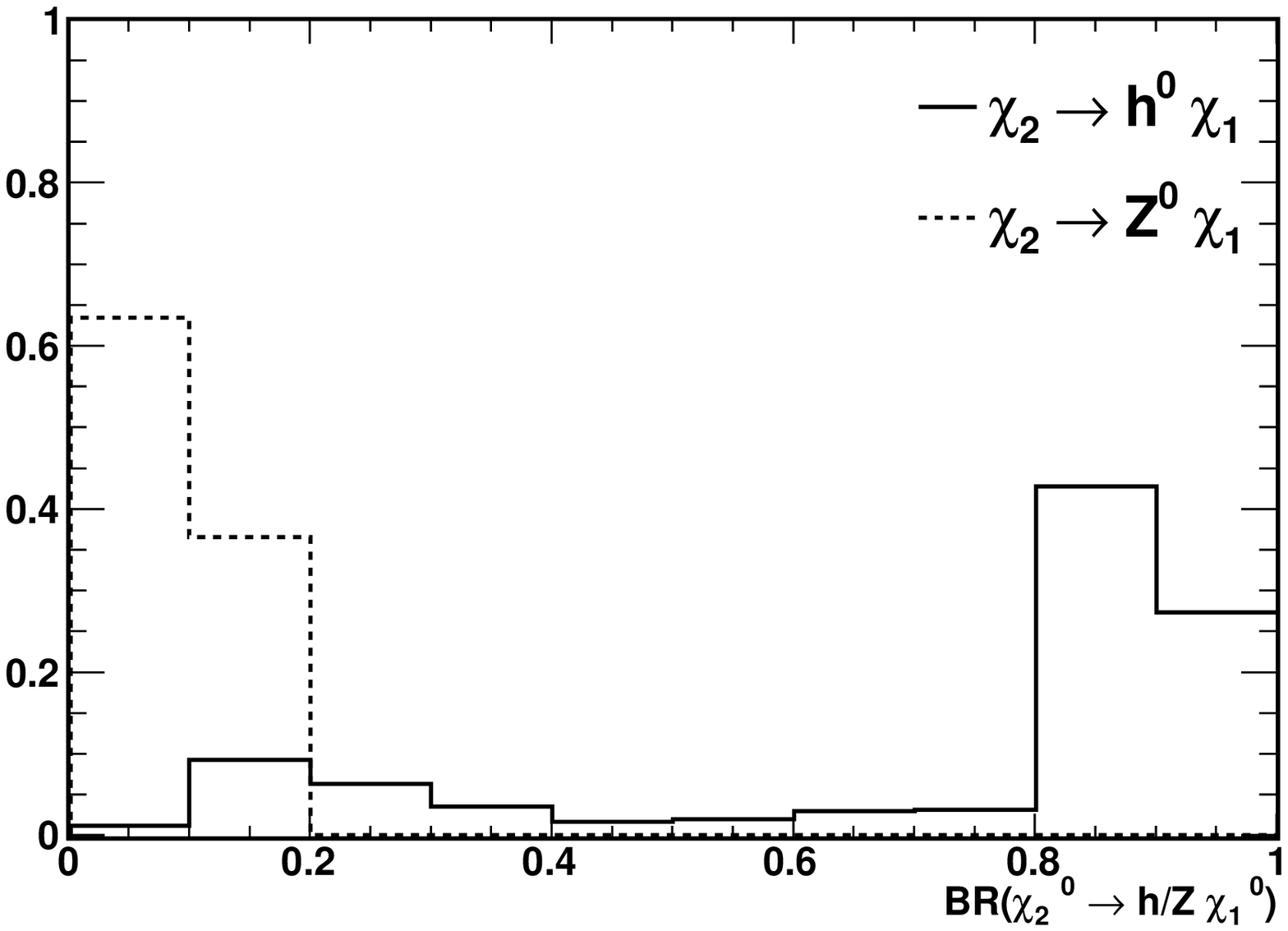} \\
\includegraphics[width=5.5cm,clip=]{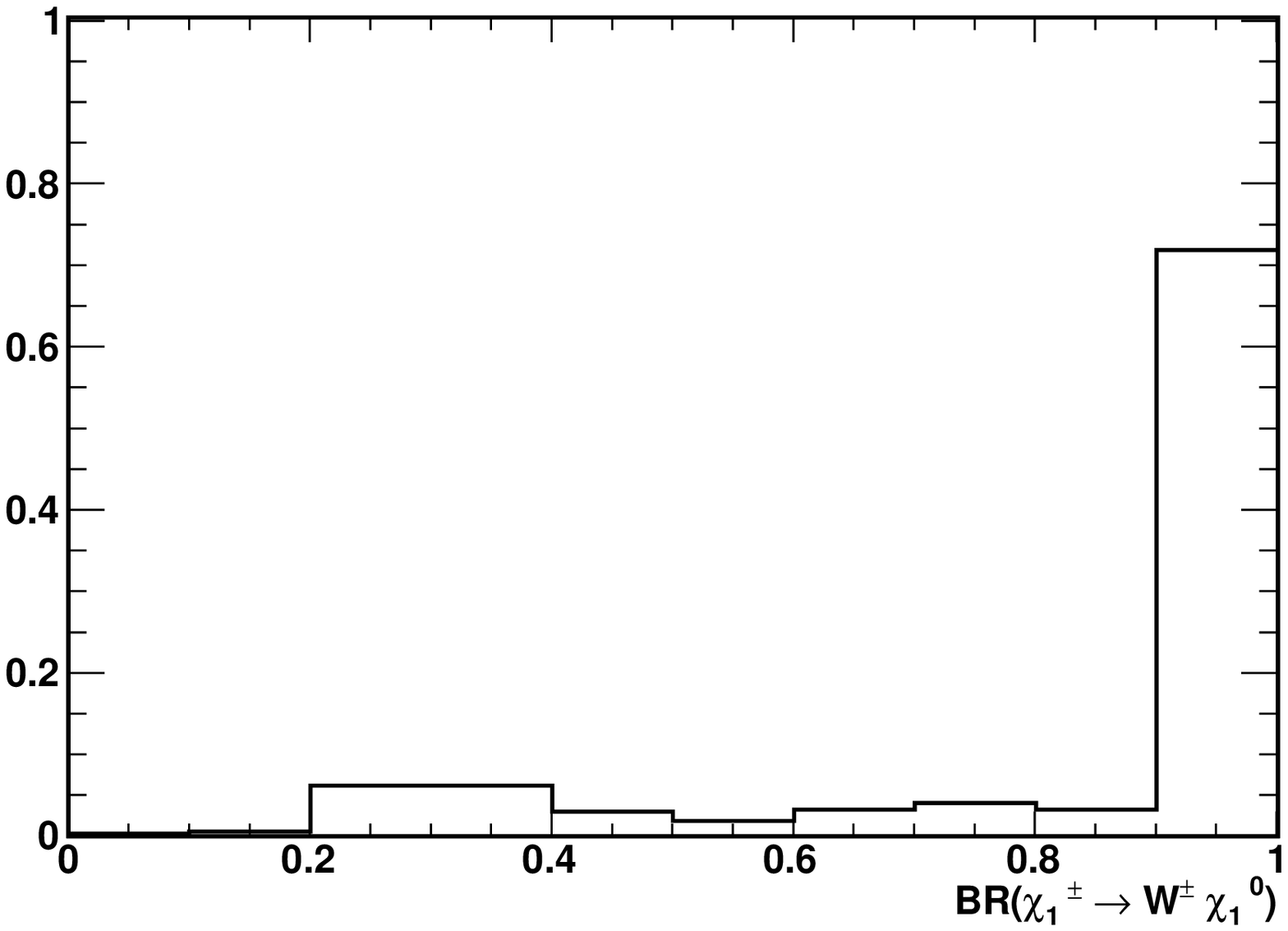} &
\includegraphics[width=5.5cm,clip=]{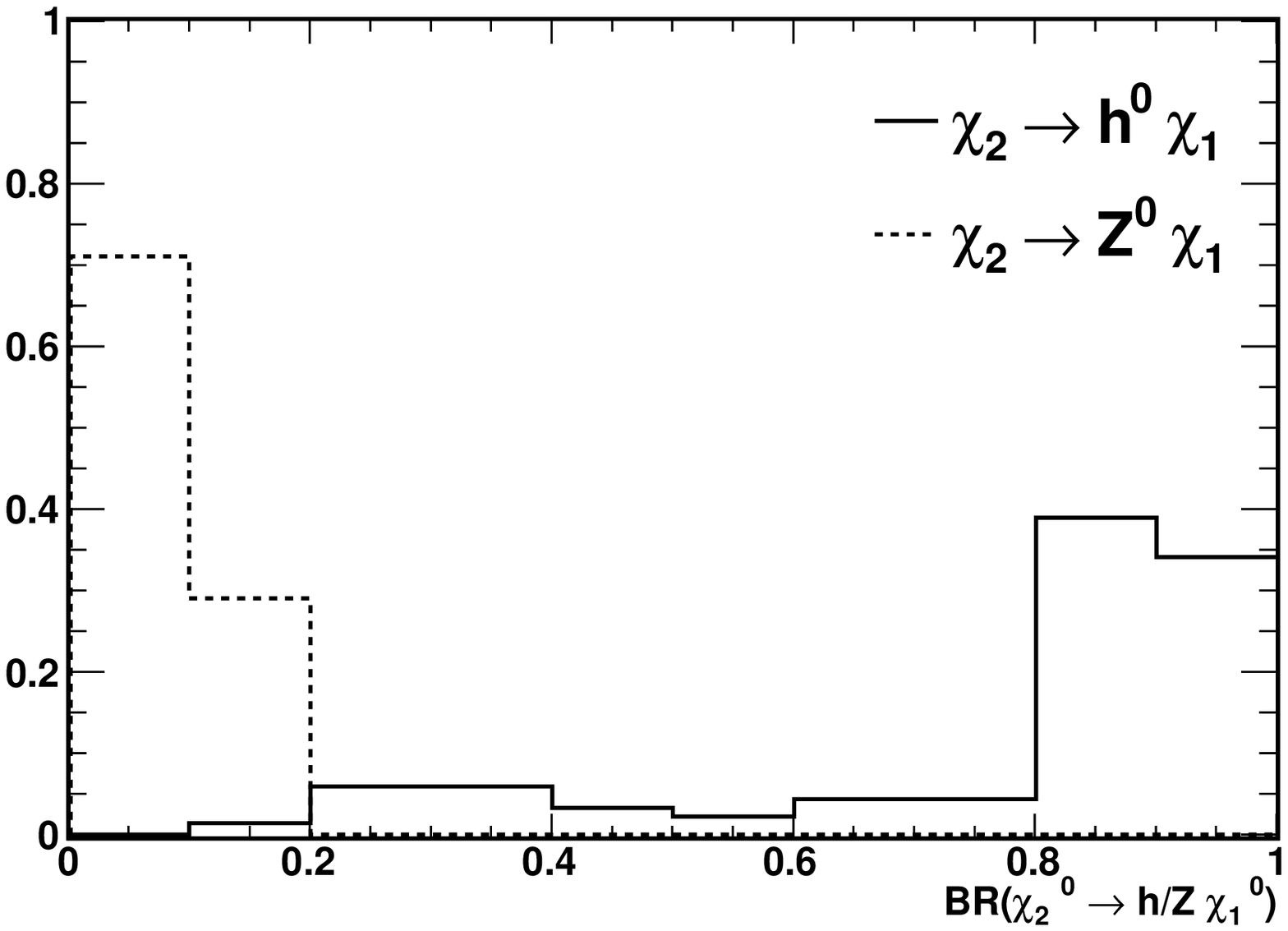} \\
\end{tabular}
\end{center}
\caption{Distribution of branching fraction values for (left) $\chi^{\pm}_1 \to W^{\pm} \chi^0_1$ 
and (right) $\chi^0_2 \to h^0 \chi^0_1$, $Z^0 \chi^0_1$  for points compatible with 
WMAP data from the cMSSM scan around (upper row) model~I and (lower row) model~II, 
showing the predominance of decays into bosons.}
\label{fig:brscan}
\end{figure}

\begin{figure}
\begin{center}
\includegraphics[width=7.0cm]{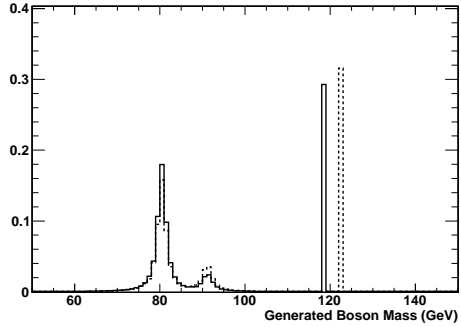}
\end{center}
\caption{Boson spectrum in inclusive SUSY events with $\ge$ 4~jets + missing energy for 
model~I (continuous line) and II (dashed line). The $h^0$ mass is 118.5 and 122.9~GeV, 
respectively.}
\label{fig:qqmass}
\end{figure}

The main features of these two models of interest in this study are the large sparticle
masses and the dominance of decays through $W^{\pm}$, $h^0$ and, to a lesser extent, $Z^0$ 
bosons.   
These features are common to most of the large-$\tan \beta$ cMSSM models~\cite{Ellis:2001msa} 
with neutralino dark matter compatible with the Cosmic Microwave Background (CMB) data. 
We verify this by performing scans of the cMSSM parameter space around both benchmark 
points to examine the mass spectrum and decay branching fractions of gauginos. 
In these scans we vary $m_0$ and $m_{1/2}$ within $\pm 300$~GeV from the benchmark 
parameters, $\tan \beta$ within $^{+5}_{-15}$, $A_0$ within $\pm 250$ and $\mu$ to 
have either sign. The sparticle spectrum corresponding to each set of parameters is
computed using {\sc SuSpect 2.2}~\cite{Djouadi:2002ze} and the decay branching ratios with 
{\sc SDecay 1.3}~\cite{Muhlleitner:2003vg}.  
We compute the neutralino relic density, $\Omega_{\chi}$, using 
{\tt Micromegas 2.2}~\cite{Belanger:2006is}. In total, we generate over 150k cMSSM points 
using a flat sampling of the parameter phase space. Of these, we retain those points 
consistent with the current limits on the lightest Higgs bosons and sparticle masses and 
yielding 0.093$< \Omega_{\chi}/h^2 <$0.129, in agreement with the WMAP seven-years 
data~\cite{Larson:2010gs}. 
\begin{figure}
\begin{center}
\begin{tabular}{cc}
\includegraphics[width=5.5cm,clip=]{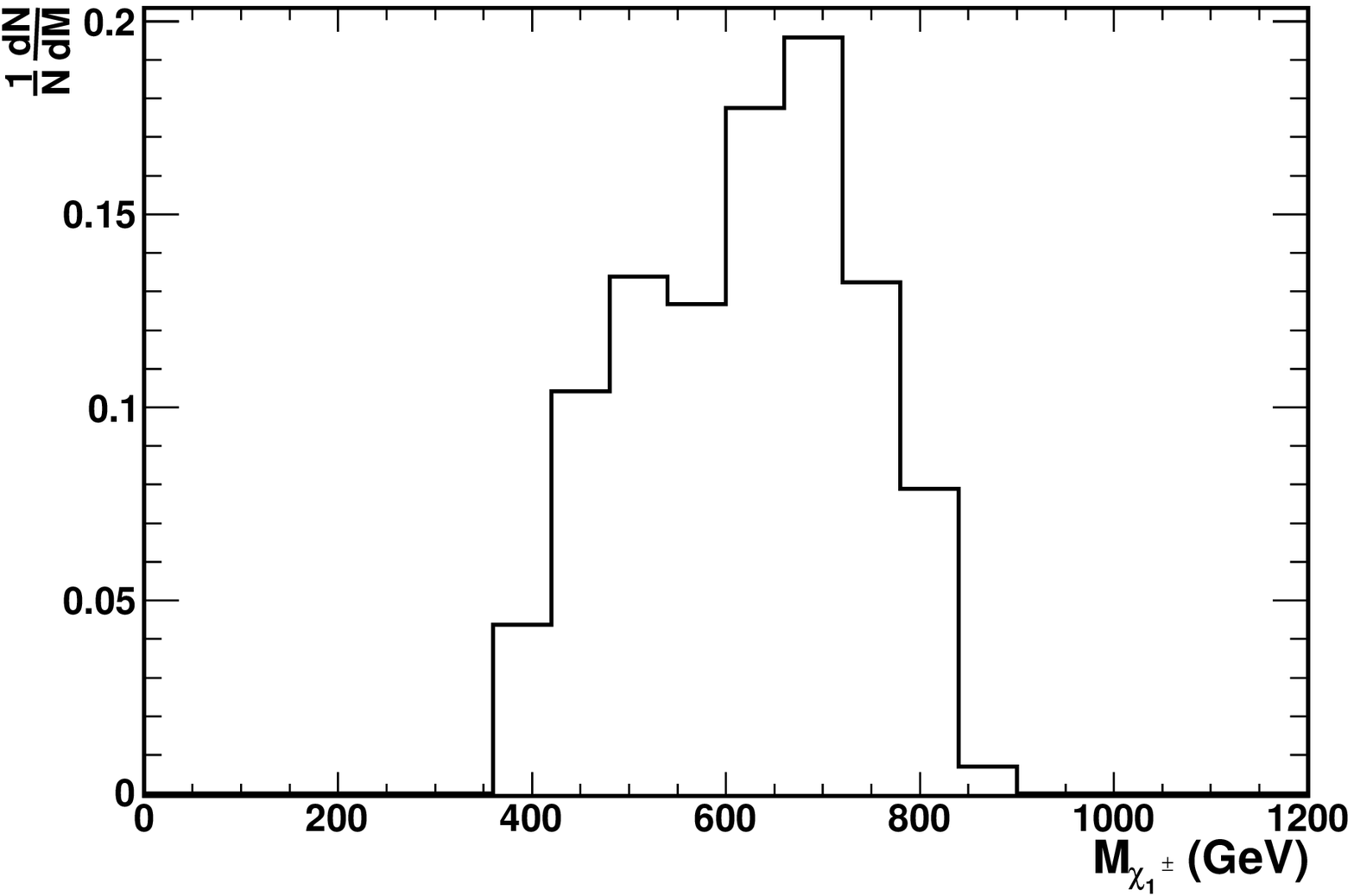} &
\includegraphics[width=5.5cm,clip=]{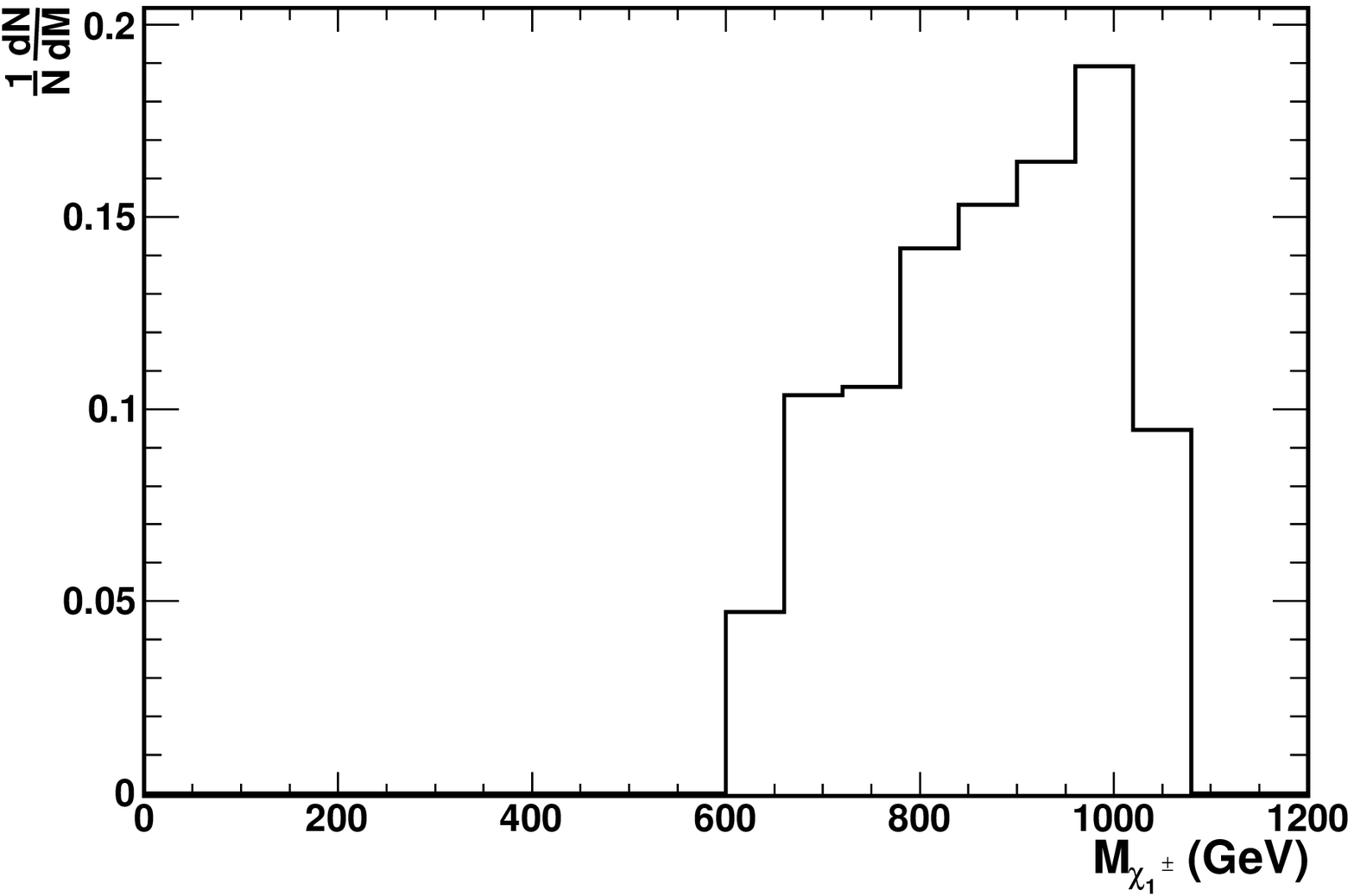} \\
\end{tabular}
\end{center}
\caption{Distribution of $\chi^{\pm}_1$ mass values for points compatible with WMAP data 
from the cMSSM scan around (left) model~I and (right) model~II.}
\label{fig:massscan}
\end{figure}
Figure~\ref{fig:massscan} shows the $\chi^{\pm}_1$ mass and Figure~\ref{fig:brscan} 
the $\chi^{\pm}_1 \to W^{\pm} \chi^0_1$ branching fraction for the accepted points. 
These results show that decays into bosons are dominant in this region of the cMSSM 
parameter space. These results can be extended to the general MSSM where a significant 
fraction of the decays of $\chi^{\pm}_1$ and $\chi^0_2$ and the majority of those of the 
heavier chargino and neutralinos are two-body processes with emission of a boson. 
In these scenarios, $W^{\pm}$, $Z^0$ and $h^0$ production is a distinctive signature 
of gaugino decays. Figure~\ref{fig:qqmass} shows the boson mass spectrum at 
generator level in inclusive SUSY events for the two models considered in this study, 
showing the $^{\pm}W$, $Z^0$ and $h^0$ boson contribution. The $W$/$Z$/$h$ discrimination 
is essential for identifying the decay processes, which sets constraints on parton energy 
resolution through the di-jet invariant mass resolution.
Precise parton energy reconstruction is similarly required to preserve the accuracy in 
the gaugino mass measurements through the determination of the endpoints of the boson 
energy spectra.

\subsection{Event Simulation}

For this study, events are generated with {\sc Pythia~6.215}~\cite{Sjostrand:2006za}. 
For model~I the SUSY spectrum and the sparticle decay branching fractions are forced 
to those obtained with {\sc SuSpect 2.2} and {\sc SDecay 1.3}, respectively.
For model~II the spectrum is generated with {\sc Isasugra~7.69}~\cite{Paige:2003mg}.
Polarised cross sections are obtained using {\tt SUSYGEN 3.0}~\cite{Katsanevas:1997fb}.
The production cross sections for unpolarised and polarised beams are summarised in 
Table~\ref{tab:xsec}.
Samples of inclusive SUSY events are processed through full detector simulation using 
the {\sc Geant-4}-based {\sc Mokka}~\cite{MoradeFreitas:2004sq} program and 
reconstructed with {\sc Marlin}-based~\cite{Gaede:2006pj} processors for a 
version of the ILD detector concept~\cite{ild}, modified for physics at CLIC~\cite{sailer}.
These events are used for a validation of the results in the 4-jet, $WW$ and $hh$ final 
states, discussed in section~3.3.5.

\section{Mass Determination by Fits to Boson Energy Spectra}

In the two body decay process $A \to B C$ into a boson $B$ and a lighter gaugino, $C$, 
which are a signature of these high-mass benchmark points, the masses of the parent and daughter 
gauginos can be extracted from the position of the kinematic edges of the boson energy spectrum. 
The technique was first proposed for squarks~\cite{Feng:1993sd} and later extended 
to other sparticles in two-body decays~\cite{Martyn:1999tc}. In the case of gaugino decays into bosons, 
their mass, $m_B$, cannot be neglected, as in the case of squark and slepton decays and the relation 
between the energy endpoint and the masses of the particle involved in the decay process are given by:
\begin{eqnarray}
\mathrm {E_{BH,BL}}= \gamma \left( E_B^{*} \pm \beta E_B^{*} \right)
\label{formula:eleh}
\end{eqnarray}
where
\begin{eqnarray}
E_B^{*} = \frac{m_A^2 + m_B^2 - m_C^2}{2 m_A} \\
\gamma = \frac{\sqrt{s}}{2 m_A} \\
\beta = \sqrt{\frac{1 - 4 m_A}{s}}
\label{formula:eleh1}
\end{eqnarray}
These formulae can be extended in a straightforward way to the case in which the gaugino $A$ is not directly 
produced in the $e^+e^-$ collisions but originates from the decay of an heavier particle, $A^{\prime}$, by 
replacing $s$ with $E_A^2$, where $E_A$ is its energy. In the case of cascading decays 
$A^{\prime} \to A B^{\prime} \to B C$, $E_A$ is obtained as 
$\sqrt{s} - E_{B^{\prime}H} < E_A < \sqrt{s} - E_{B^{\prime}L}$.

\subsection{Channels}

We study the following processes~\footnote{Throughout the paper the charge conjugate of the 
given state is also implied} for model~I:
\begin{itemize}
\item $e^+e^- \to \chi^+_1 \chi^-_1 \to W^+ \chi^0_1 W^- \chi^0_1$; $W \to q \bar q'$,
\item $e^+e^- \to \chi^0_2 \chi^0_2 \to h^0 \chi^0_1 h^0 \chi^0_1$; $h \to b \bar b$,
\item $e^+e^- \to \chi^+_2 \chi^+_2 \to W^+ \chi^0_2 W^- \chi^0_1 \to W^+ h^0 \chi^0_1 W^- \chi^0_1$; 
$h \to b \bar b$, $W \to q \bar q'$,
\item $e^+e^- \to \chi^+_2 \chi^+_2 \to h^0 \chi^+_1 W^- \chi^0_1 \to h^0 W^+ \chi^0_1 W^- \chi^0_1$; 
$h \to b \bar b$, $W \to q \bar q'$,
\item $e^+e^- \to \chi^+_2 \chi^+_2 \to Z^0 \chi^+_1 W^- \chi^0_1 \to Z^0 W^+ \chi^0_1 W^- \chi^0_1$; 
$Z \to q \bar q$, $W \to q \bar q'$,
\item $e^+e^- \to \chi^0_4 \chi^0_3 \to W^+ \chi^-_1 W^- \chi^+_1 \to W^+ W^- \chi^0_1 W^- W^+ \chi^0_1$;  
$W \to q \bar q'$.
\end{itemize}
For model~II we study:
\begin{itemize}
\item $e^+e^- \to \chi^+_1 \chi^-_1 \to W^+ \chi^0_1 W^- \chi^0_1$; $W \to q \bar q'$
\item $e^+e^- \to \chi^0_2 \chi^0_2 \to h^0 \chi^0_1 h^0 \chi^0_1$; $h \to b \bar b$.
\end{itemize}
They explore increasing event complexities from 4-jet with two bosons to eight jets with 
four bosons. Mass values and their statistical uncertainties are extracted by a binned $\chi^2$ fit
using the {\tt Minuit} package~\cite{James:1975dr}. 
The boson energy spectrum  from simulation is compared to reference spectra generated 
according to Eq.(\ref{formula:eleh}) above, changing the sparticle masses. An integrated luminosity of 
2~ab$^{-1}$ is assumed. The $\sqrt{s}$ value in Eq.(\ref{formula:eleh1}) is either kept to the nominal value 
of 3~TeV or smeared to account for ISR and beamstrahlung effects, as discussed below. Since the
fit procedures requires to generate a large number of reference spectra, which are statistically independent, 
in the fitter iterations, each of these is filled with $2.5 \times 10^6$ random entries distributed according 
to Eq.(\ref{formula:eleh}), to minimise effects from their statistical fluctuations in the {\tt Minuit} 
calculations, in particular in the determination of the derivatives.

\subsection{Effect of Beam Spectra and Energy Resolution}

The effects of initial state radiation (ISR), beamstrahlung (BS) and finite resolution 
in parton energy reconstruction are taken into account. 
Beamstrahlung effects on the luminosity spectrum are included using results of the CLIC 
beam simulation for the 2008 accelerator parameters~\cite{Braun:2008zzb}. Initial state 
radiation is included in the event generation in {\sc Pythia}. The beamstrahlung spectrum 
obtained is then used for smearing the $\sqrt{s}$ value in Eq.~\ref{formula:eleh1}.
\begin{figure}[hb!]
\begin{center}
\begin{tabular}{cc}
 \subfloat[ISR]{\includegraphics[width=6.5cm,clip]{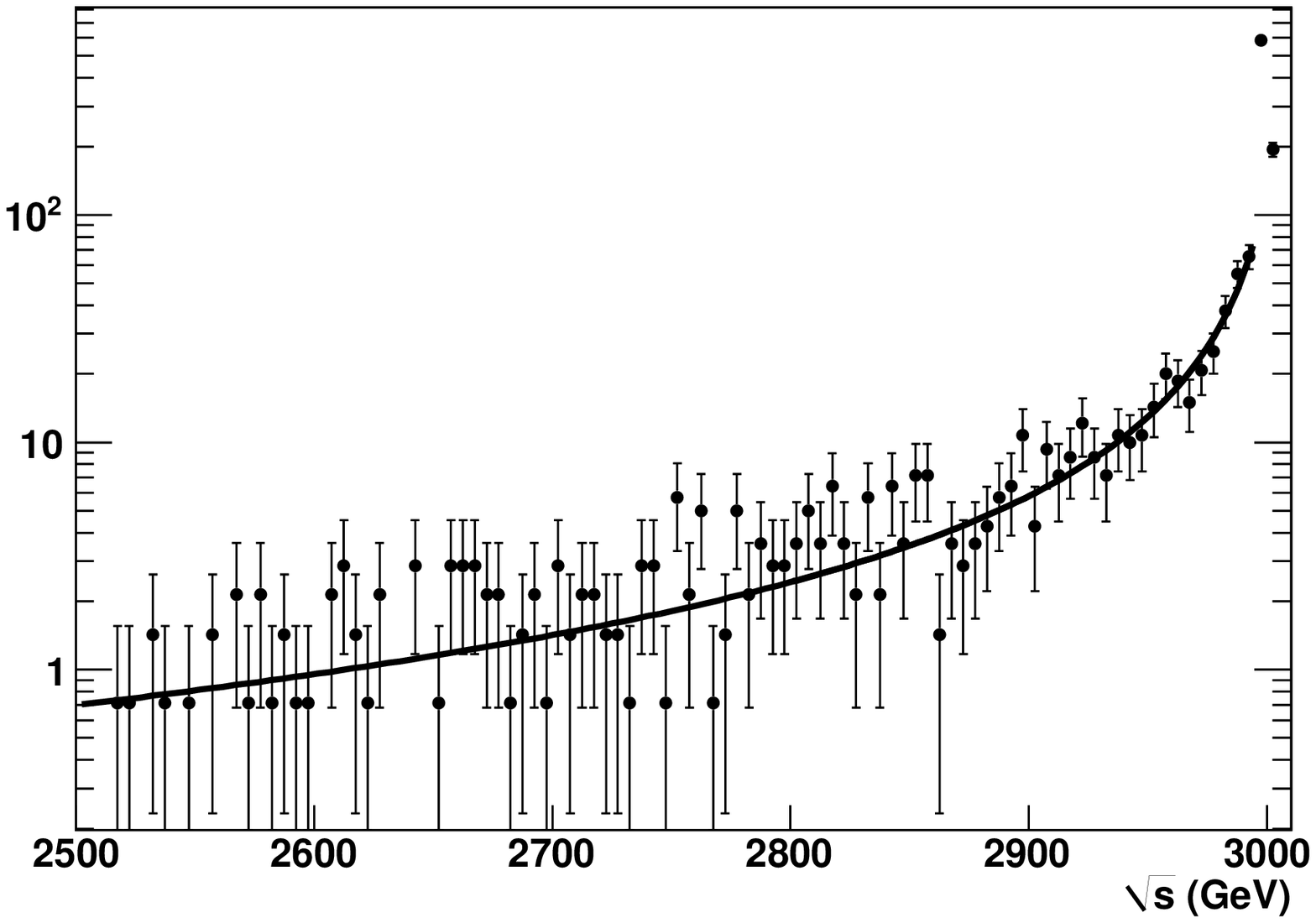}} &
 \subfloat[Beamstrahlung]{\includegraphics[width=6.5cm,clip]{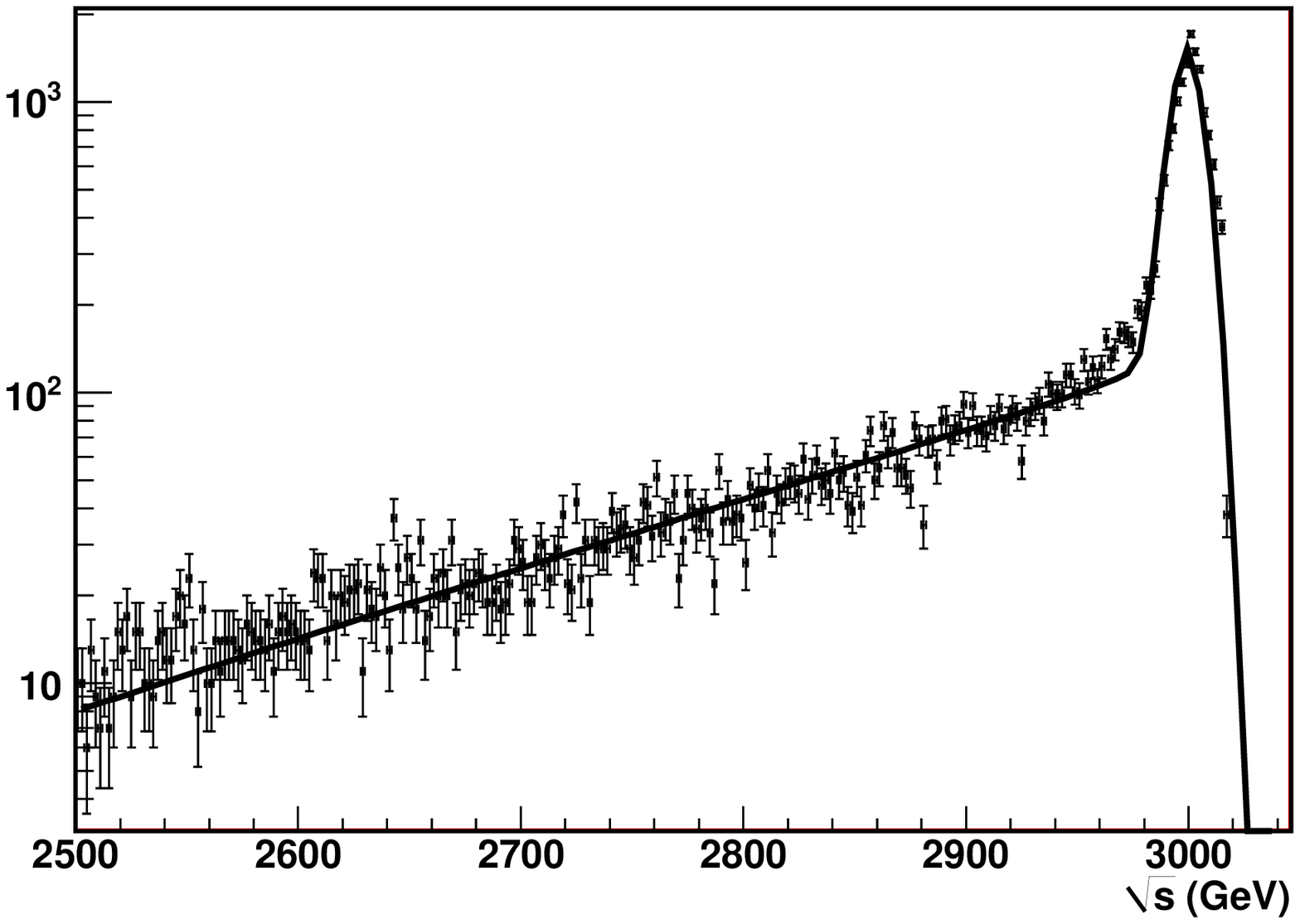}} \\
\end{tabular}
\end{center}
\caption{Centre-of-mass energy distribution including (a) ISR and (b) ISR and beamstrahlung. 
The points represent the simulation and the lines the phenomenological functions used for 
describing their shape.}
\label{fig:isr_circe}
\end{figure}
\begin{figure}[ht!]
\begin{center}
\begin{tabular}{cc}
\subfloat[Model~I]{\includegraphics[width=7.0cm]{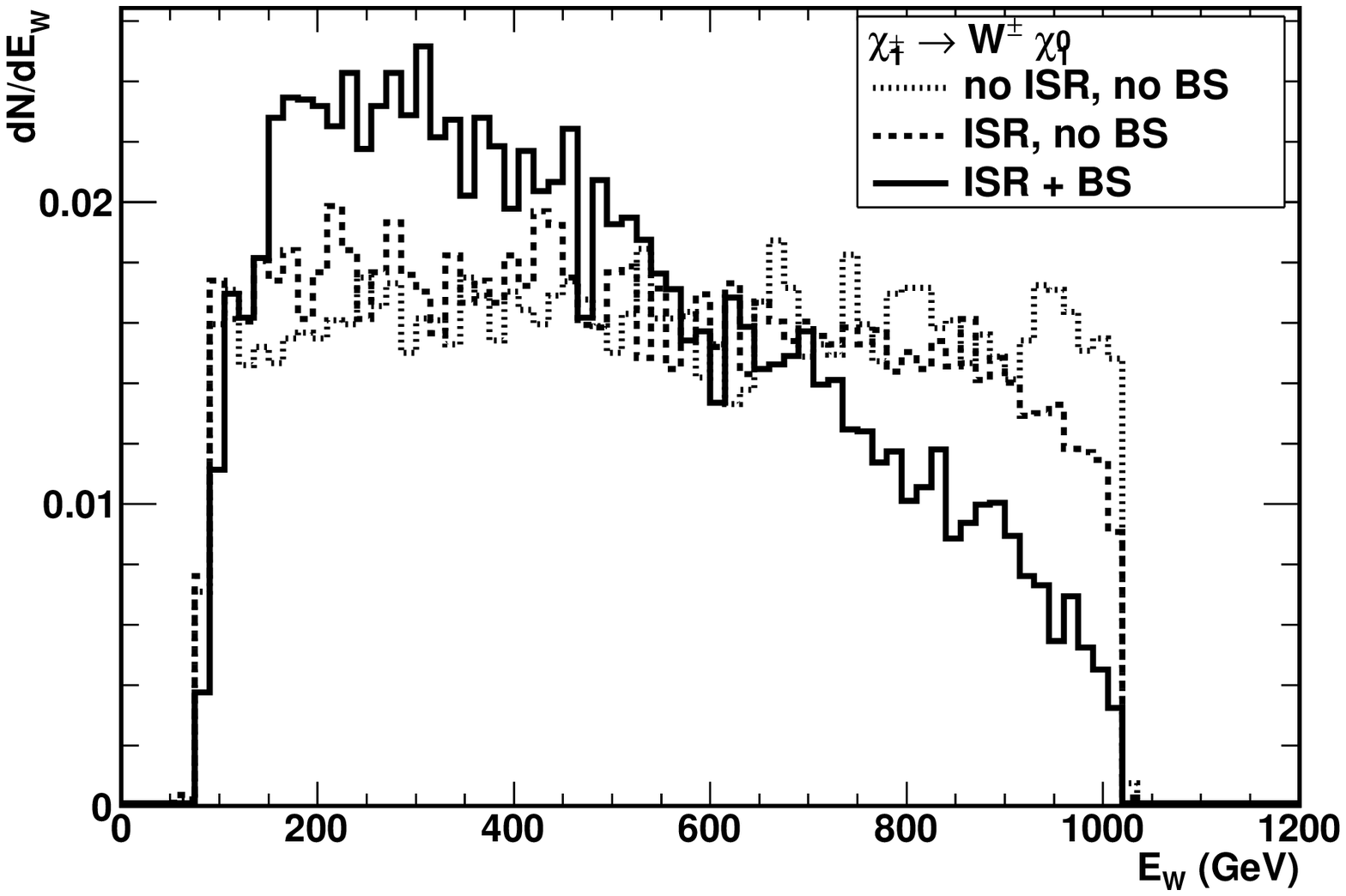}} &
\subfloat[Model~II]{\includegraphics[width=7.0cm]{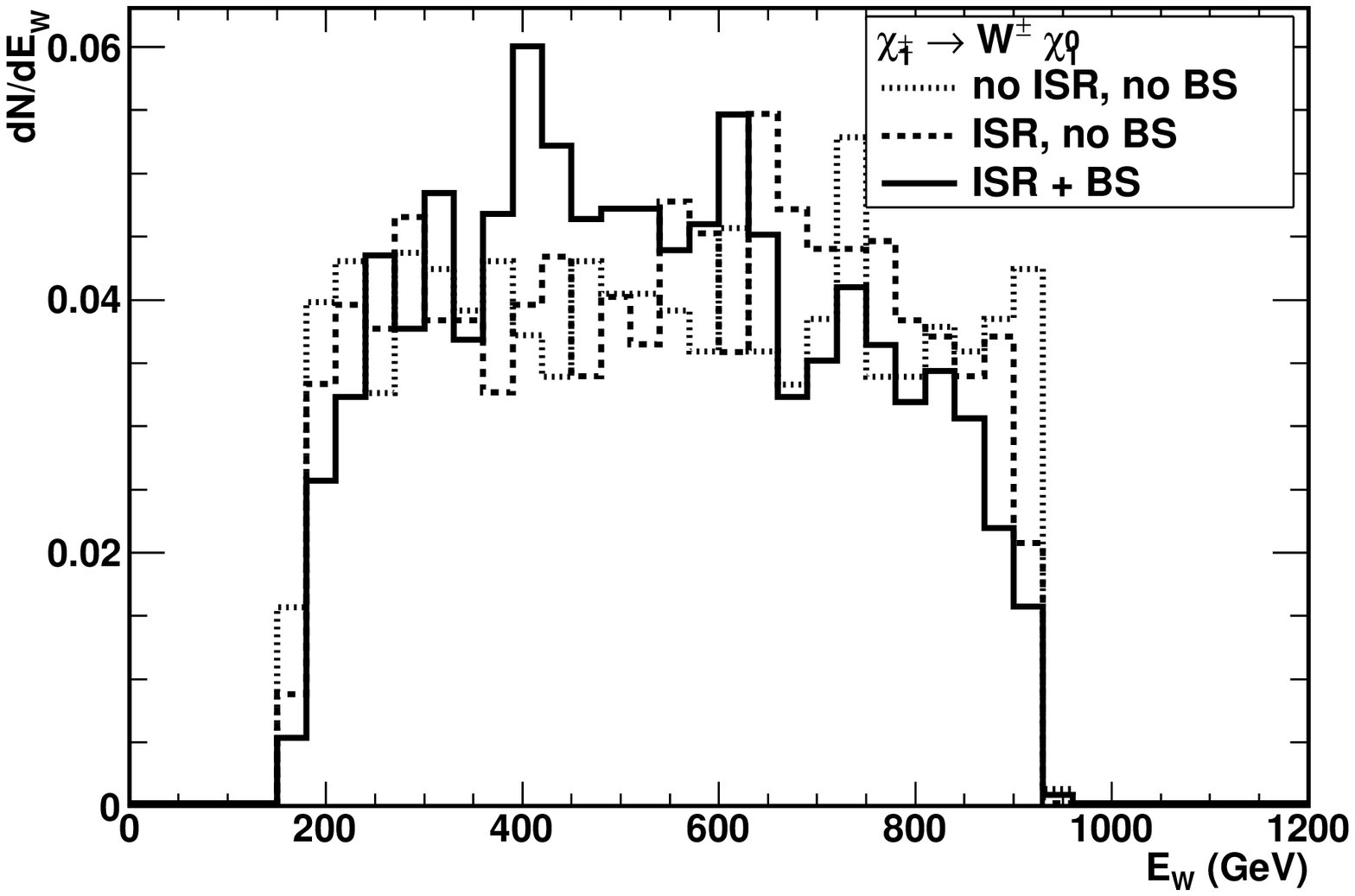}} \\
\end{tabular}
\end{center}
\caption{$W^{\pm}$ energy spectrum in the $\chi^{\pm}_1 \to W^{\pm} \chi^0_1$ process 
without radiation effects (dashed), with ISR only (dotted) and with both ISR and 
beamstrahlung (continuous) for (a) Model~I and (b) Model~II.}
\label{fig:comp}
\end{figure}
We model the ISR spectrum by an approximate solution to the Gribov-Lipatov equation, 
proposed in~\cite{Skrzypek:1990qs}. In the formula we leave free the $\eta$ parameter 
and the fraction of events off the full energy peak. We determine them by a fit to the 
ISR spectrum obtained for {\sc Pythia} signal events (see Figure~\ref{fig:isr_circe}). 
We study the accuracy on the mass measurements in some of the channels for no radiation 
effects, ISR only and ISR + BS. Figure~\ref{fig:comp} shows these effects on the $W^{\pm}$ 
energy spectrum for the $\chi^{\pm}_1 \to W^{\pm} \chi^0_1$ process.
In this study we consider only fully hadronic decays of bosons, since we need to reconstruct their
energy. The effect of the finite resolution in the determination of the boson energy is included by 
applying a Gaussian smearing to the energy of the partons produced in the boson decay. The smearing of 
the energy of the two partons is independent and we test the effect of various energy resolution values 
in the range 0 $< \delta E/E<$ 0.125. The energy smearing does not account for energy missing in neutrinos. 
This is particularly important in the reconstruction of $h^0 \to b \bar b$ decays, where either a $b$ or 
a $c$ hadron decays semi-leptonically.
 
\subsection{Analysis and Results}

We perform the study by selecting for each of the production and decay processes given above the
final state parton topologies and boson contents  which have the most favourable signal contribution.
We consider only fully hadronic final states and estimate the signal purity and the dominant SUSY 
background processes at generator level assuming perfect jet clustering and gauge boson identification.
The fraction of SUSY final states yielding a 2-$q$ topology is 4.7\% (2.3\%), a 4-$q$ is  13.3\% 
(1.8\%), a 6-$q$ is  3.3\% (0.5\%) and an 8-$q$  is 4.0\% (0.8\%) for model~I (II), respectively. 
In model~II, the branching fractions of decays into a $\tilde \tau^{\pm} \to \tau^{\pm} \chi^0_1$ are 
important, which explains the reduced rate of multi-quark final states. However, these modes are 
not considered here, since our study focuses on decays into bosons. 

We assume to operate the collider at 3~TeV for an integrated luminosity of 2~ab$^{-1}$ with 
unpolarised beams.
Mass fits are performed on samples of bosons in the selected topology populated with either signal only
or inclusive SUSY events and we study the evolution of the statistical accuracy on the masses with the 
smearing of the energy spectrum due to beam radiation and detector energy resolution effects. We estimate 
the change in signal purity with the parton energy resolution and the identification of the topology 
using only reconstructed quantities in section~3.4.

\subsubsection{$\chi^{\pm}_1 \to W^{\pm} \chi^0_1$}

\begin{figure}
\begin{center}
\begin{tabular}{cc}
\includegraphics[width=6.0cm,clip=]{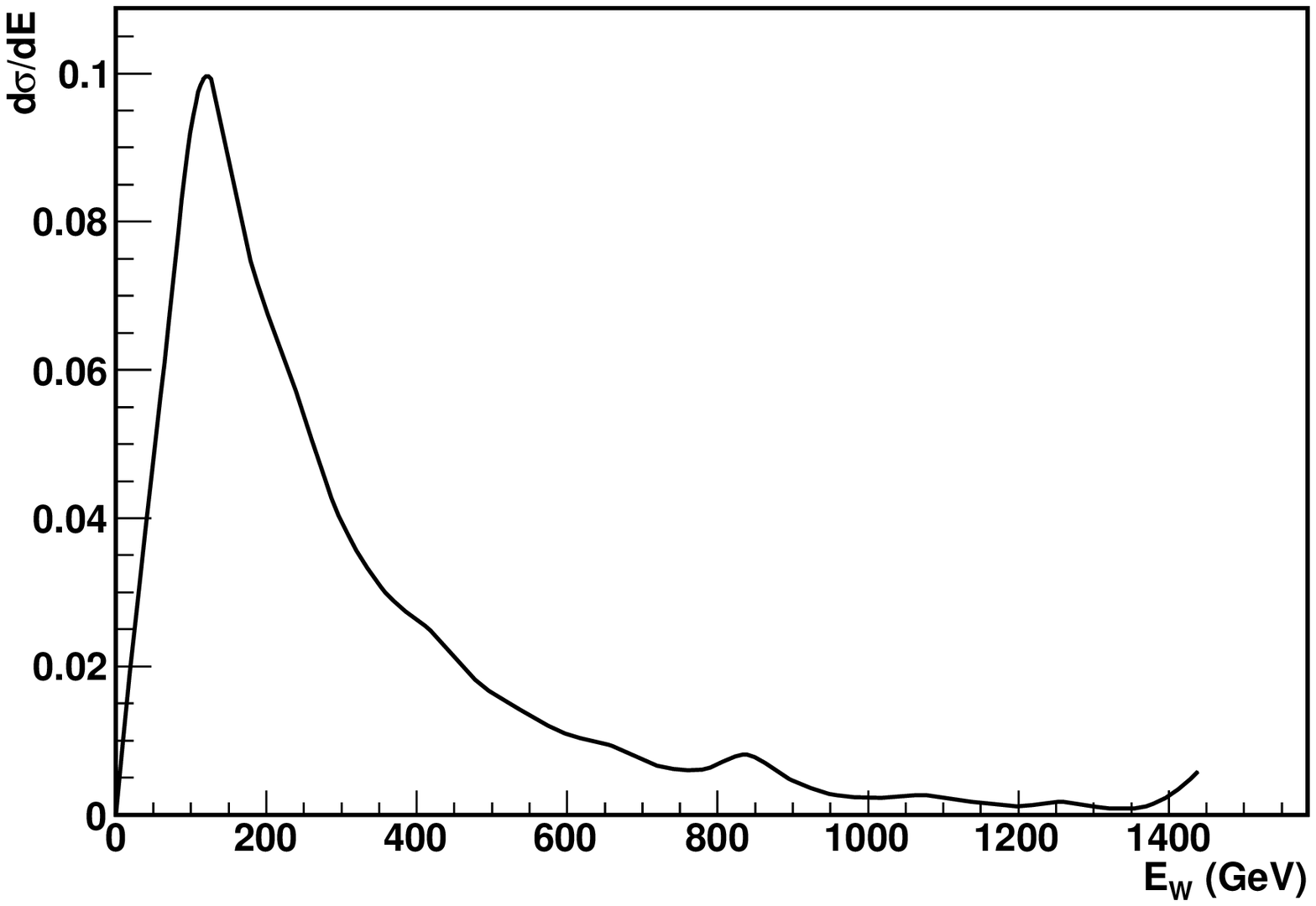} &
\includegraphics[width=6.0cm,clip=]{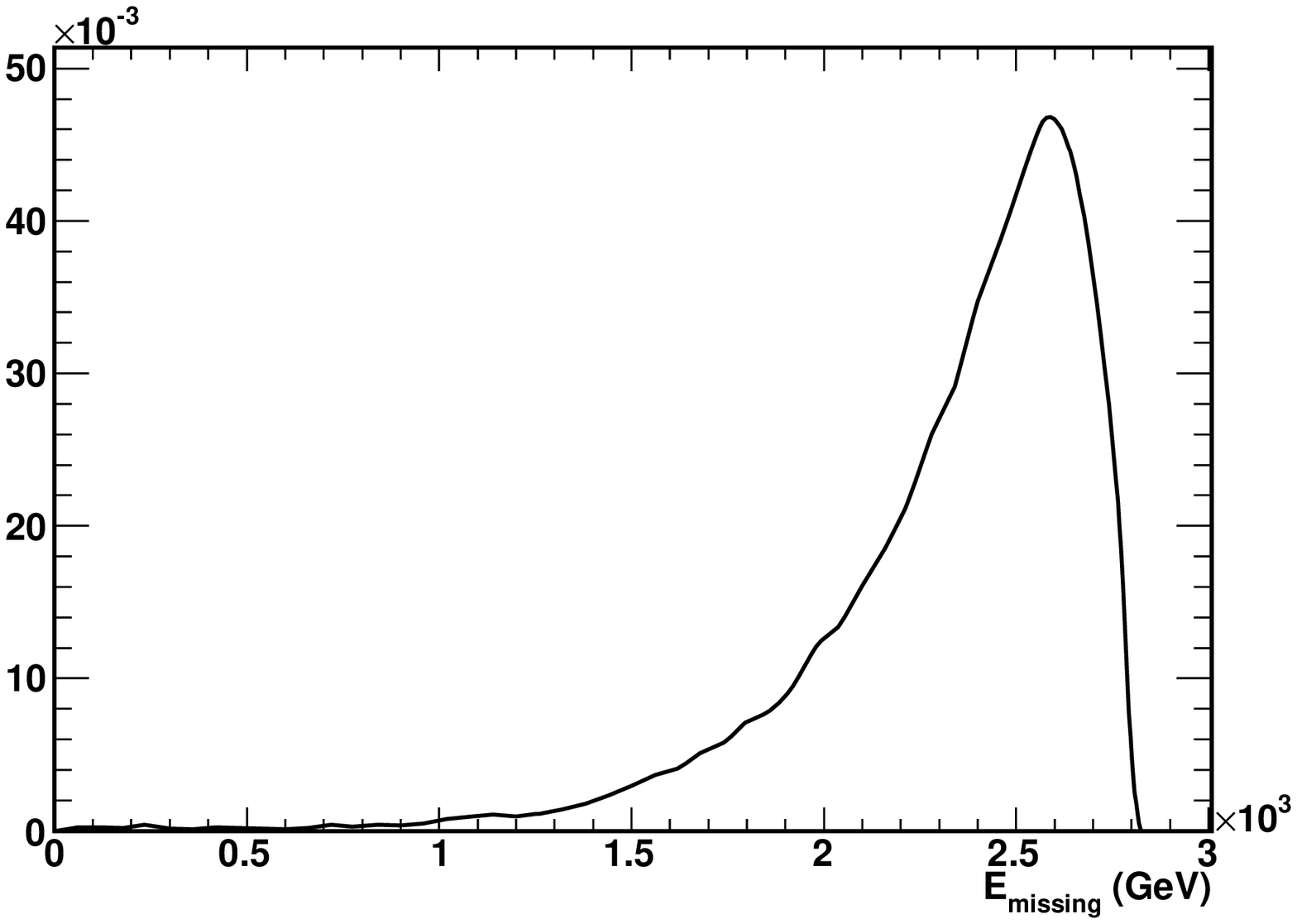} \\
\includegraphics[width=6.0cm,clip=]{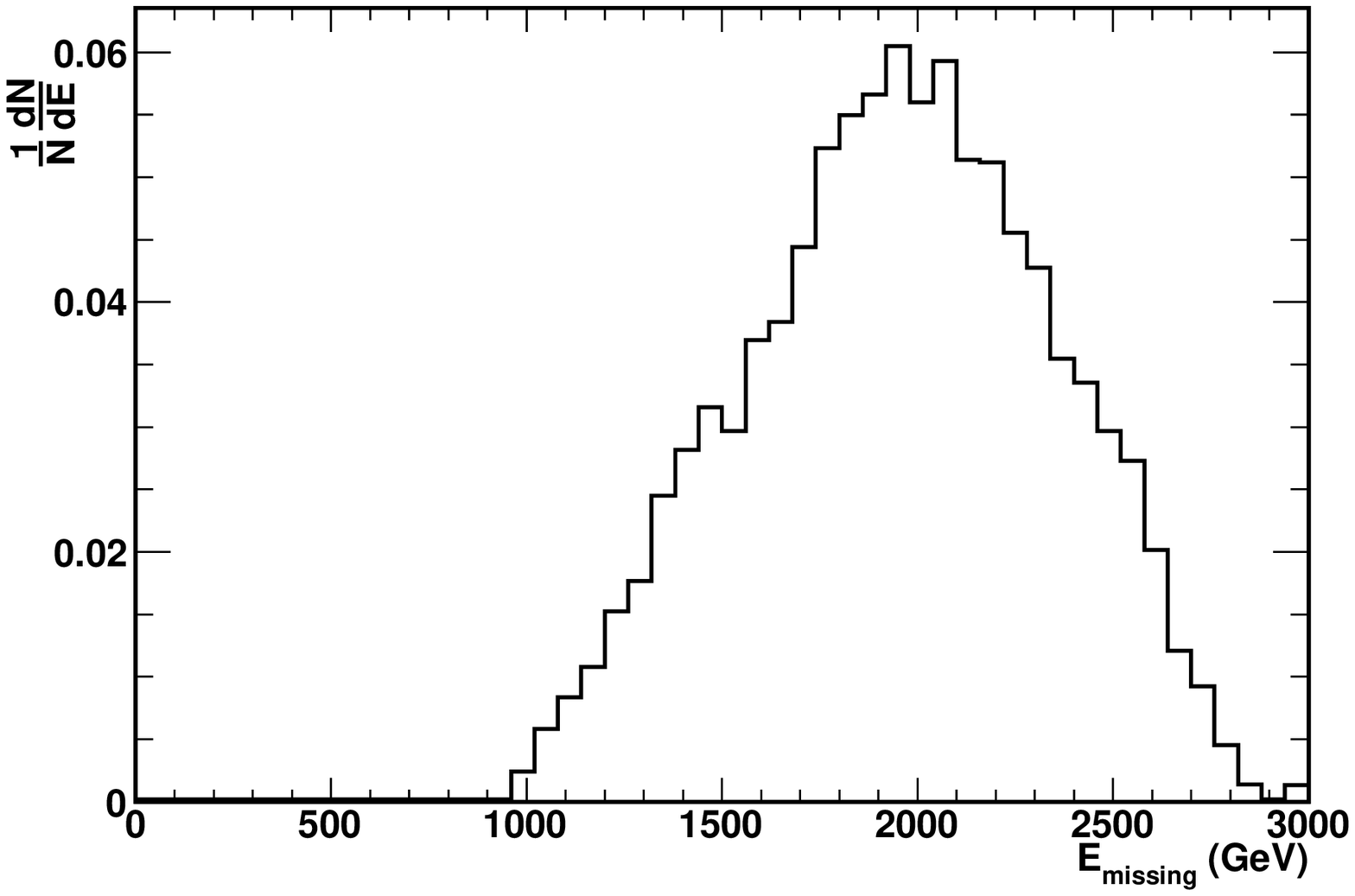} &
\includegraphics[width=6.0cm,clip=]{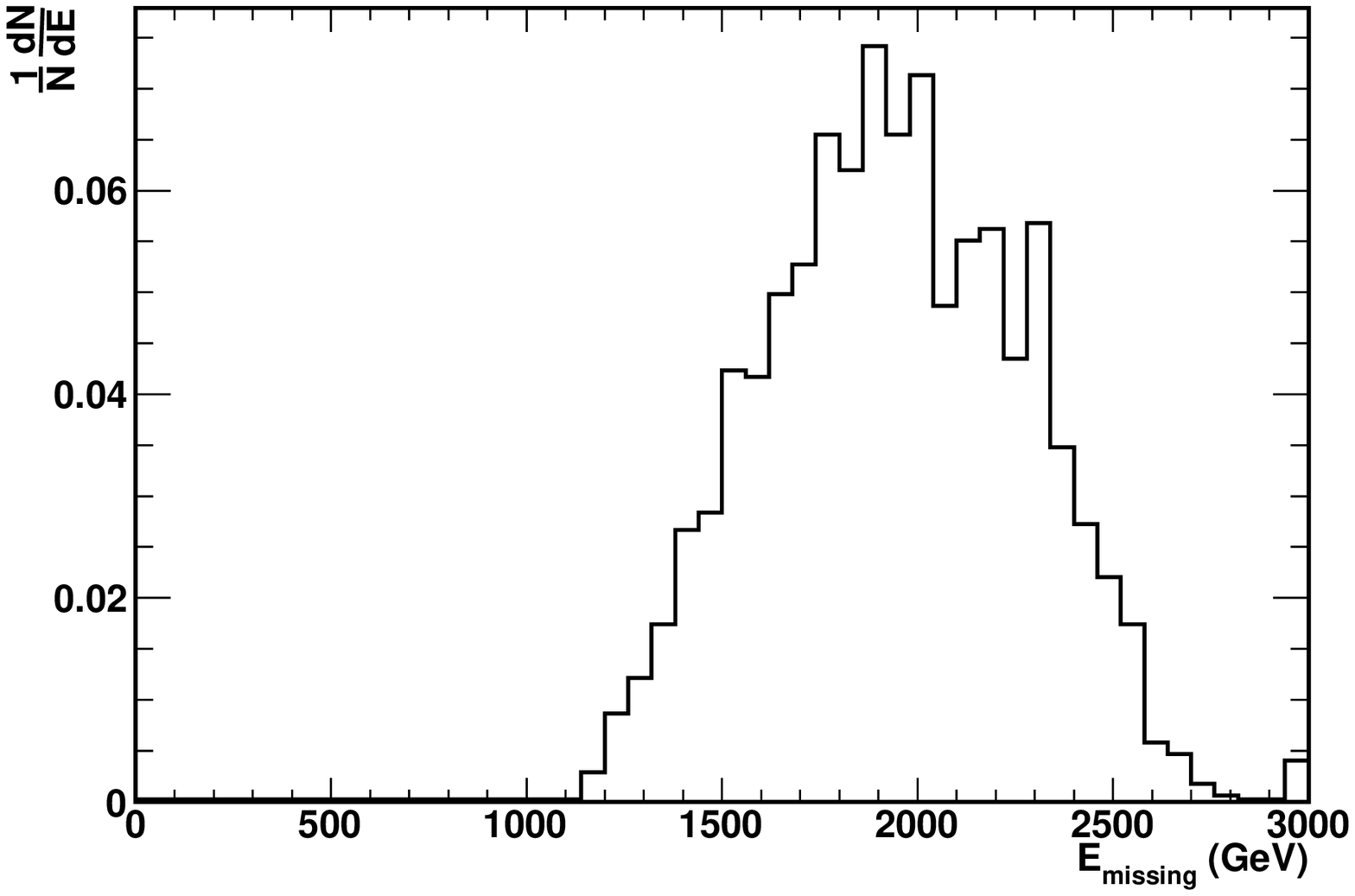} \\
\end{tabular}
\end{center}
\vspace*{-0.5cm}
\caption{Kinematics of $e^+e^- \to W^+W^- \nu_e \bar \nu_e$ and events with $|\cos \theta_{W}| < 0.85$ 
obtained with {\tt Comphep 4.5.1}: (upper left) $W^{\pm}$ energy spectrum and (upper right) missing energy 
spectrum. Missing energy distribution for signal $e^+e^- \to \chi^+_1 \chi^-_1$ events for (lower left) 
model~I  and (lower right) model~II.}
\label{fig:wwnn}
\end{figure}

The $e^+e^- \to \chi^+_1 \chi^-_1 \to W^+ \chi^0_1 W^- \chi^0_1$; $W \to q \bar q'$ 
process leads to a distinctive final state with four hadronic jets and missing energy. 
The SUSY background in model~I is almost entirely due to charginos produced through 
$e^+ e^- \to e^+_L e^-_L$, $e^{\pm}_L \to \chi^{\pm}_1 \nu_e$, which account for 12\% of 
the inclusive $WW$ + missing energy SUSY sample. Since the chargino energy in this 
process is lower than the beam energy the $W$ energy is correspondingly shifted to lower values.
In the case of model II, the signal $e^+e^- \to \chi^+_1 \chi^-_1$ accounts for 87\% of the 
$W^+W^-$ + missing energy final state with background contributions from $\chi^+_2 \chi^-_2$ 
and $\chi^+_2 \chi^-_1$ + c.c.
The main standard model irreducible background is due to $e^+e^- \to W^+W^- \nu_e \bar \nu_e$, 
which has a cross section of 124~fb. However, the $W$ production in this process is forward 
peaked while it is central in SUSY s-channel signal events. Requiring each $W$ boson to be 
produced within $|\cos \theta| < 0.85$, where $\theta$ is its polar angle reduces the $WW\nu\nu$ 
cross section to 28.7~fb. The energy distribution of $W^{\pm}$ bosons in $WW\nu\nu$ events within 
the angular acceptance, generated with {\tt Comphep 4.5.1}~\cite{Comphep}, is shown in 
Figure~\ref{fig:wwnn}. This background affects only the lower endpoint of the $W$ energy spectrum.
Further, the bulk of the signal SUSY events can be separated from this background
based on the event missing energy. 
The results of the 1-par fit of the $\chi^{\pm}_{1}$ mass performed for spectra with no 
radiation, only ISR and also beamstrahlung effects are given in Table~\ref{tab:chi1pm}.
\begin{figure}[hb!]
\begin{center}
\begin{tabular}{ccc}
\subfloat[no Rad.]{\includegraphics[width=4.5cm,clip=]{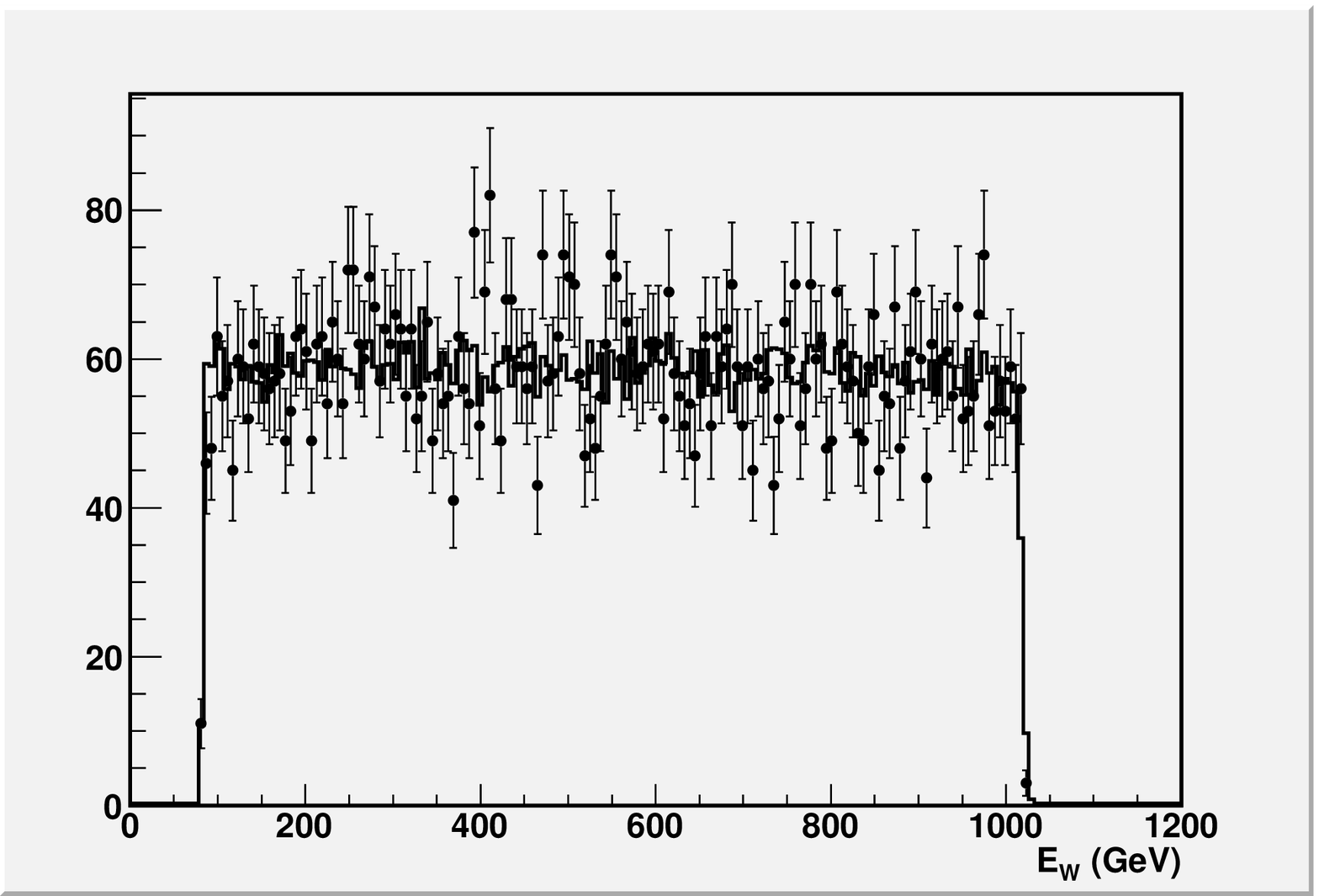}}  & 
\subfloat[ISR Only]{\includegraphics[width=4.5cm,clip=]{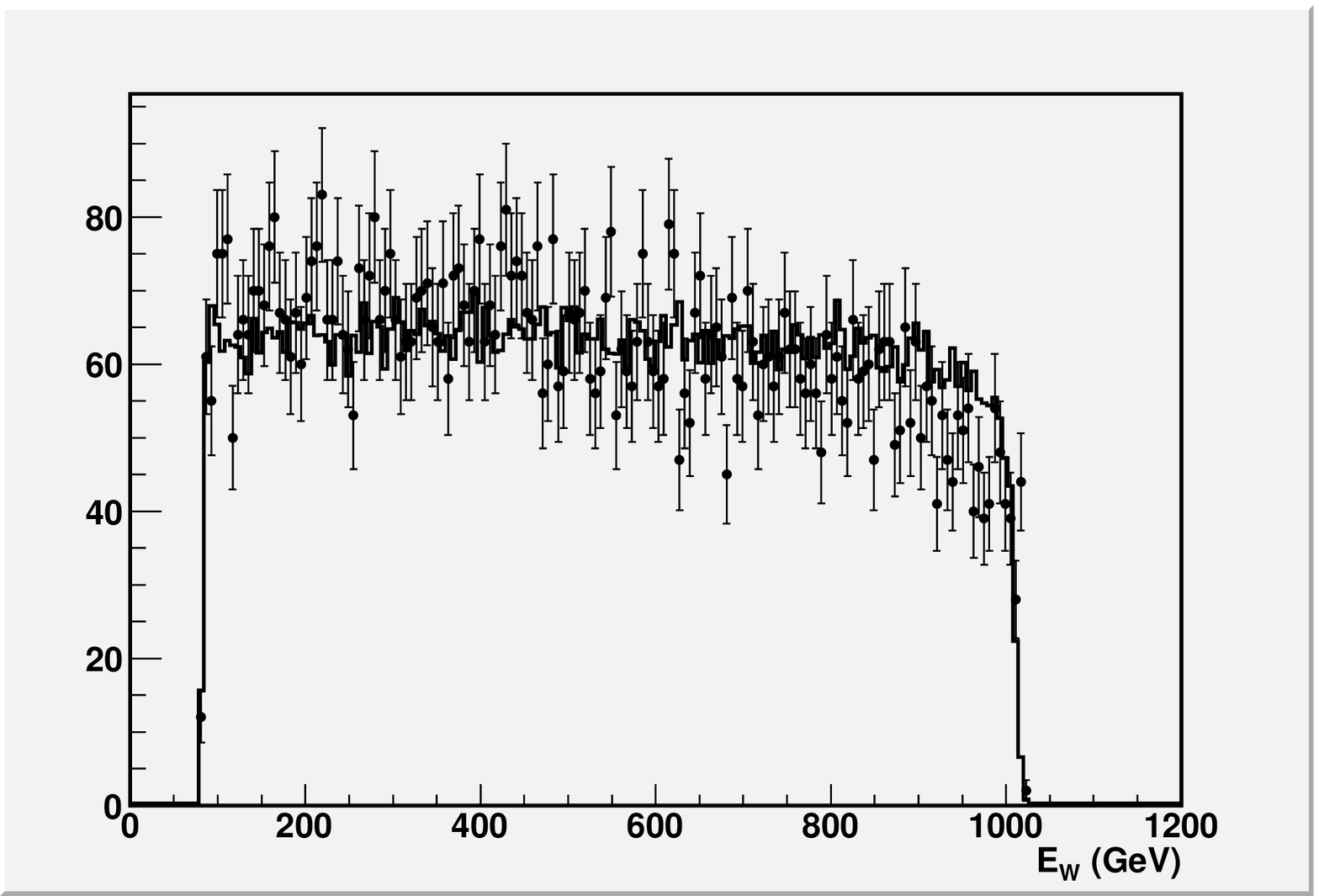}}  & 
\subfloat[ISR + BS]{\includegraphics[width=4.5cm,clip=]{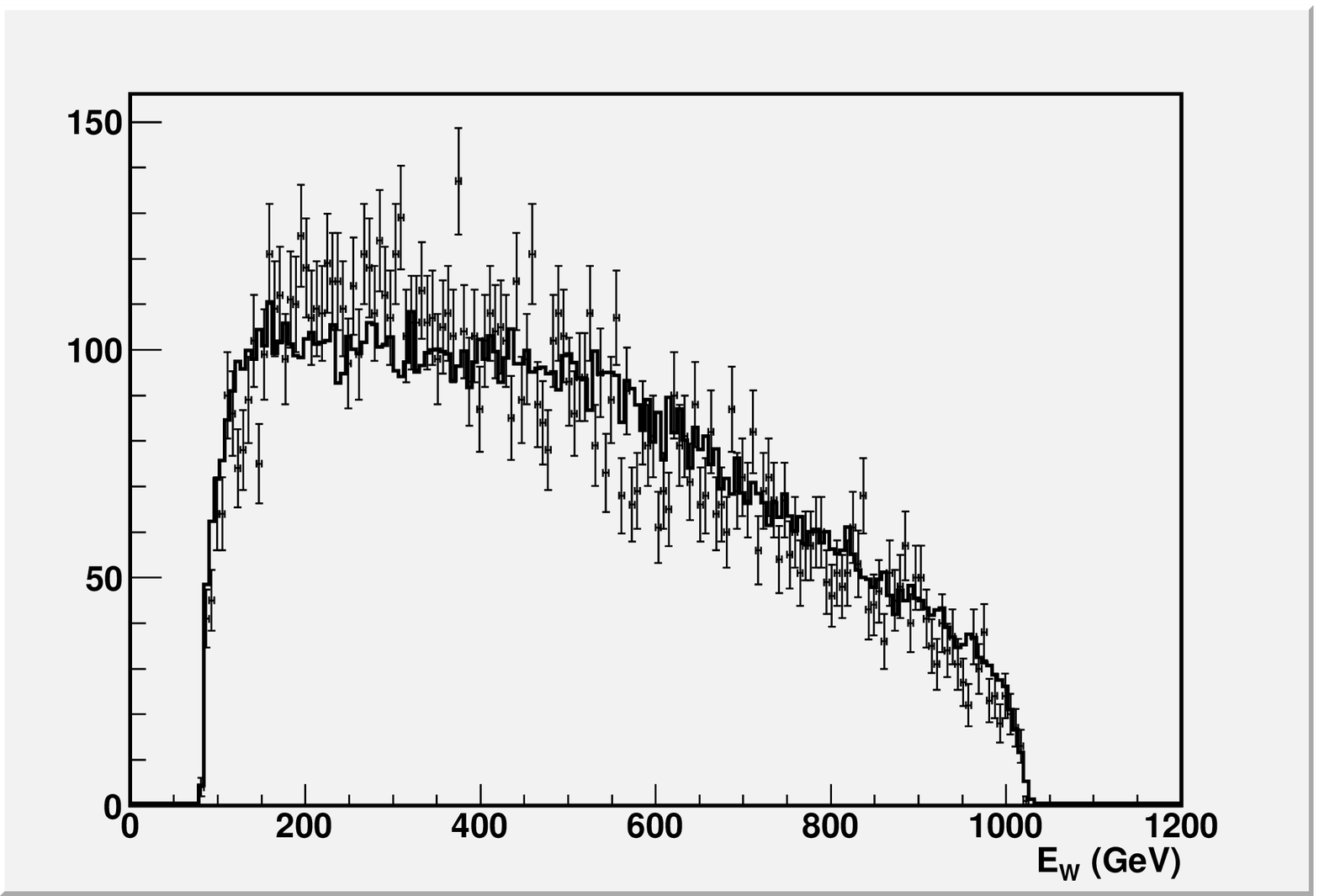}}  \\
\end{tabular}
\end{center}
\vspace*{-0.5cm}
\caption{$W^{\pm}$ energy spectrum in the four-jet topology, 2~$W$ + missing energy final state 
for $\chi^+_1 \chi^-_1$ signal events in model~I at 3~TeV with (a) no radiation. (b) only ISR and 
(c) ISR+BS. The points with error bars are the simulation and the line the fitted spectrum.}
\label{fig:efitchi1pm}
\end{figure}
Here, we keep the lightest neutralino mass, $M_{\chi^0_1}$, fixed to its model value, since 
it should be independently determined in the 
$e^+e^- \to \tilde \mu^+_R \tilde \mu^-_R \to \mu^+ \mu^- \chi^0_1 \chi^0_1$ process and the 
corresponding $e^+e^- \to \tilde e^+_R \tilde e^-_R$ to an accuracy of $\pm$1.0~GeV~\cite{smuons}. 
The fitted spectra are shown in Figure~\ref{fig:efitchi1pm}.
We estimate the systematic uncertainty by varying $M_{\chi^0_1}$ within this range and repeating 
the fit, it amounts to $\pm$2~GeV on the determination of $M_{\chi^{\pm}_1}$. 
\begin{figure}
\begin{center}
\begin{tabular}{cc}
\subfloat[ISR Only]{\includegraphics[width=7.0cm]{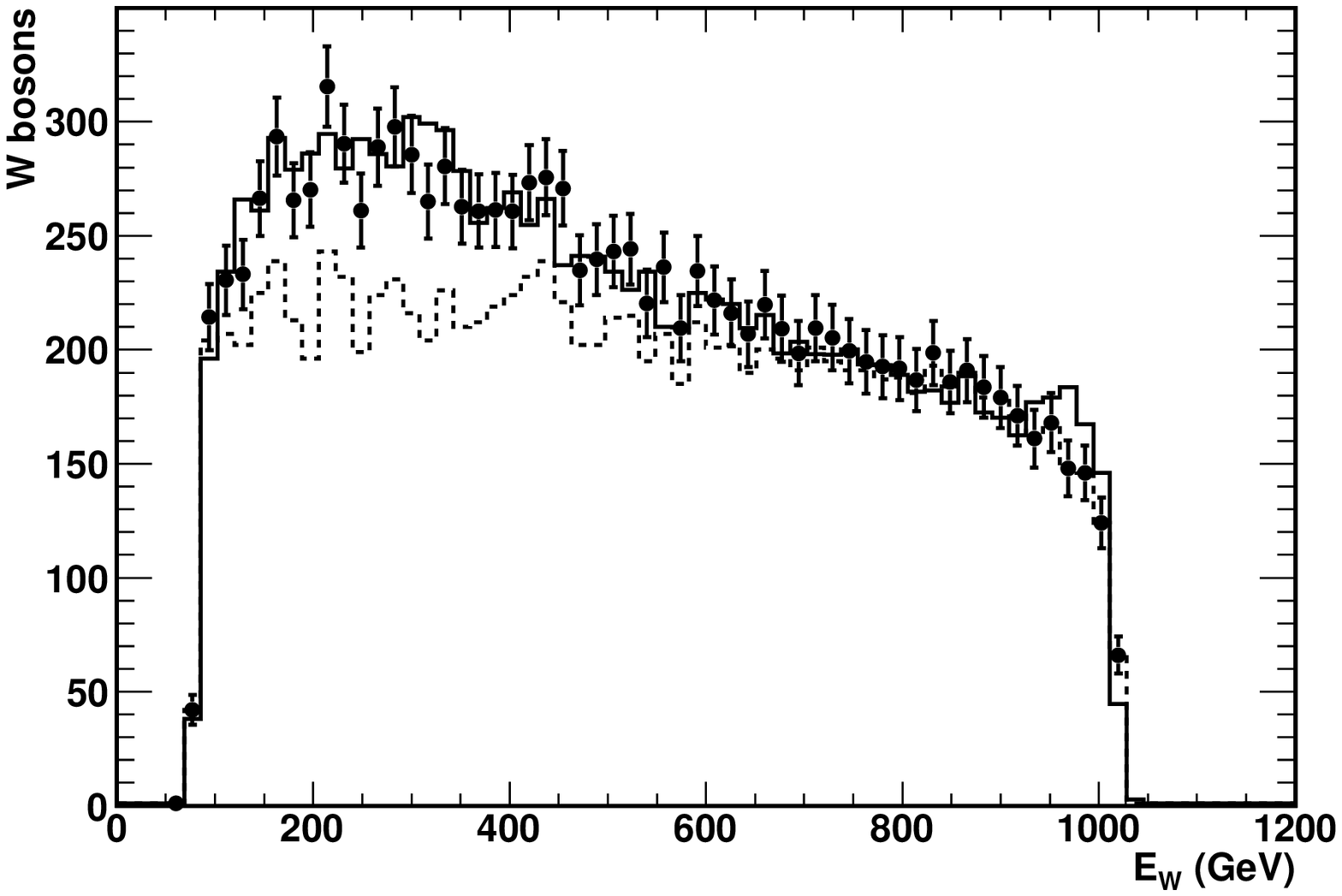}}  & 
\subfloat[ISR + BS]{\includegraphics[width=7.0cm]{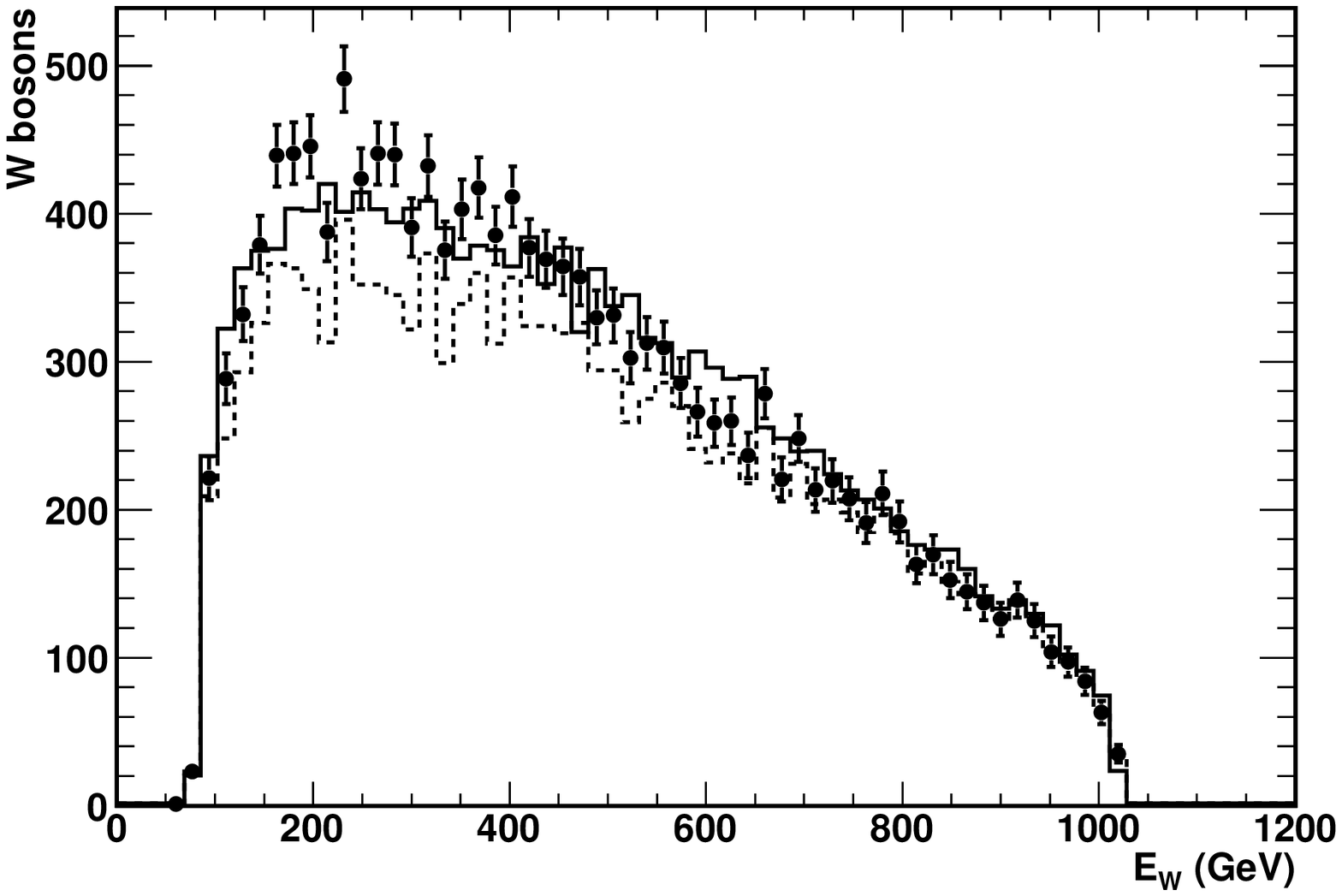}}  \\
\end{tabular}
\end{center}
\vspace*{-0.5cm}
\caption{$W^{\pm}$ energy spectrum in the four jet topology, 2~$W$ + missing energy final state 
for inclusive SUSY events in model~I at 3~TeV with only ISR (a) and ISR+BS (b). The points 
with error bars are the simulation, the continuous line the fitted spectrum and the dashed line 
the fitted contribution from $\chi^+_1 \chi^-_1$ signal events.}
\label{fig:efitchi1pmeL}
\end{figure}

\begin{table}
\caption{Statistical accuracy on $\chi^{\pm}_1$ mass from a fit to the $E_{W^{\pm}}$ endpoints 
 for $\chi^+_1 \chi^-_1$ signal events and 2~ab$^{-1}$ of integrated luminosity at 3~TeV, under 
 different assumptions}
\begin{center}
\begin{tabular}{|l|c|ccccc|}
\hline
Particle        & Mass  & No Rad       & ISR      & ISR+BS            &  ISR+BS   &  ISR+BS    \\
                & (GeV) &            &            & $\delta E/E$=0    &  =0.025   &  =0.05  \\
\hline
Model~I         &       &           &             &            &      &   \\
$\chi^{\pm}_1$  & 643.2 & $\pm$~0.91 & $\pm$~1.39 & $\pm$~2.09 &     $\pm$~2.89 &     $\pm$~3.60 \\
\hline
Model~II        &       &           &             &            &      &   \\
$\chi^{\pm}_1$  &1062.2 & $\pm$~6.10   & $\pm$~8.25 & $\pm$10.11 &  $\pm$11.0 &  $\pm$12.0\\
\hline
\end{tabular}
\end{center}
\label{tab:chi1pm}
\end{table}
The relative mass accuracy for the two models scales with production cross section as expected.
We notice that the deterioration of the mass accuracy due to BS is larger for model~I (+50.4\%), 
where the $\chi^{\pm}_1$ mass is significantly smaller compared to the beam energy, compared 
to model~II (+28.4\%), since the   $\chi^{+}_1 \chi^{-}_1$ threshold sits close to $\sqrt{s}$
and the BS effects are dumped by the fall of the production cross section near threshold. 
\begin{table}
\caption{Statistical accuracy on $\chi^{\pm}_1$ mass from a fit to the $E_{W^{\pm}}$ endpoints 
 for 2~$W$ + missing energy inclusive SUSY events and 2~ab$^{-1}$ of integrated luminosity at 3~TeV, 
 under different assumptions}
\begin{center}
\begin{tabular}{|l|c|ccccc|}
\hline
Particle        & Mass  & No Rad       & ISR      & ISR+BS         & ISR+BS       & ISR+BS      \\
                & (GeV) &            &            & $\delta E/E$=0 & =0.025       & =0.050      \\   
\hline
Model~I         &        &           &            &                &              &   \\
\hline
2-par Fit       &        &           &            &                &              &   \\
$\chi^{\pm}_1$  & ~643.2 & $\pm$~1.4 & $\pm$~1.7  & $\pm$~2.6      &  $\pm$~3.8   & $\pm$~4.1 \\
\hline
3-par Fit       &        &           &            &                &              &   \\
$\chi^{\pm}_1$  & ~643.2 & $\pm$~2.0 & $\pm$~2.3  & $\pm$~3.5      &  $\pm$~4.8   & $\pm$~5.2  \\
$\tilde e_L$    & 1102.2 & $\pm$48   & $\pm$50    & $\pm$56        &  $\pm$~63   & $\pm$66  \\
\hline
\end{tabular}
\end{center}
\label{tab:chi1pmeL}
\end{table}
Then, we repeat the fit accounting for both the direct chargino production contribution to 
the $W$ energy spectrum and that from $\tilde e_L^{\pm} \to \chi^{\pm} \nu_e$ decays. These decays 
still offer some sensitivity to the chargino mass but cannot be distinguished from chargino 
pair production in the event selection. At this specific benchmark point the selectron left 
mass $M_{e_L}$ cannot be measured with the electron energy spectrum in the 
$e^+ \chi^0_1 e^- \chi^0_1$ final state since the spectrum is dominated by the 
$\tilde e_R \tilde e_R$ process. 
It could be measured using the mixed $e^+ \chi^0_1 h^0 \chi^0_1 h^0$ and possibly 
$e^{\pm} \chi^0_1 W^{\mp} \nu \chi^0_1$ modes at 3~TeV and in a threshold scan. We perform 
fits where  $M_{\tilde e_L}$ is either kept at its model value or it is treated as a 
free parameter, while the $\chi^{\pm}_1$ mass and the relative contribution of the two 
processes to the $W$ spectrum are kept free. Results are summarised in Table~\ref{tab:chi1pmeL}
and Figure~\ref{fig:efitchi1pmeL}.

\subsubsection{$\chi^0_2 \to h^0 \chi^0_1$}

The $\chi^0_2 \to h^0 \chi^0_1$ decay of the chargino pair yields two Higgs bosons plus missing 
energy. The same final states originates also from 
$e^+e^- \to \tilde \nu_{\ell} \tilde \nu_{\ell} \to \chi^0_2 \nu_{\ell} \chi^0_2 \nu_{\ell}$. 
For the parameters of model~I $M_{\tilde \nu_{\ell}}$ = 1097.2~GeV, the 
$e^+e^- \to \tilde \nu_{\ell} \tilde \nu_{\ell}$ production cross section is 14.5~fb.
\begin{table}[hb!]
\caption{Statistical accuracy on $\chi^{0}_{2}$ mass from a fit to the $E_{h^{0}}$ endpoints 
 for $\chi^0_2 \chi^0_2$ signal events and 2~ab$^{-1}$ of integrated luminosity at 3~TeV, under 
 different assumptions}
\begin{center}
\begin{tabular}{|l|c|ccccc|}
\hline
Particle        & Mass   & No Rad     & ISR        & ISR+BS       & ISR+BS    & ISR+BS   \\
                & (GeV)  &            &            & $\delta E/E$=0  &  =0.025  & =0.05 \\
\hline
Model~I         &        &            &            &      &    &       \\
$\chi^{0}_2$  & 643.2    & $\pm$~1.01 & $\pm$~1.17 & $\pm$~2.58 &  $\pm$~3.59 & $\pm$~4.54 \\
\hline
Model~II        &        &            &            &      &    &       \\
$\chi^{0}_2$    & 1064.2 & $\pm$10.64 & $\pm$11.12 & $\pm$16.71 & $\pm$19.04 & $\pm$23.42  \\
\hline
\end{tabular}
\end{center}
\label{tab:chi20}
\end{table}
\begin{figure}[ht!]
\begin{center}
\begin{tabular}{cc}
\subfloat[ISR Only]{\includegraphics[width=6.5cm]{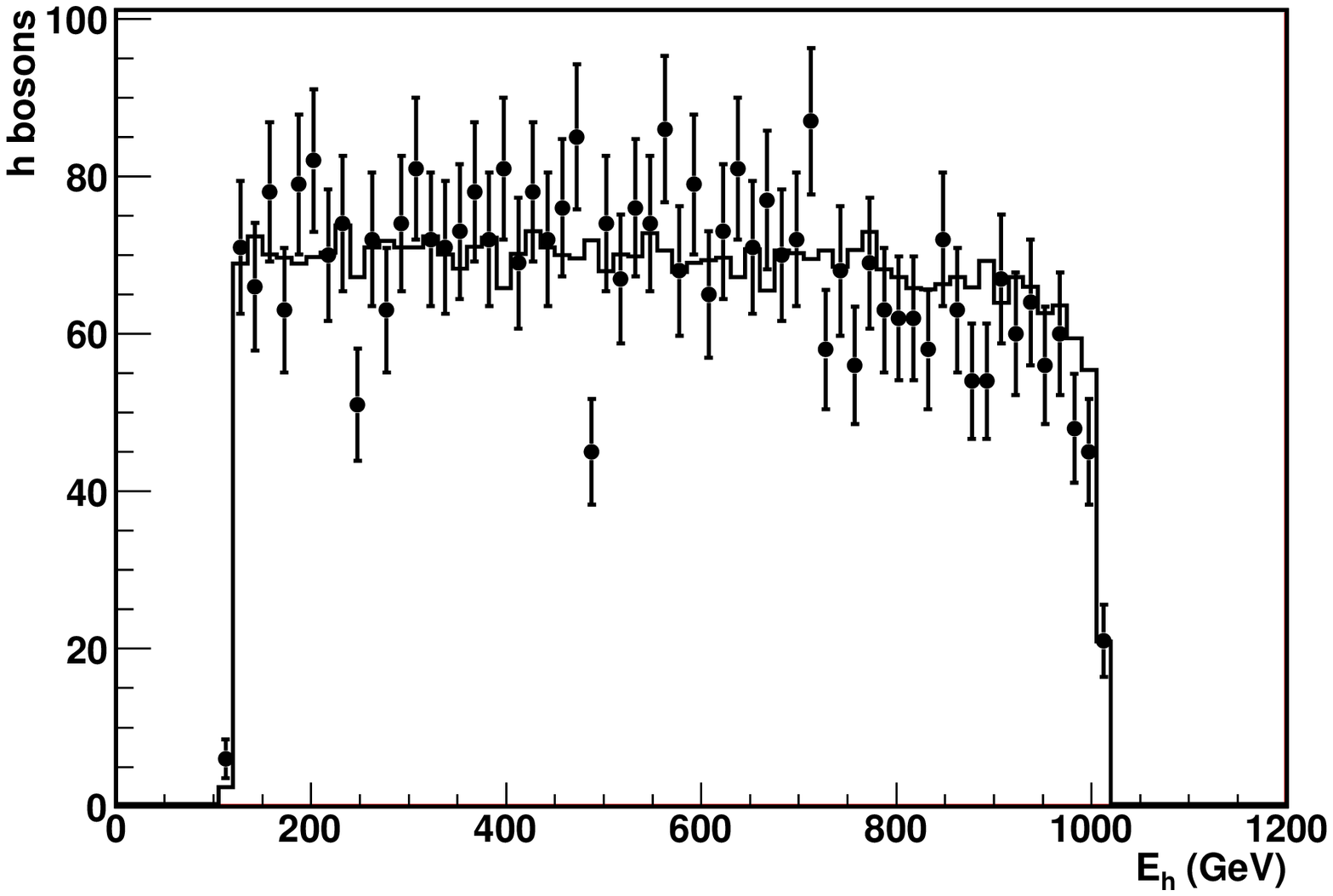}}  & 
\subfloat[ISR + BS]{\includegraphics[width=6.5cm]{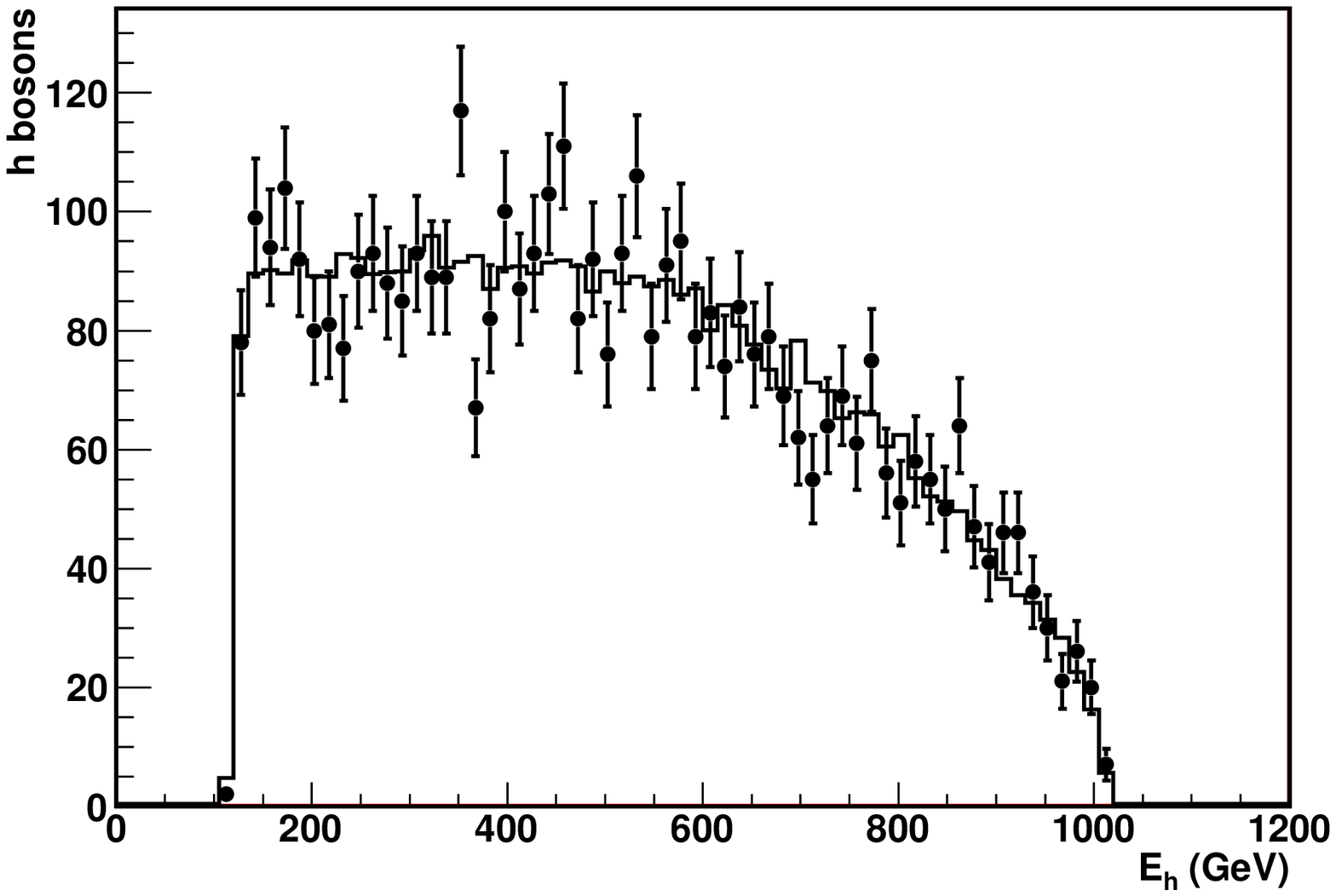}}  \\
\end{tabular}
\end{center}
\vspace*{-0.5cm}
\caption{$h^{0}$ energy spectrum in the 4-jet, 2~$h$ + missing energy final state for 
$\chi^0_2 \chi^0_2$ signal events in model~I at 3~TeV with only ISR (left) and ISR+BS (right). The 
points with error bars are the simulation and the continuous line the fitted spectrum.}
\label{fig:efitchi20}
\end{figure}
\begin{table}
\caption{Statistical accuracy on $\chi^{0}_2$ mass from a fit to the $E_{W^{\pm}}$ endpoints 
 for 2~$h$ + missing energy inclusive SUSY events and 2~ab$^{-1}$ of integrated luminosity at 
 3~TeV, under different assumptions}
\begin{center}
\begin{tabular}{|l|c|ccccc|}
\hline
Particle            & Mass  & No Rad      & ISR       & ISR+BS       & ISR+BS    & ISR+BS  \\
                    & (GeV) &             &           &  $\delta E/E$=0 & =0.025 & =0.050     \\
\hline
Model~I             &        &            &            &             &             &   \\
$\chi^{0}_2$        & ~643.2 & $\pm$ ~2.5 & $\pm$ ~3.2 & $\pm$ ~5.3  &  $\pm$ ~6.8 &  $\pm$ ~8.4  \\
$\tilde \nu_{\ell}$ & 1097.2 & $\pm$ 43.4 & $\pm$ 51.8 & $\pm$ 52.7  &   $\pm$ 60.4 &  $\pm$ 69.3 \\
\hline
\end{tabular}
\end{center}
\label{tab:chi20sn}
\end{table}
\begin{figure}[h!]
\begin{center}
\begin{tabular}{cc}
\subfloat[ISR Only]{\includegraphics[width=6.5cm]{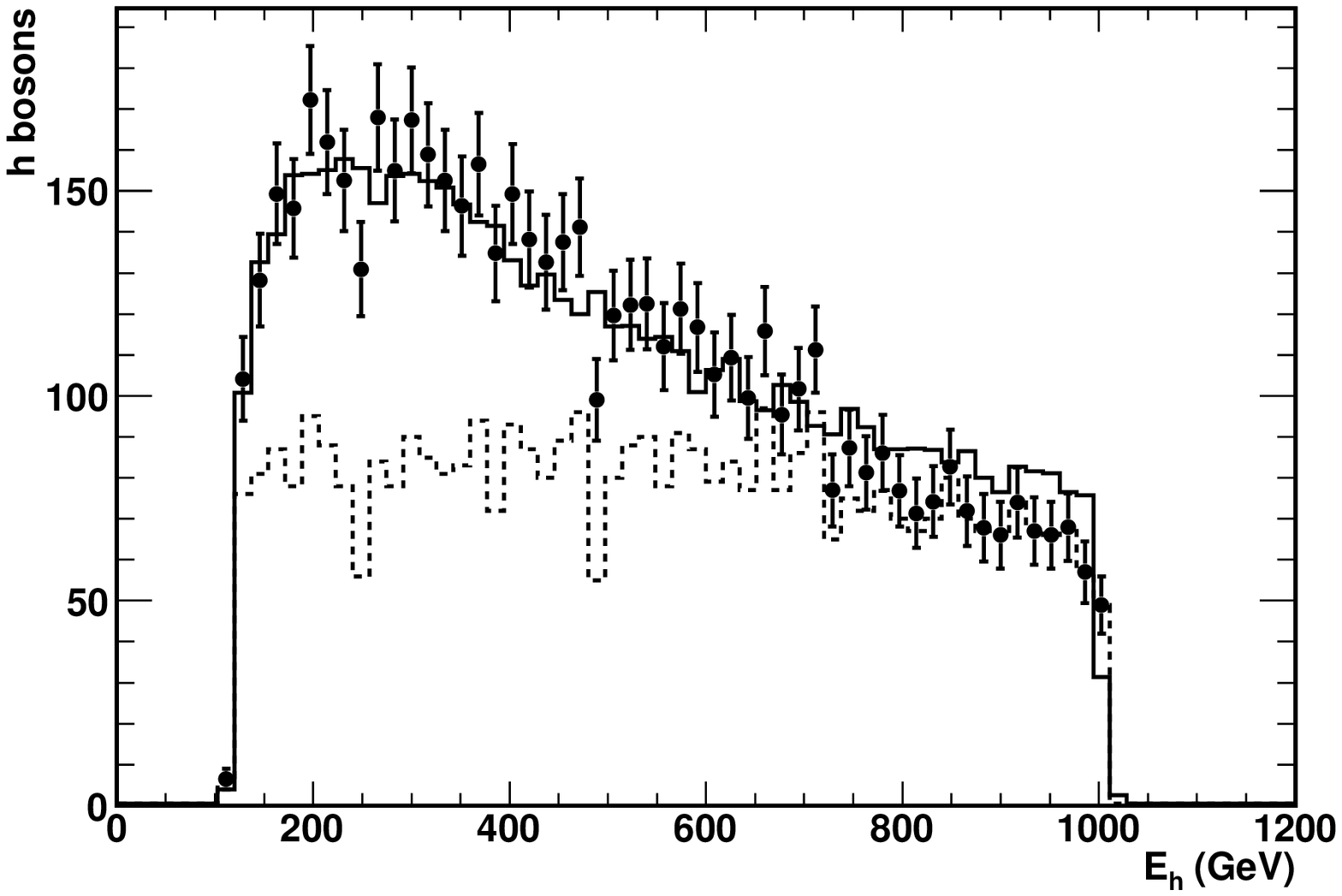}}  & 
\subfloat[ISR + BS]{\includegraphics[width=6.5cm]{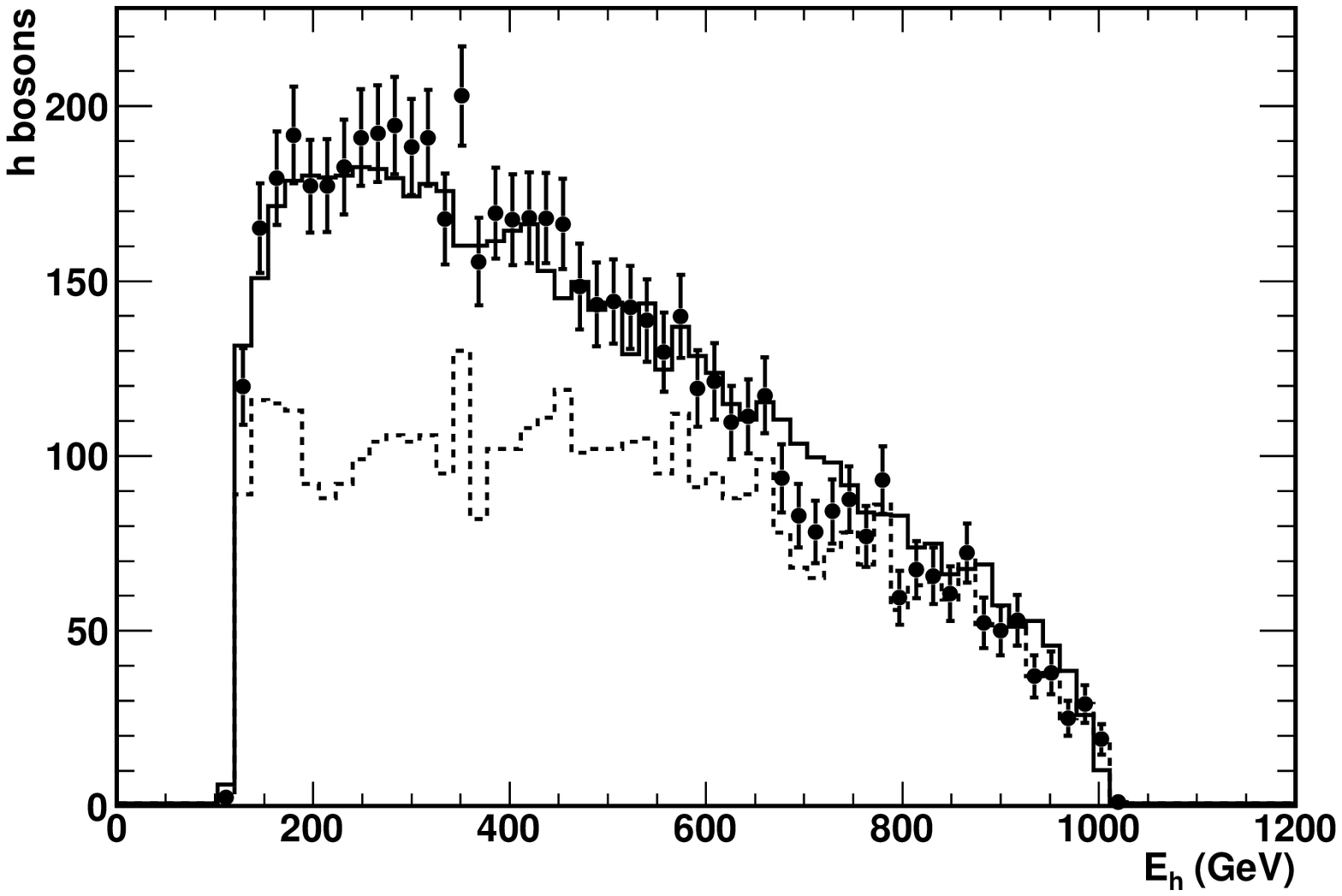}}  \\
\end{tabular}
\end{center}
\vspace*{-0.5cm}
\caption{$h^{0}$ energy spectrum in the 4-jet, 2~$h$ + missing energy final state for 
inclusive SUSY events in model~I at 3~TeV with only ISR (left) and ISR+BS (right). The points with 
error bars are the simulation, the continuous line the fitted spectrum and the dashed line the fitted 
contribution from $\chi^0_2 \chi^0_2$ events.}
\label{fig:efitchi20Ssn}
\end{figure}
The sneutrino channel accounts for 68\% of the 4-jet, $hh$ + missing energy inclusive SUSY sample. 
In model II, the tree-level sneutrino production cross section is 6~fb which drops to 2.3~fb 
accounting for ISR and BS and the signal purity is 89\% .
This final state is expected to be virtually immune from irreducible SM backgrounds, since the
cross section for double WW fusion light Higgs production, $e^+e^- \to h^0 h^0 \nu_e \bar \nu_e$ 
is only $\simeq$1~fb for 115~$<M_h<$~130~GeV~\cite{ghhh}. Fit results on signal events are 
summarised in Table~\ref{tab:chi20} and the fitted spectra given in Figure~\ref{fig:efitchi20}. 
Again, the relative increase of the statistical uncertainty due to BS is larger for model~I 
compared to model~II. 
We repeat the fit accounting for both the direct $\chi^0_2$ production and the irreducible  
$\tilde \nu_{\ell}$ SUSY background contribution to the $W$ energy spectrum for model~I where its
contribution is more important. Since the sneutrino decays through a $\chi^0_2$, the $h^0$ energy  
spectrum retains some sensitivity to the $\chi^0_2$ mass. A multi-parameter fit with the $\chi^0_2$ 
and $\tilde \nu_{\ell}$ masses treated as free parameters, together with the relative contribution 
of the two processes to the $h^0$ spectrum is performed (see Figure~\ref{fig:efitchi20Ssn}). 
Results are summarised in Table~\ref{tab:chi20sn}.

\subsubsection{$\chi^{\pm}_2 \to W^{\pm} \chi^{0}_1$, $\chi^{\pm}_2 \to W^{\pm} \chi^{0}_2$, 
$\chi^{\pm}_2 \to h^0 \chi^{\pm}_1$}

The $\chi^{\pm}_2$ chargino has one of largest production cross section for supersymmetric particles.
Contrary to the lighter states discussed above, there is no dominant decay channel and the analysis 
has to include several final states. We consider here $W^{\pm}$, $Z^0$ and $h^0$ spectra from combinations 
of decays with with both 6- and 8-parton final states. The channels we have considered are not 
exhaustive but are representatives of the topologies offered by decays of $\chi^+_2 \chi^-_2$ pairs.
\begin{figure}[hb!]
\begin{center}
\begin{tabular}{cc}
\subfloat[6-jet $WWh$]{\includegraphics[width=6.5cm]{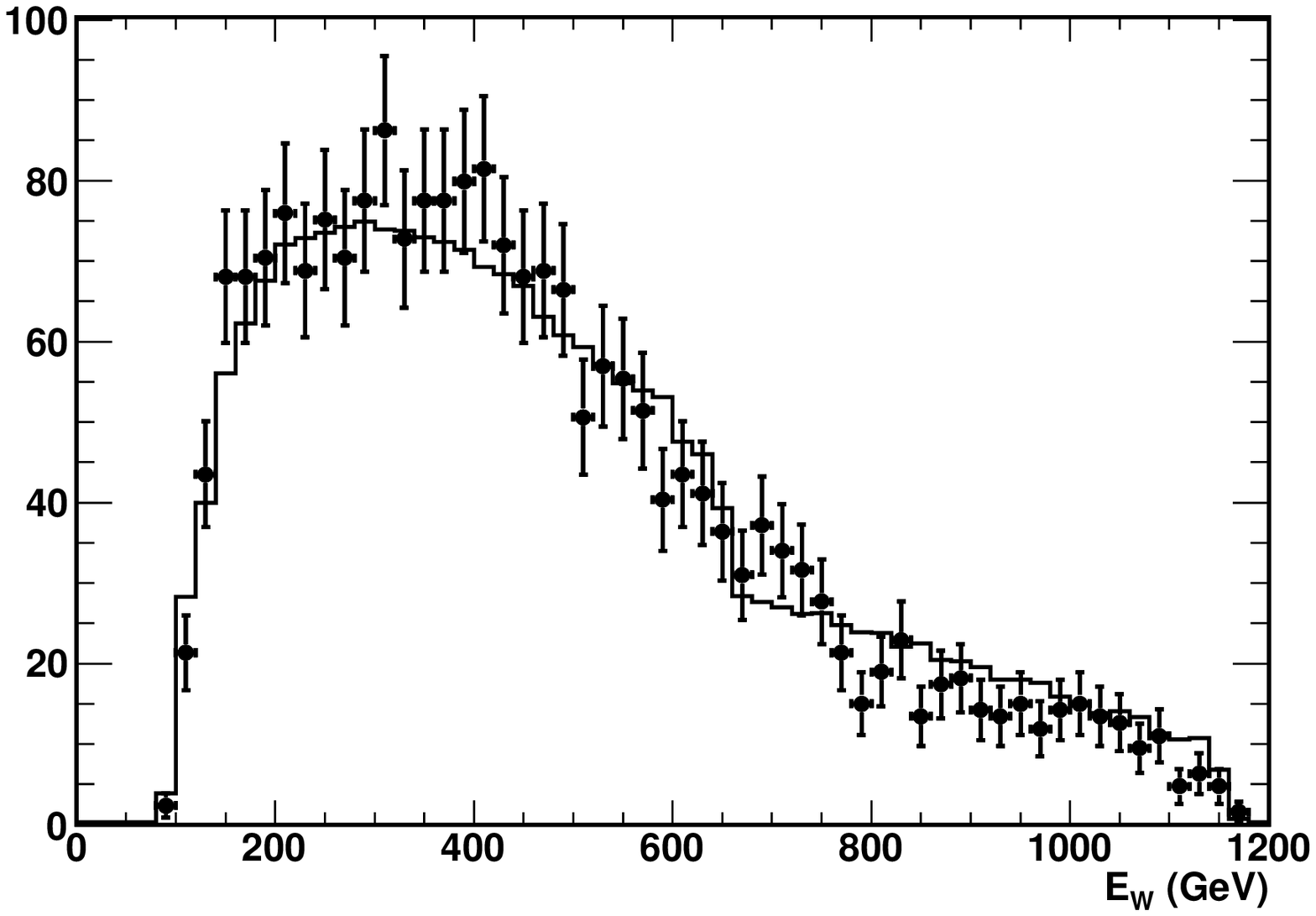}}  &  
\subfloat[6-jet $WWh$]{\includegraphics[width=6.5cm]{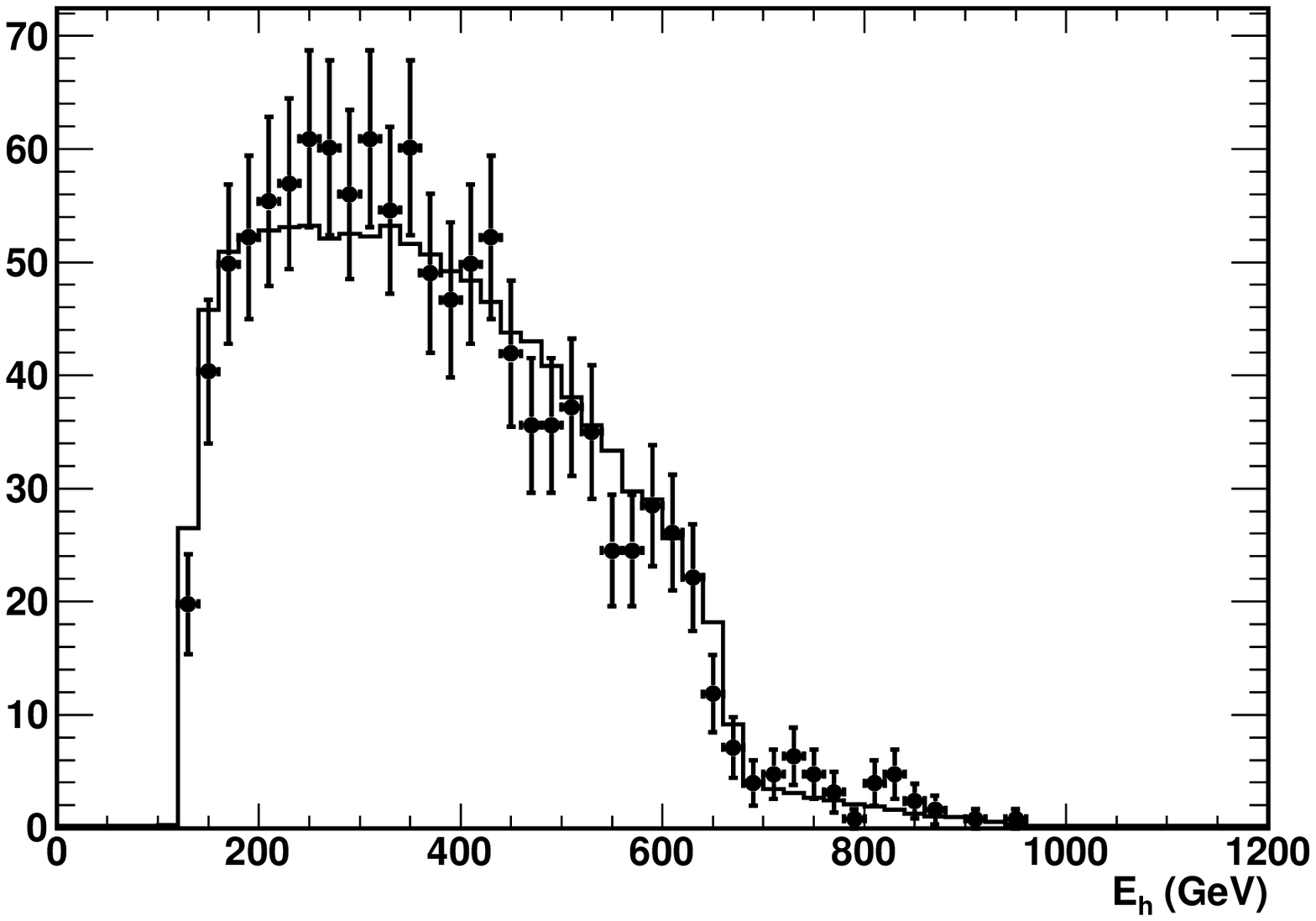}}  \\
\subfloat[8-jet $WWZh$]{\includegraphics[width=6.5cm]{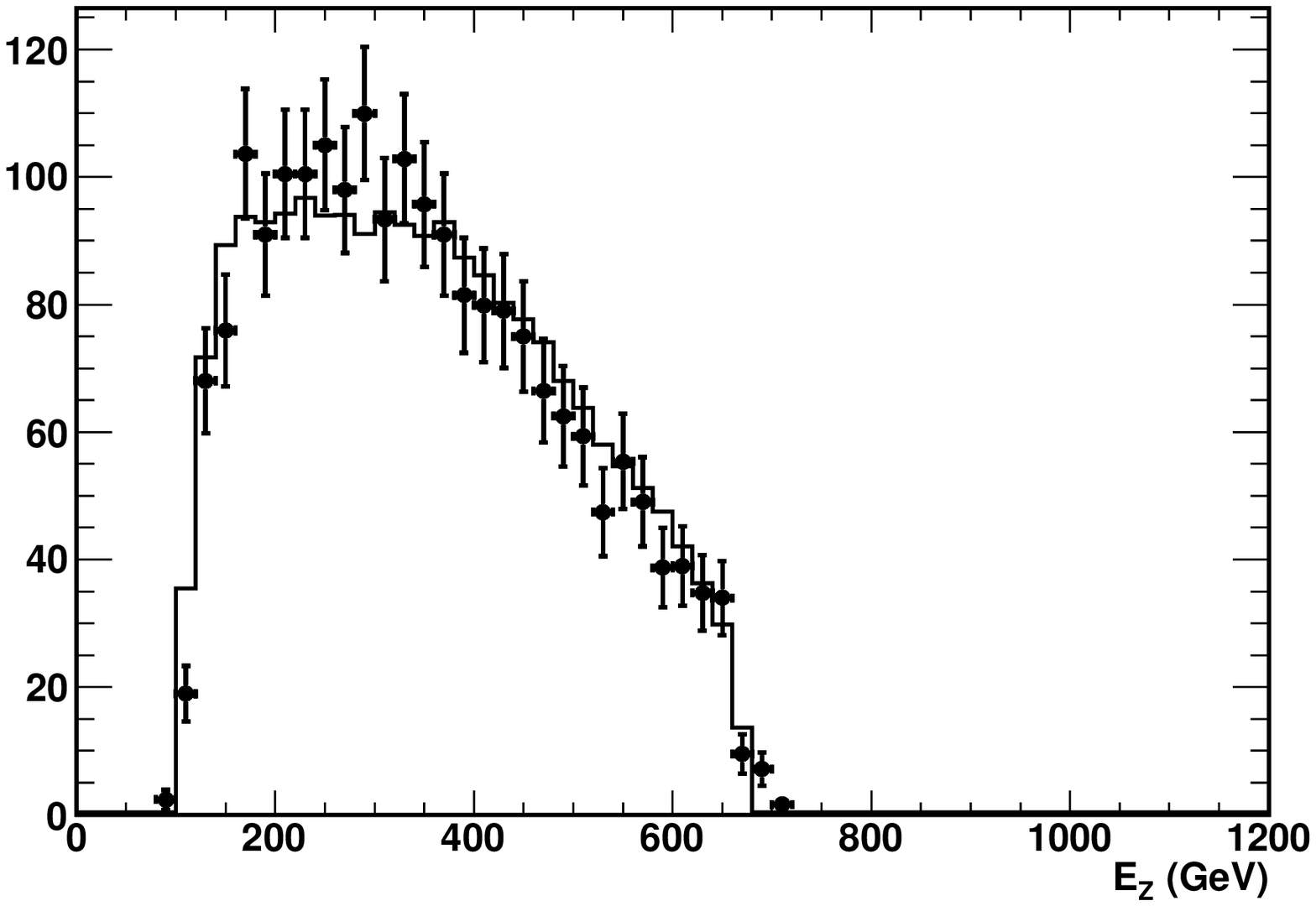}}  &  
\subfloat[8-jet $WWZh$]{\includegraphics[width=6.5cm]{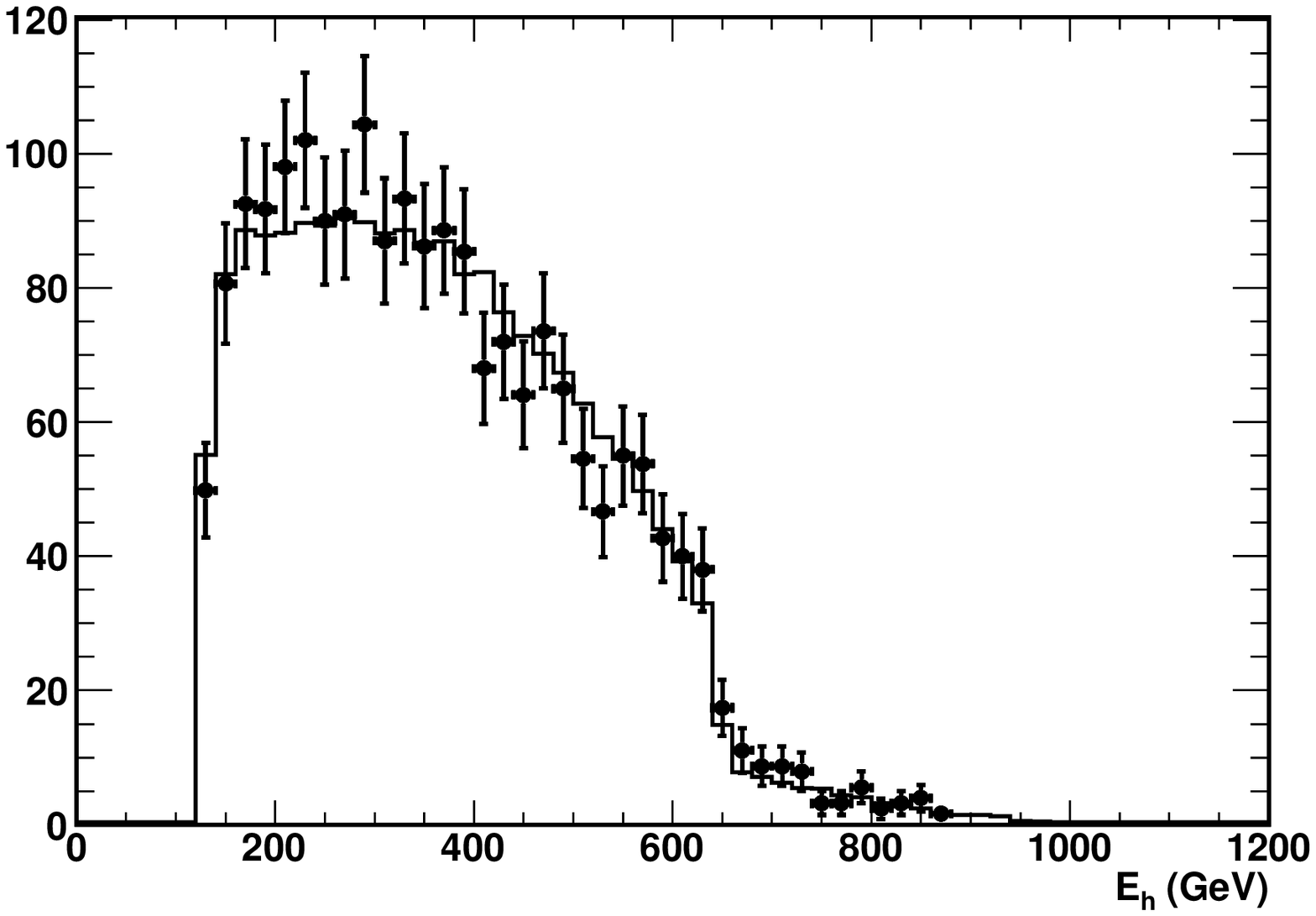}}  \\
\end{tabular}
\end{center}
\vspace*{-0.5cm}
\caption{Energy spectra for $\chi^+_2 \chi^-_2$ signal events: (upper row) in the six jet $WWh$ final 
state and (lower row) in the eight jet $WWZh$ final state with ISR and BS effects for model~I at 3~TeV. 
The points with error bars are the simulation and the line the fitted spectrum.}
\label{fig:efitchi2pm}
\end{figure}
\begin{table}[h!]
\caption{Statistical accuracy on $\chi^{\pm}_{2}$ mass from the combination of fits to the $E_{W^{\pm}}$ 
and $E_{h^0}$ spectra for  $\chi^+_2 \chi^-_2$ signal events in 6-jet $WWh$ and 8-jet $WWZh$ topologies 
and 2~ab$^{-1}$ of integrated luminosity at 3~TeV, under different assumptions}
\begin{center}
\begin{tabular}{|l|c|ccccc|}
\hline
Particle        & Mass  & No Rad       & ISR        & ISR+BS            & ISR+BS     & ISR+BS    \\
                & (GeV) &              &            &  $\delta E/E$=0   &  =0.025    &  =0.050   \\
\hline
Model~I         &       &              &            &                   &            &            \\
$\chi^{\pm}_2$  & 916.7 & $\pm$~2.2 & $\pm$~2.8   & $\pm$~3.6         & $\pm$~3.9  & $\pm$~4.2  \\
\hline
\end{tabular}
\end{center}
\label{tab:chi2pm}
\end{table}
The 6-jet topology accounts for 31\% of the total yield of $\chi^+_2 \chi^-_2$ pairs. The 
$W^+W^-h^0$ + missing energy final state receives two contributions from signal $\chi^+_2 \chi^-_2$
pair production: $e^+e^- \to \chi^+_2 \chi^+_2 \to W^+ \chi^0_2 W^- \chi^0_1 \to W^+ h^0 \chi^0_1 W^- \chi^0_1$
with $h \to b \bar b$, $W \to q \bar q'$ and  $e^+e^- \to \chi^+_2 \chi^+_2 \to h^0 \chi^+_1 W^- \chi^0_1 
\to h^0 W^+ \chi^0_1 W^- \chi^0_1$ with $h \to b \bar b$, $W \to q \bar q'$. Each event has two $W$ and one 
Higgs boson. The purity in $\chi^+_2 \chi^-_2$ pairs is 77~\% with other SUSY contribution to this final state 
coming from $\chi^0_3 \chi^0_4$ and $\chi^{\pm}_1 \chi^{\mp}_2$. The $W$ spectrum combines $W$s 
produced in the direct $\chi^{\pm}_2$ decay into two different states, $\chi^0_2$ and $\chi^0_1$.
In the final state also the Higgs boson is sensitive to the $\chi^{\pm}_2$ mass, through the contribution 
from the $\chi^{\pm}_2 \to h^0 \chi^{\pm}_1$ decays, while part of the bosons come from the $\chi^{0}_2$ decay.
In both cases we perform a 2-par fit leaving both the $M_{\chi^{\pm}_2}$ and the fraction of the two contributions 
free (see Figure~\ref{fig:efitchi2pm}). Then, we consider processes with four bosons yielding the 8-jet topology, 
which accounts for 19\%  of the $\chi^+_2 \chi^-_2$ final states. 
The $Z^0 h^0 W^+ W^-$ final state has 85\% contribution from $\chi^+_2 \chi^-_2$ pairs with the remaining mostly 
due to $\chi^0_3 \chi^0_4$. We use both the $Z^0$ and the $h^0$ spectrum to perform the mass fits. In this topology 
the $Z^0$ energy spectrum from the process $e^+e^- \to \chi^+_2 \chi^-_2 \to Z^0 \chi^+_1 W^- \chi^0_2 \to Z^0 W^+ 
W^- h^0 \chi^0_1 \chi^0_1 $ receives almost exclusively contributions from the direct 
$\chi^{\pm}_2 \to Z^0 \chi^{\pm}_1$ decay.  For the fit to the $h^0$ spectrum we include the contribution from the 
$\chi^{\pm}_2 \to h^0 \chi^{\pm}_1$ and that from $\chi^{0}_2 \to h^0 \chi^{0}_1$, leaving the contribution of the 
two processes free, as done for the 6-jet topology (see Figure~\ref{fig:efitchi2pm}). Results are combined and the 
statistical uncertainties are summarised in Table~\ref{tab:chi2pm}.

\subsubsection{$\chi^0_{3,4} \to W^{\pm} \chi^{\mp}_1$}

\begin{figure}[ht!]
\begin{center}
\begin{tabular}{cc}
\subfloat[ISR Only]{\includegraphics[width=6.5cm]{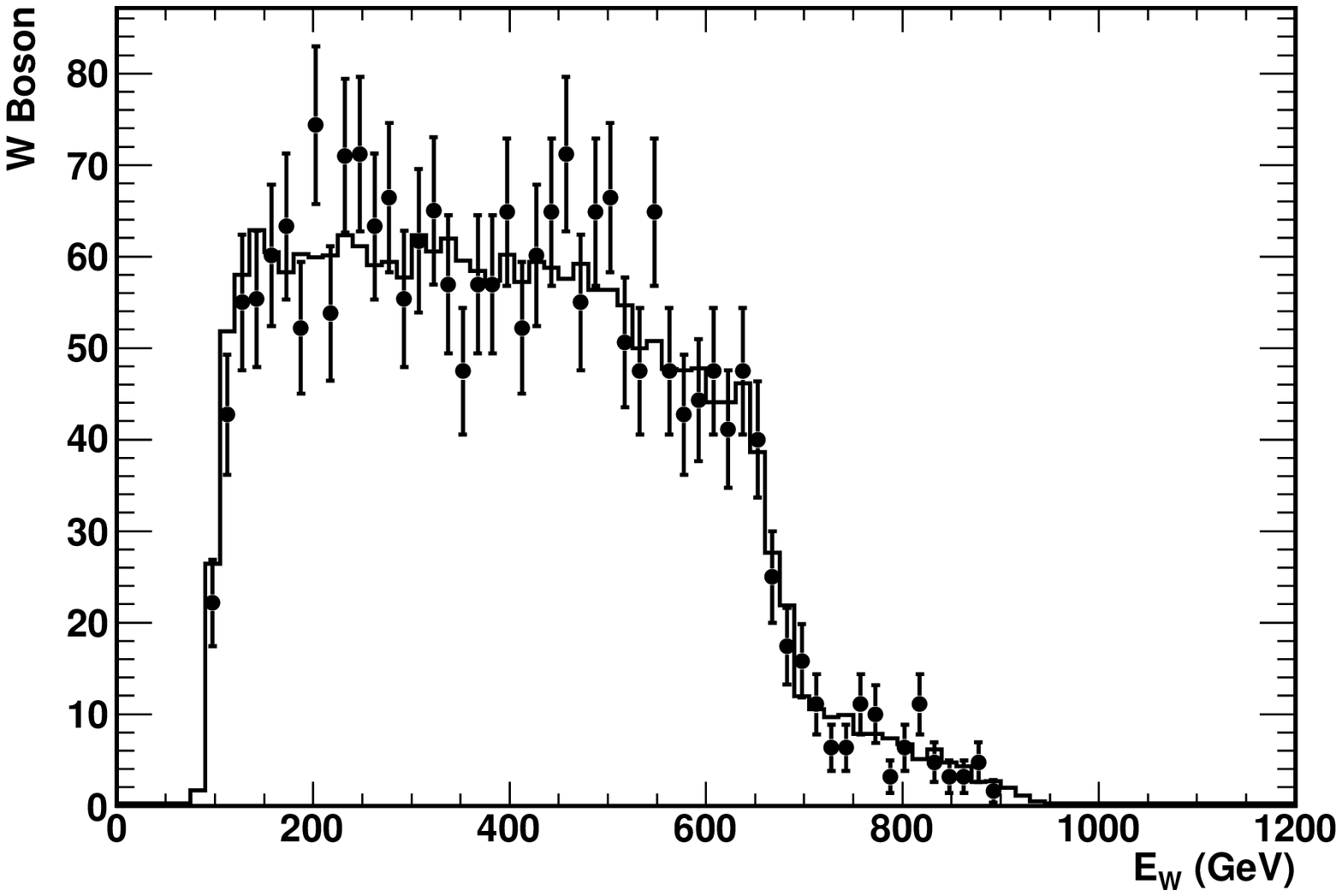}} &
\subfloat[ISR + BS]{\includegraphics[width=6.5cm]{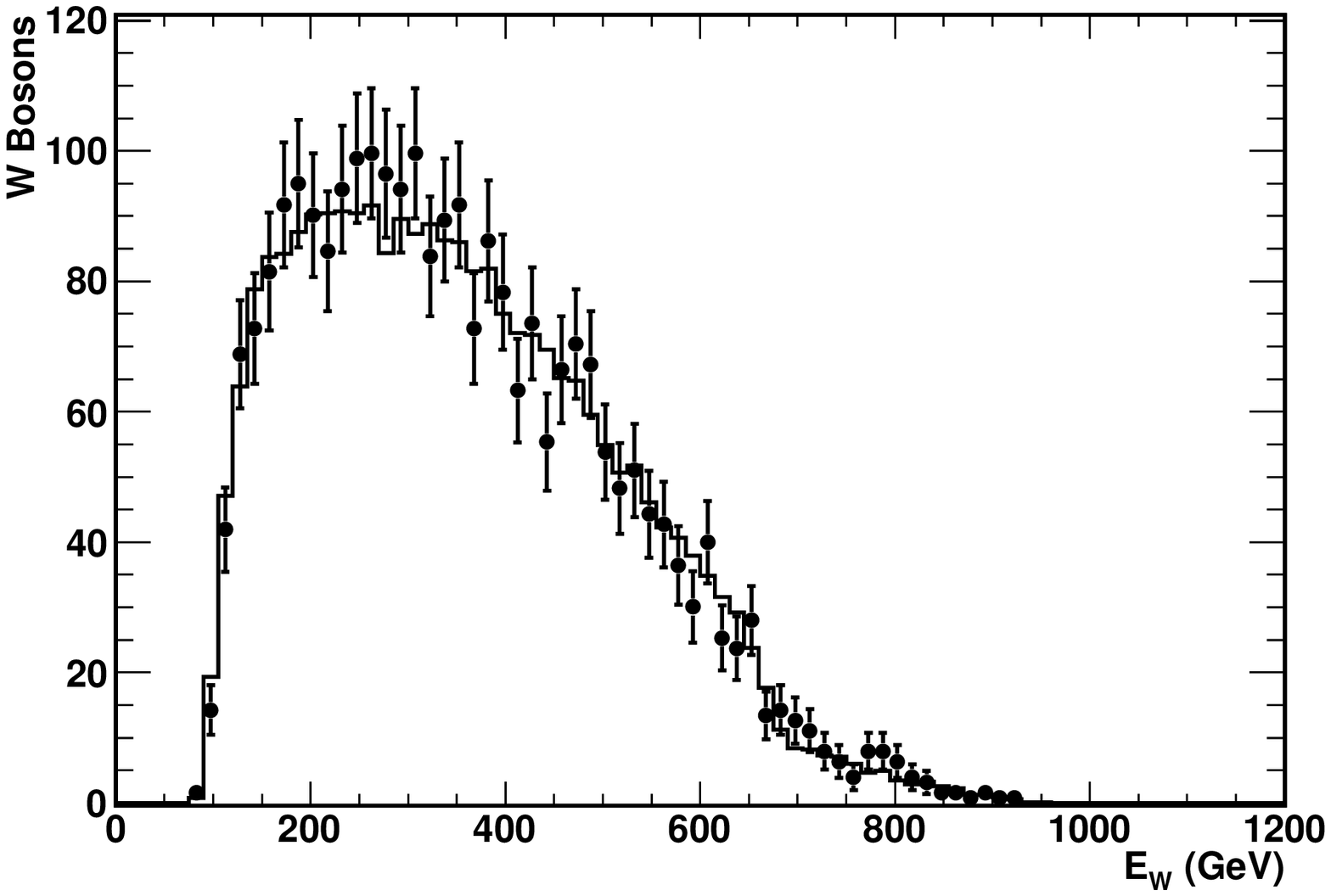}} \\
\end{tabular}
\end{center}
\vspace*{-0.5cm}
\caption{$W^{\pm}$ energy spectrum in the 8-jet topology, 4~$W$ + missing energy final state at 3~TeV 
with only ISR (left) and ISR+BS (right) for $\chi^0_3 \chi^0_4$ signal events in model~I. The points 
with error bars are the simulation and the line the fitted spectrum.}
\label{fig:efitchi34}
\end{figure}

The $e^+e^- \to \chi^0_3 \chi^0_4$ process has a sizable cross section in model~I and a good number of 
signal events can be reconstructed. The decay channel considered here is $\chi^0_{3,4} \to W^{\pm} \chi^{\mp}_1 
\to W^{\pm} W{\mp} \chi^0_1$ which gives an 8-jet topology with, 4 $W$ bosons and missing energy. This 
final state is challenging, due to its large jet multiplicity, but the signature is striking and there are 
essentially no SUSY or SM irreducible background processes contributing to it. In each event two of the $W$ 
are sensitive to the $\chi^0_{3,4}$ mass and the others to the $\chi^{\pm}_1$ mass. Since this can be 
precisely measured in the 2~$W$ + missing energy channel, it is safe to assume here that its mass is 
known. Results are given in Table~\ref{tab:chi34} and Figure~\ref{fig:efitchi34}.
\begin{table}[h!]
\caption{Statistical accuracy on $\chi^{0}_{3,4}$ mass from a fit to the $E_{W^{\pm}}$ endpoints 
 for $\chi^0_3 \chi^0_4$ signal events and 2~ab$^{-1}$ of integrated luminosity at 3~TeV, 
 under different assumptions}
\begin{center}
\begin{tabular}{|l|c|cccc|}
\hline
Particle        & Mass  & No Rad     & ISR        & ISR+BS         & ISR+BS   \\
                & (GeV) &            &            & $\delta E/E$=0 & =0.05   \\
\hline
Model~I           &       &           &             &       &    \\
$\chi^{0}_{3}$    & 905.5 & $\pm$ 7.1 & $\pm$ 7.9   & $\pm$12.6 & $\pm$15.1 \\
$\chi^{0}_{4}$    & 916.7 & $\pm$ 8.7 & $\pm$ 8.9   & $\pm$13.0 & $\pm$15.4 \\
\hline
\end{tabular}
\end{center}
\label{tab:chi34}
\end{table}
The structure of the cMSSM implies that the two heaviest neutralinos, $\chi^0_3$ and $\chi^0_4$ are nearly 
degenerate. However, this is not a general feature of supersymmetry and it does not apply to less constrained 
scenarios, such as the MSSM, where the $\chi^0_4$ - $\chi^0_3$ mass splitting can be  $\sim$20\% of their 
masses, or larger. Therefore, we repeat the fit, leaving the masses of $\chi^0_3$ and $\chi^0_4$ free and 
test the result for a mass splitting of 10~GeV, as in our model~I, and one of 40~GeV obtained by decreasing 
the $\chi^0_3$ mass. We find a resolution on the $\chi^0_3$ - $\chi^0_4$ mass splitting of $\sim$15-18~GeV 
and masses which are apart $\sim$35~GeV and more can be resolved. 

\subsection{Energy Resolution Effects}

\begin{figure}[ht!]
\begin{center}
\begin{tabular}{cc}
\subfloat[Model I]{\includegraphics[width=6.0cm,clip=]{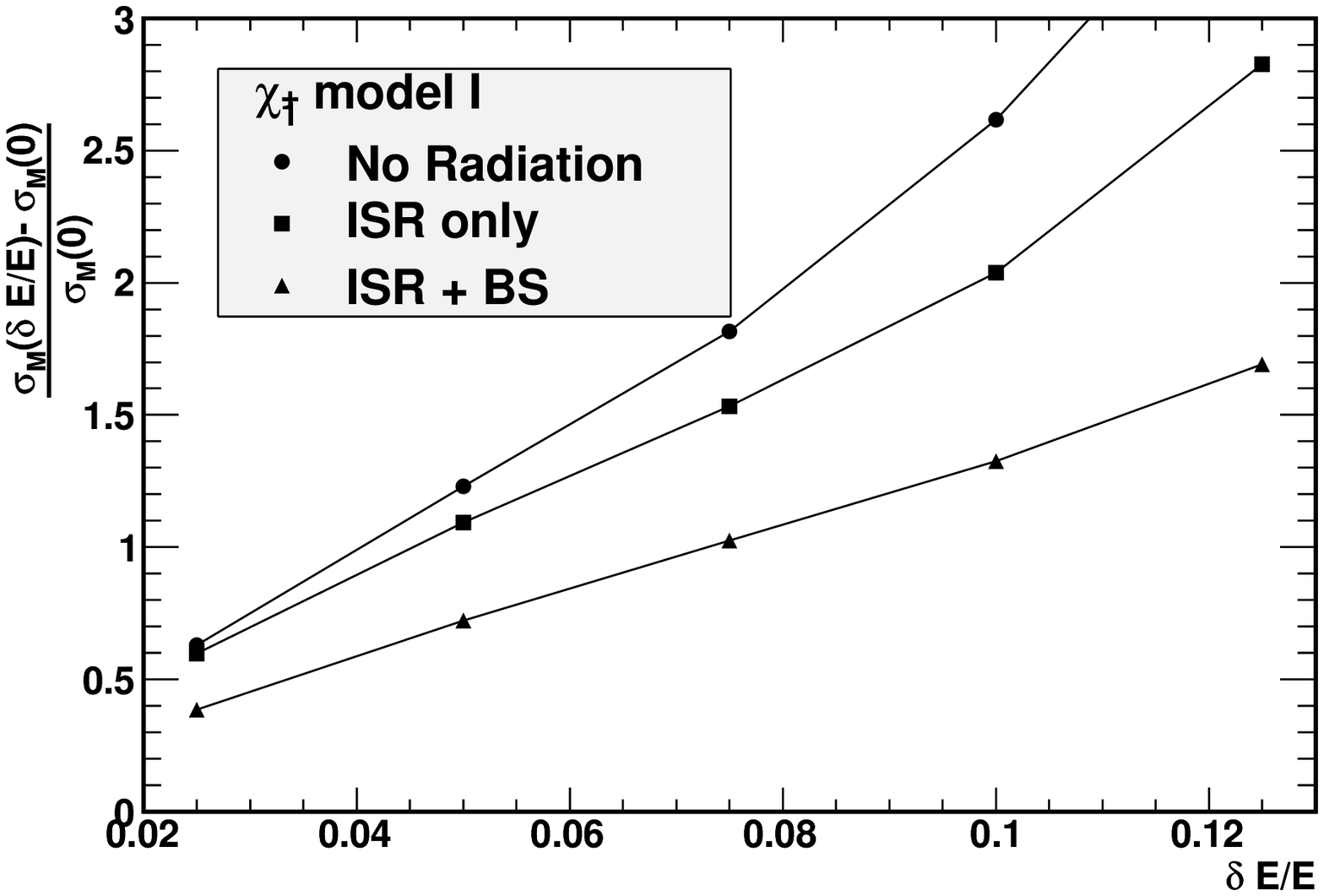}}  & 
\subfloat[Model II]{\includegraphics[width=6.0cm,clip=]{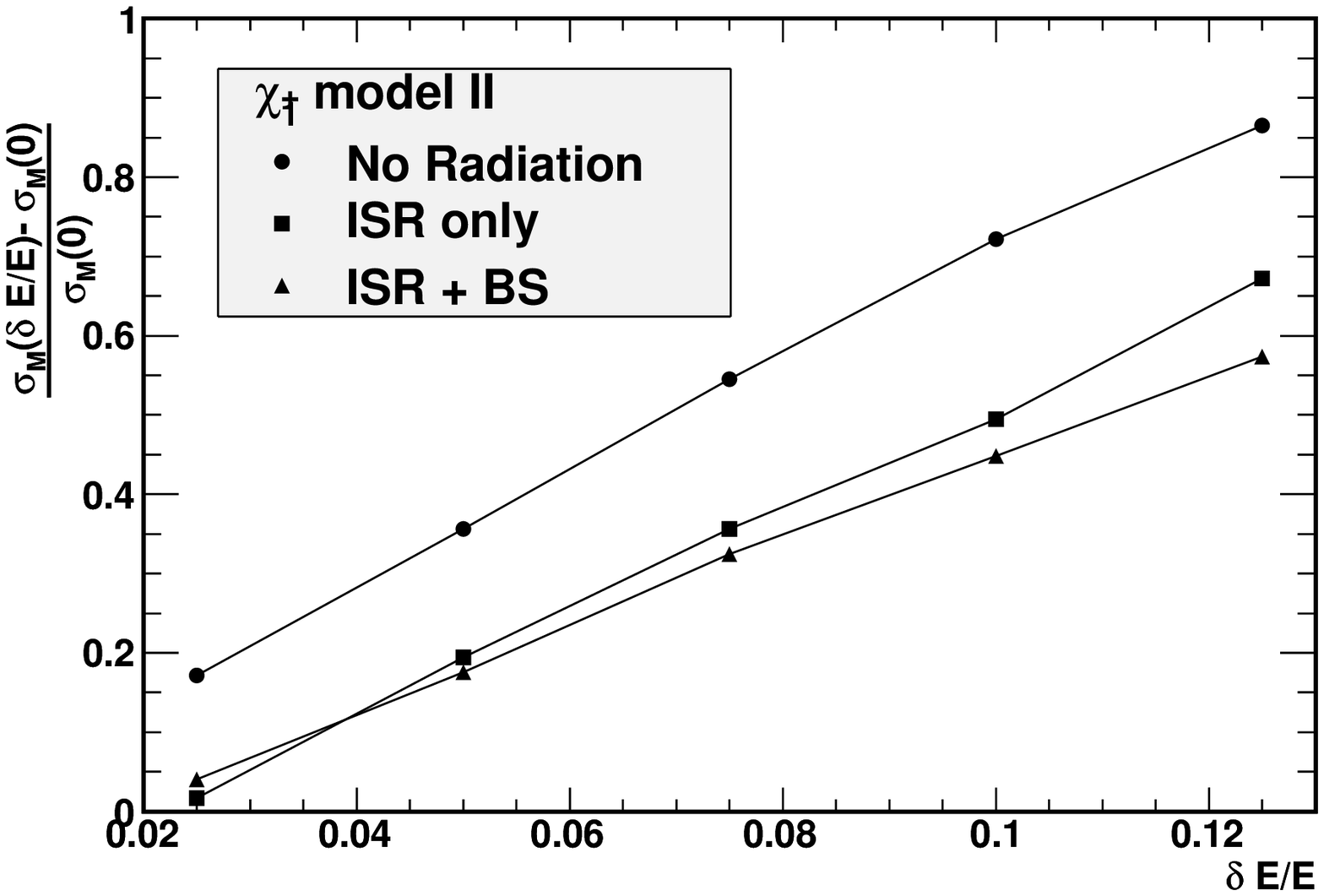}}  \\
\end{tabular}
\end{center}
\caption{Relative change of the statistical accuracy on the $\chi^{\pm}_{1}$ mass as a function 
of the jet energy resolution from one-parameter $\chi^2$ fits to the energy spectrum for signal events in 
model I (a) and II (b).}
\label{fig:chi1eres}
\end{figure}

\begin{figure}[h!]
\begin{center}
\includegraphics[width=6.5cm,clip=]{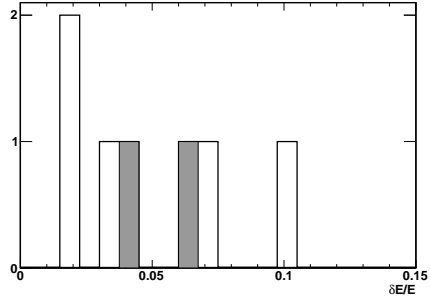}
\end{center}
\vspace*{-0.5cm}
\caption{Values of the relative energy resolution $\delta E/E$ at which its contribution on the 
mass statistical uncertainties equals that of ISR and beamstrahlung for the channels considered
in this study. Model~I is shown by the open histogram and model~II by the grey histogram.}
\label{fig:Elim}
\end{figure}

The effect of the jet energy resolution on the $\chi^{\pm}_1$ masses for 
both models are shown in Figure~\ref{fig:chi1eres} in terms of the relative change of the statistical 
accuracy of the mass determination as a function of the parton energy resolution $\delta E/E$. 
As expected, the effect is larger when beam radiation is not considered. We establish a quantitative 
criterion for the energy resolution requiring that the contribution from the parton energy resolution 
to the statistical mass accuracy should not exceed the combined effect of ISR and beamstrahlung. 
We assume that these contributions adds quadratically and derive the limits to $\delta E/E$ for the 
different channels which are summarised in Figure~\ref{fig:Elim}.
\begin{figure}[h!]
\begin{center}
\begin{tabular}{ccc}
\subfloat[$\delta E/E$=0.025]{\includegraphics[width=4.5cm]{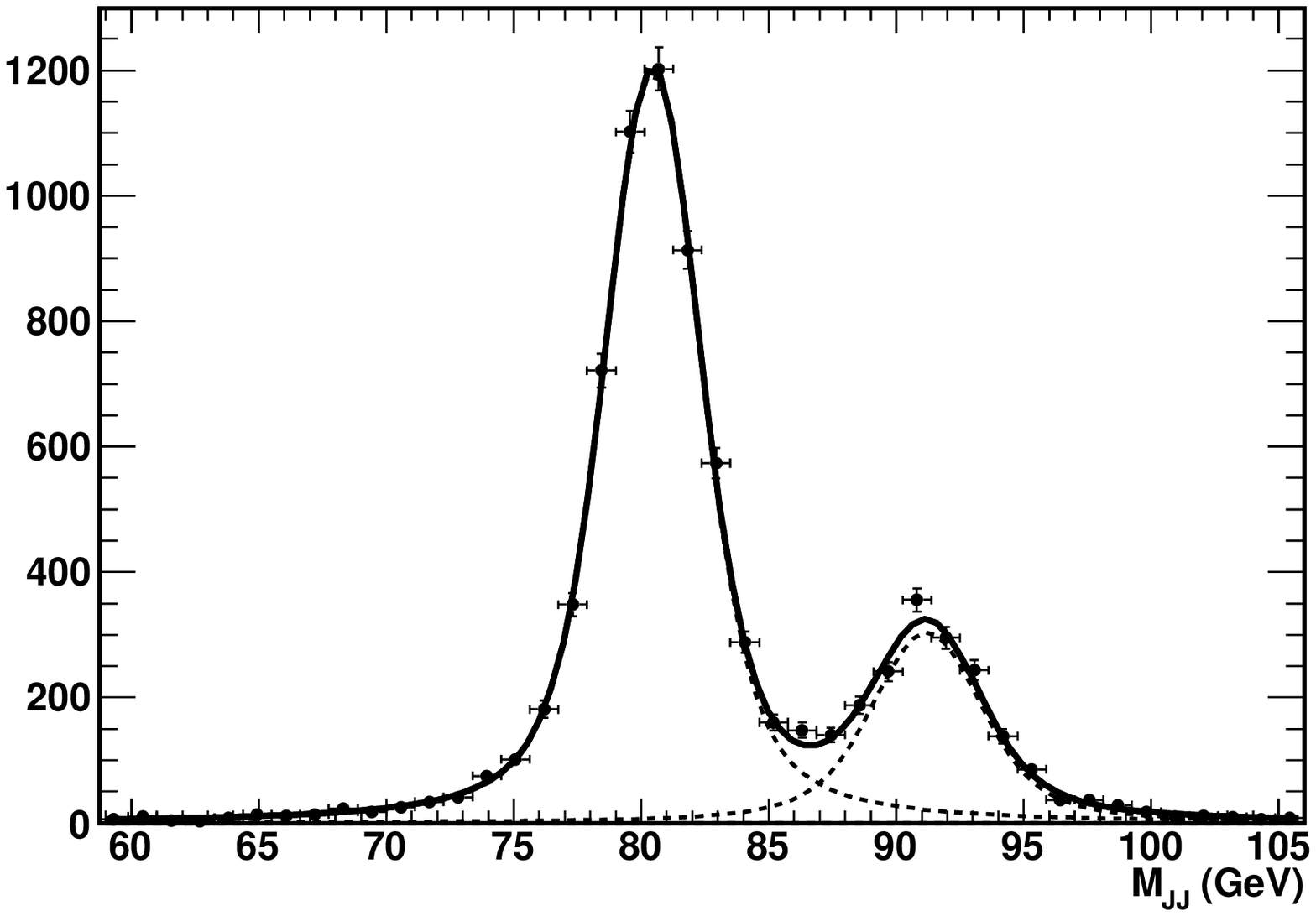}} &
\subfloat[$\delta E/E$=0.050]{\includegraphics[width=4.5cm]{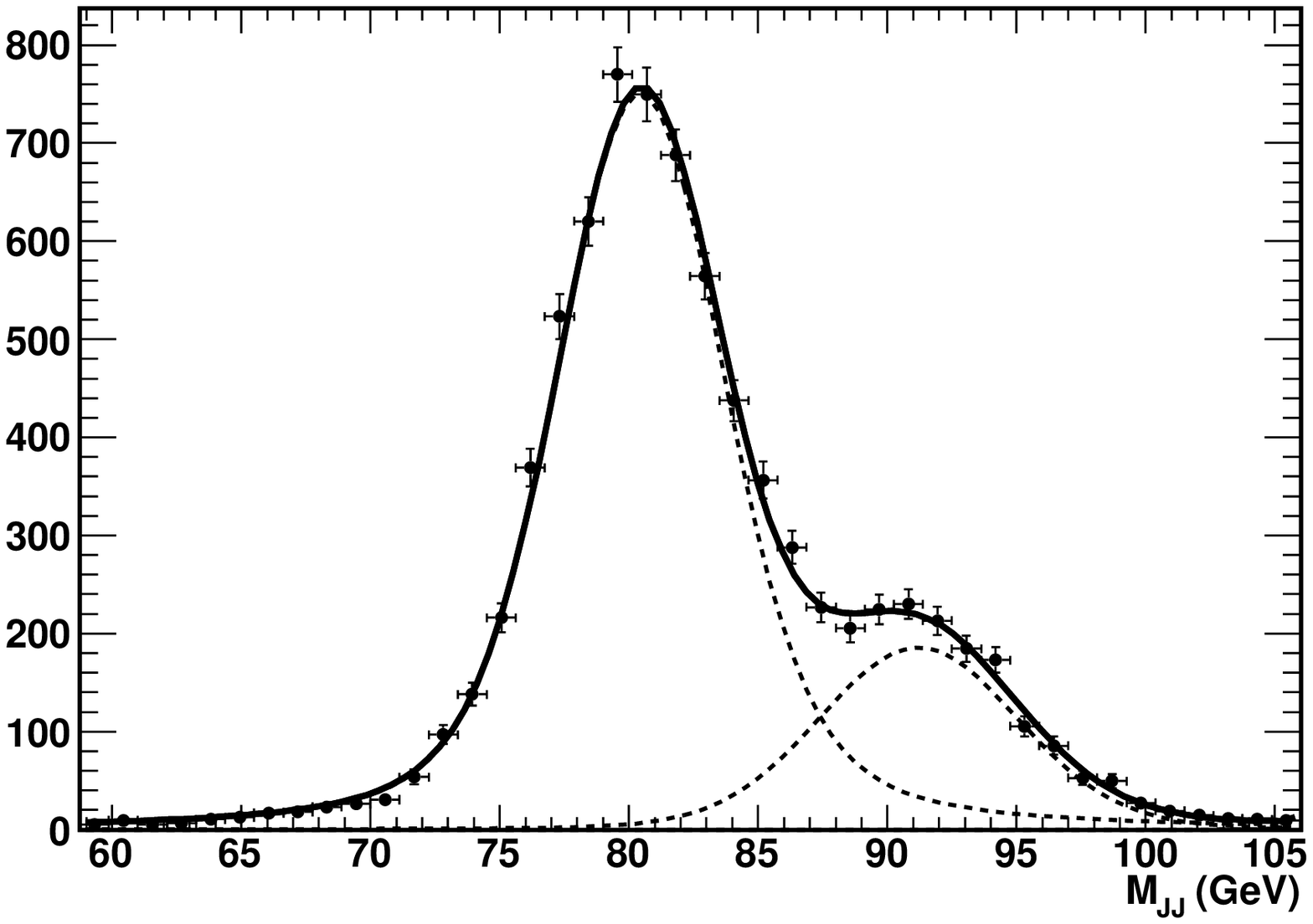}} &
\subfloat[$\delta E/E$=0.075]{\includegraphics[width=4.5cm]{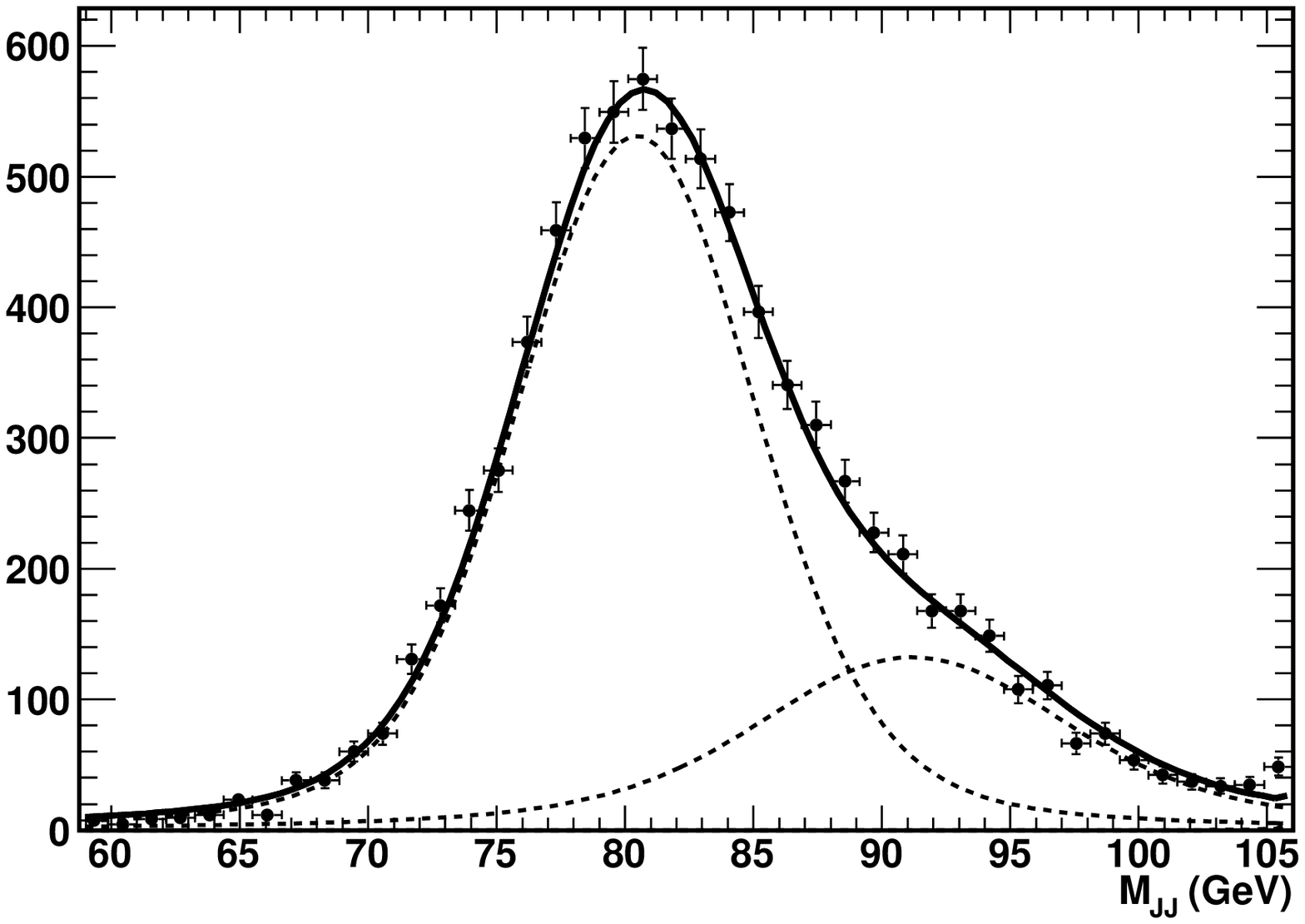}} \\
\end{tabular}
\end{center}
\vspace*{-0.5cm}
\caption{Di-jet invariant mass in 8-jet inclusive SUSY events in model~I for various values of 
parton energy resolution.}
\label{fig:massz}
\end{figure}
\begin{table}[h!]
\caption{Gaussian width of the $Z^0$ peak and purity of $Z^0$ decays selected from their 
compatibility with the $Z^0$ mass in 8-jet inclusive SUSY events in model~I for various values 
of parton energy resolution.}
\begin{center}
\begin{tabular}{|l|c|c|c|}
\hline
$\delta E/E$ & $\sigma_Z$ & Purity \\
             & (GeV)      &        \\
\hline
0.           &  1.64      & 0.877  \\
0.025        &  2.42      & 0.720  \\
0.040        &  3.33      & 0.418  \\
0.050        &  4.14      & 0.290  \\
0.075        &  5.34      & 0.220  \\
\hline   
\end{tabular}
\end{center}
\label{tab:massz}
\end{table}
Then, we consider the effect of the parton energy resolution on the di-jet invariant mass. We study 
the specific case of the $Z^0$ identification in the 8-jet $e^+e^- \to \chi^+_2 \chi^+_2 \to Z^0 \chi^+_1 
W^- \chi^0_1 \to Z^0 W^+ \chi^0_1 W^- \chi^0_1$; $Z \to q \bar q$, $W \to q \bar q'$ process, which is 
the process most sensitive to di-jet mass resolution in model~I, due to the large $W$ yield. 
We identify the $Z^0$ boson requiring that the di-jet mass is compatible with the nominal mass, 
$M_Z$=91.2~GeV, within 3~$\sigma_Z$, where $\sigma_Z$ is the peak Gaussian width measured on genuine 
$Z^0 \to q \bar q$ di-jets. We vary the Gaussian parton energy resolution (see Figure~\ref{fig:massz}) 
and study the purity in real $Z^0$ bosons selected by this selection. Since $W^{\pm}$ bosons are dominant 
in 8-jet topology SUSY events (53\%) and $Z^0$ bosons make only 15.5\% of the di-jets, the leakage from 
the $W^{\pm}$ peak is important already for moderate values of the energy resolution, as shown in 
Table~\ref{tab:massz}. 

Operating the collider with polarised beams may be important to improve the 
statistical accuracy in the determination of heavier states such as $\chi^0_3$ 
and $\chi^0_4$. However, given the broad scope of the research program at a
multi-TeV collider various states of polarisation will be likely selected, 
making our assumption a fair estimate for the total cumulative statistics of 
signal events.

\subsubsection{Validation with Full Simulation and Reconstruction}

In order to validate the results obtained above at generator level, accounting only for a 
simple energy smearing, the $\chi^+_1 \chi^-_1$  and $\chi^0_2 \chi^0_2$ analyses are 
repeated on fully simulated and reconstructed events to verify the accuracy when accounting 
for the reconstruction effects in full.

\begin{figure}[h!]
\begin{center}
\includegraphics[width=7.0cm]{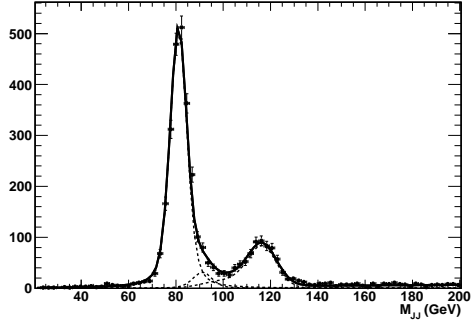}
\end{center}
\caption{Di-jet invariant mass for di-boson inclusive SUSY fully simulated and reconstructed 
events for model~I with the fitted contributions from $W^{\pm}$, $Z^0$ and $h^0$}
\label{fig:mfitDST1}
\end{figure}

A sample of inclusive SUSY events for model~I, generated with ISR+BS, corresponding 
to 0.5~ab$^{-1}$ of integrated luminosity, is fully simulated and reconstructed using the 
CLIC version of the ILD detector concept. First, the 4-jet + missing energy events are 
reconstructed. Events are pre-selected requiring a visible energy  250 $< E_{tot} <$ 1800~GeV, 
an energy in charged particles larger than 150~GeV, transverse energy larger than 200~GeV, 
a jet multiplicity 2$\le N_{jets} <$5 and at least 20~charged reconstructed particles. Jets
clustering is performed using the Durham jet algorithm~\cite{Catani:1991hj}, with $y_{cut}$ 
= 0.0025, on the reconstructed particle flow objects of the Pandora particle flow 
package~\cite{Thomson:2009rp}.
These events are then forced into four jets and the di-jet invariant mass for all the three 
possible pairings is computed. The jet pairing minimising the difference between the di-jet 
invariant masses is selected, provided the mass difference is below 20~GeV. The resulting 
mass distribution on generated SUSY di-boson events is shown in Figure~\ref{fig:mfitDST1}. 
\begin{table}[h!]
\caption{Fraction of $W^{\pm}$, $Z^0$ and $h^0$ bosons from the fit to the di-jet invariant mass 
distribution in 4-jet + missing energy inclusive SUSY events for model~I for 0.5~ab$^{-1}$ of 
integrated luminosity, compared to the generated values.}
\begin{center}
\begin{tabular}{|l|cc|}
\hline
Boson           & Fitted             & Simulated        \\
                & Fraction of Evts.\ & Fraction of Evts \\
\hline
$W^{\pm}$       & 0.650 $\pm$ 0.011  & 0.645 $\pm$ 0.005 \\
$Z^{0}$         & 0.040 $\pm$ 0.009  & 0.020 $\pm$ 0.002 \\
$h^{0}$         & 0.215 $\pm$ 0.010  & 0.243 $\pm$ 0.003 \\
\hline
\end{tabular}
\end{center}
\label{tab:wzh}
\end{table}
The fraction of $W^+W^-$, $Z^0Z^0$ and $h^0h^0$ events is extracted by a $\chi^2$ fit to 
the di-jet mass distribution. The $W^{\pm}$ and $Z^0$ mass peaks are parametrised as Breit-Wigner 
functions convoluted with a Gaussian term describing the experimental resolution. The mass
and width values of the Breit-Wigner functions are fixed to their generated values, while 
the total area and the width of the Gaussian resolution terms are left free in the fit. 
\begin{figure}
\begin{center}
\includegraphics[width=6.0cm]{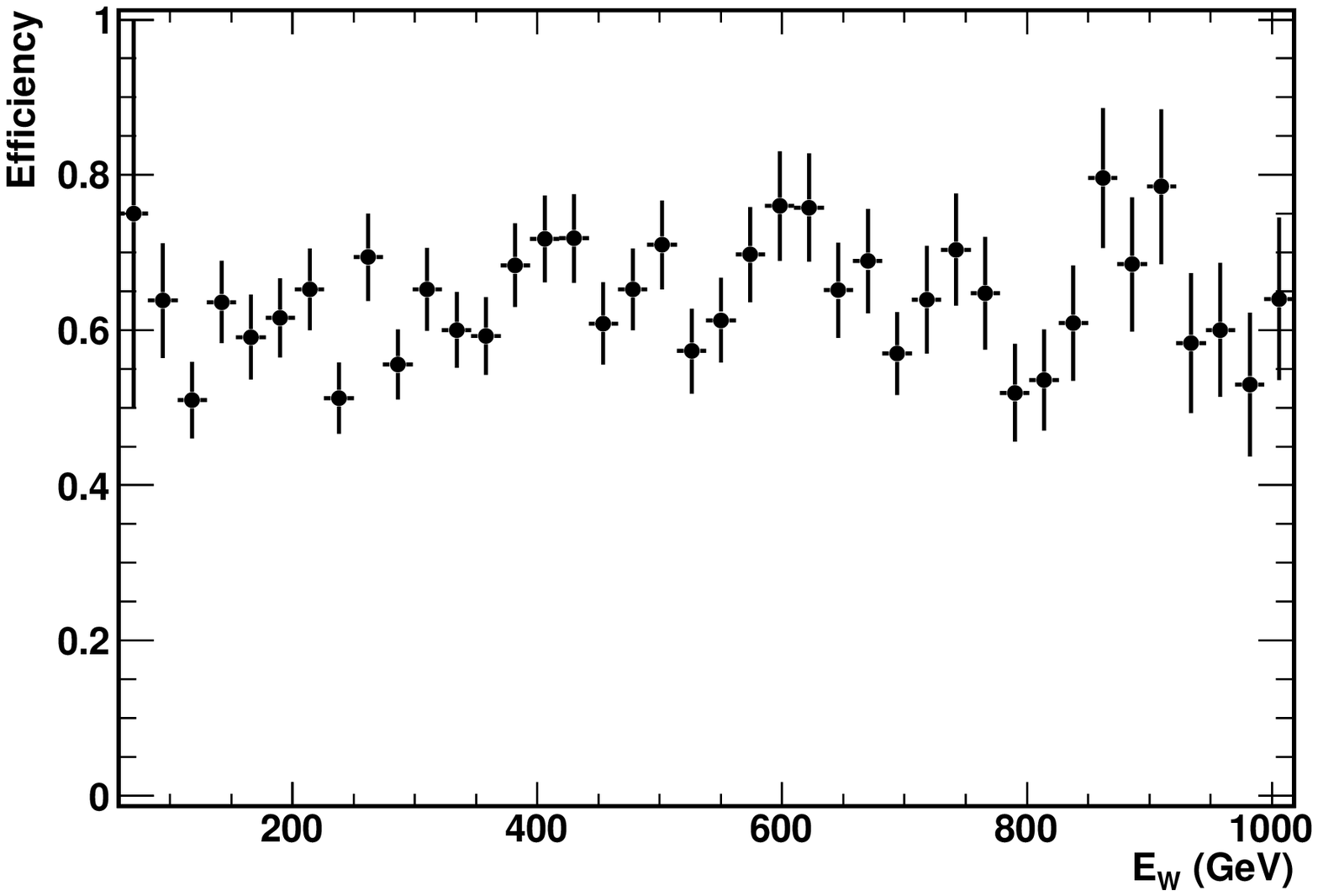}
\end{center}
\caption{Efficiency for event reconstruction and selection on fully simulated and reconstructed 4-jet 
$\chi^+_1 \chi^-_1 \rightarrow W^+ \chi^0_1 W^- \chi^0_1 \to q \bar q^{\prime} \chi^0_1 q \bar 
q^{\prime} \chi^0_1 $ signal events as a function of the $W$ energy.}
\label{fig:effW}
\end{figure}
\begin{figure}[h!]
\begin{center}
\begin{tabular}{cc}
\includegraphics[width=7.0cm]{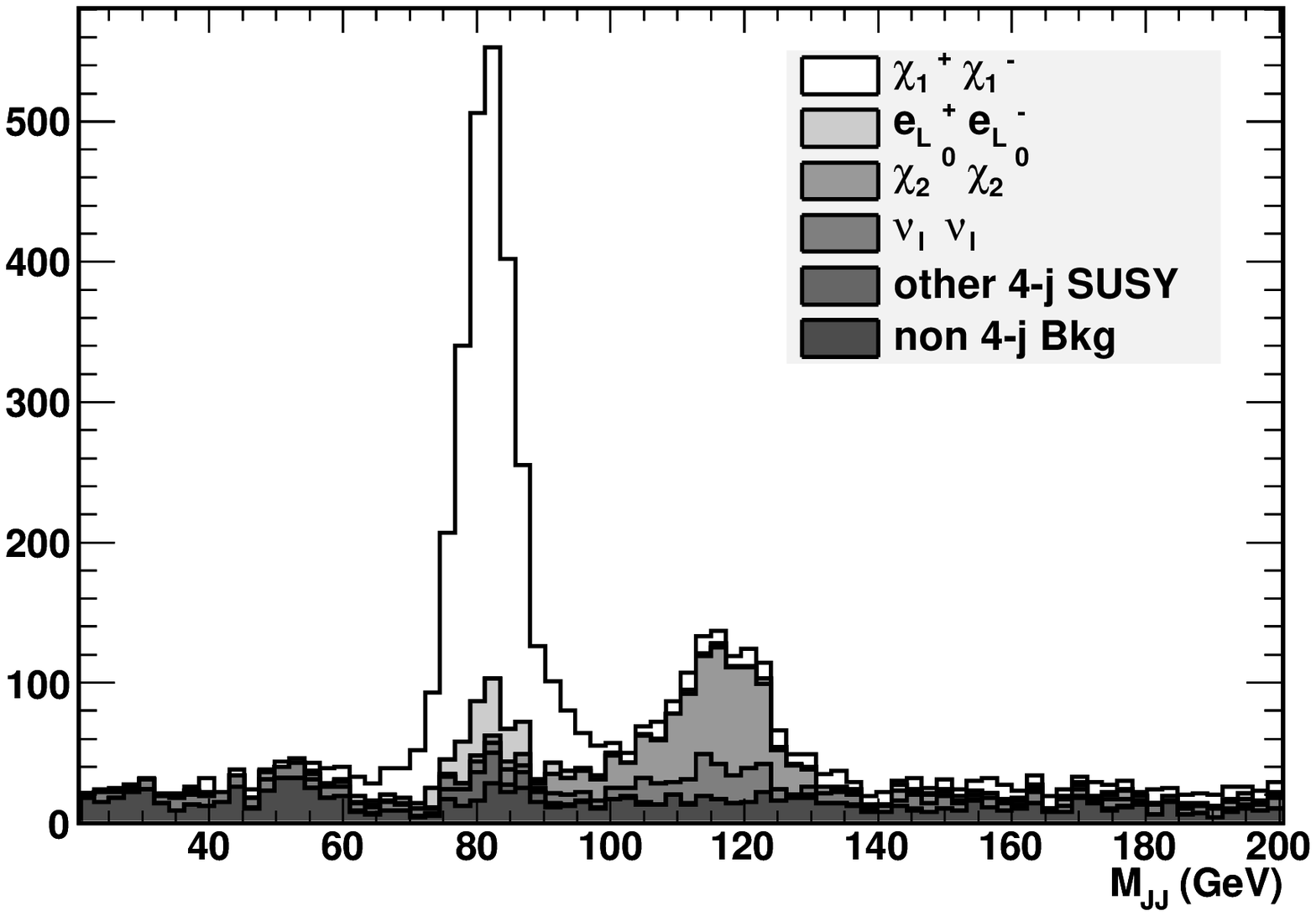} &
\includegraphics[width=7.0cm]{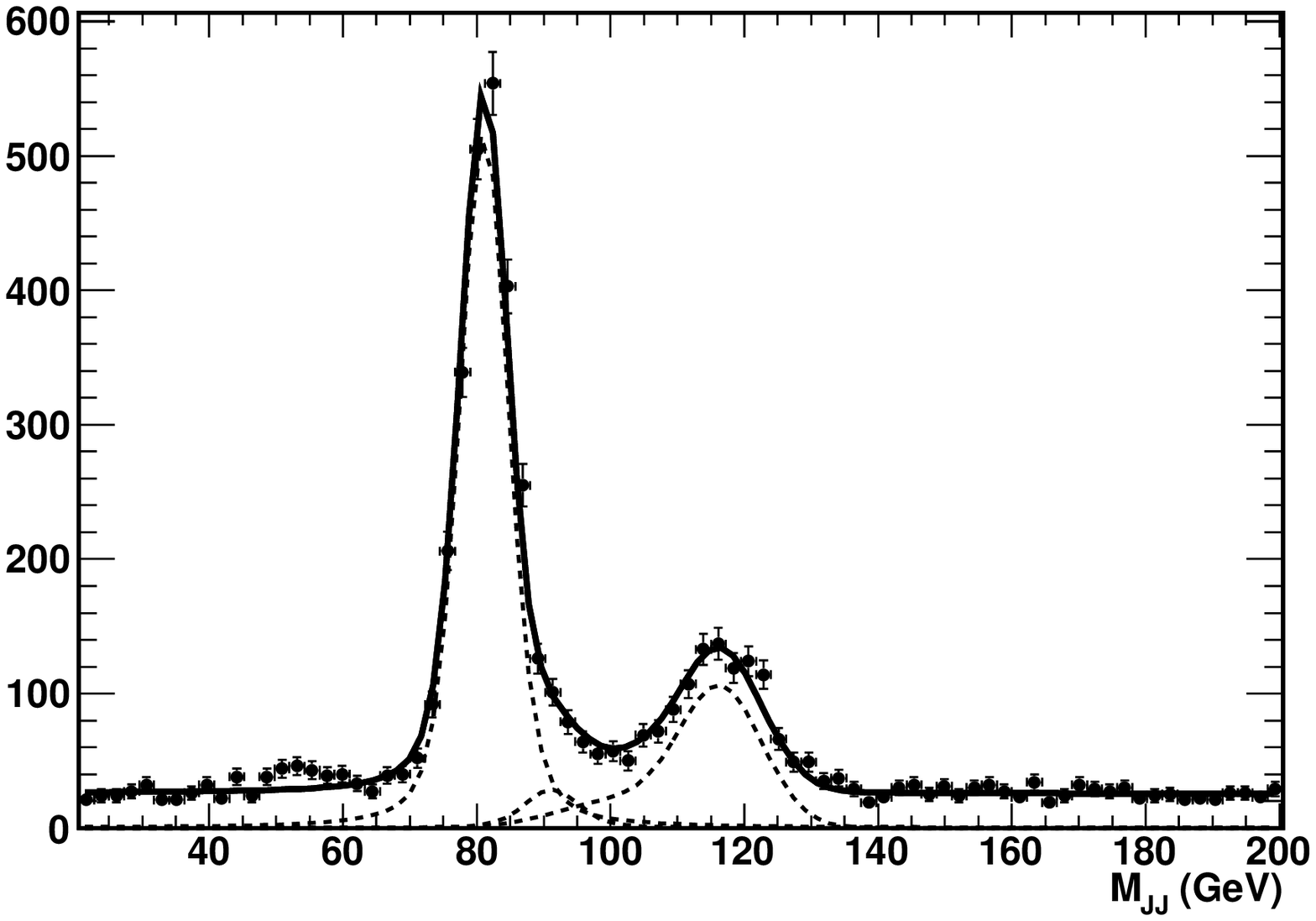} \\
\end{tabular}
\end{center}
\caption{Di-jet invariant mass for selected inclusive SUSY fully simulated and reconstructed 
events for model~I (left) with the different sources highlighted and (right) with the background 
and signal components fitted.}
\label{fig:mfitDST2}
\end{figure}
The $h^0$ peak, which has negligible natural width, is modelled as the sum of two 
Gaussian curves, one representing the correctly reconstructed signal events, centred at the nominal 
$M_h$ value, the second describing decays where the mass has a lower reconstructed value due to 
semi-leptonic $b$ decays. The central value, width and fraction of events in this second Gaussian 
is extracted by a fit to a pure sample of decays into $h^0$ bosons and fixed in the fit, while the 
Gaussian width of the main peak is kept free. Results are given in Table~\ref{tab:wzh} and the 
fitted functions are overlayed to the reconstructed spectrum in Figure~\ref{fig:mfitDST1}. 

Then, events with di-jets compatible with the $WW$ hypothesis are selected. The total selection efficiency 
is 60\% for $\chi^+_1 \chi^-_1 \rightarrow W^+ \chi^0_1 W^- \chi^0_1 \to q \bar q^{\prime} \chi^0_1 q \bar 
q^{\prime} \chi^0_1 $ signal events.  
\begin{figure}[ht!]
\begin{center}
\begin{tabular}{cc}
\includegraphics[width=7.0cm]{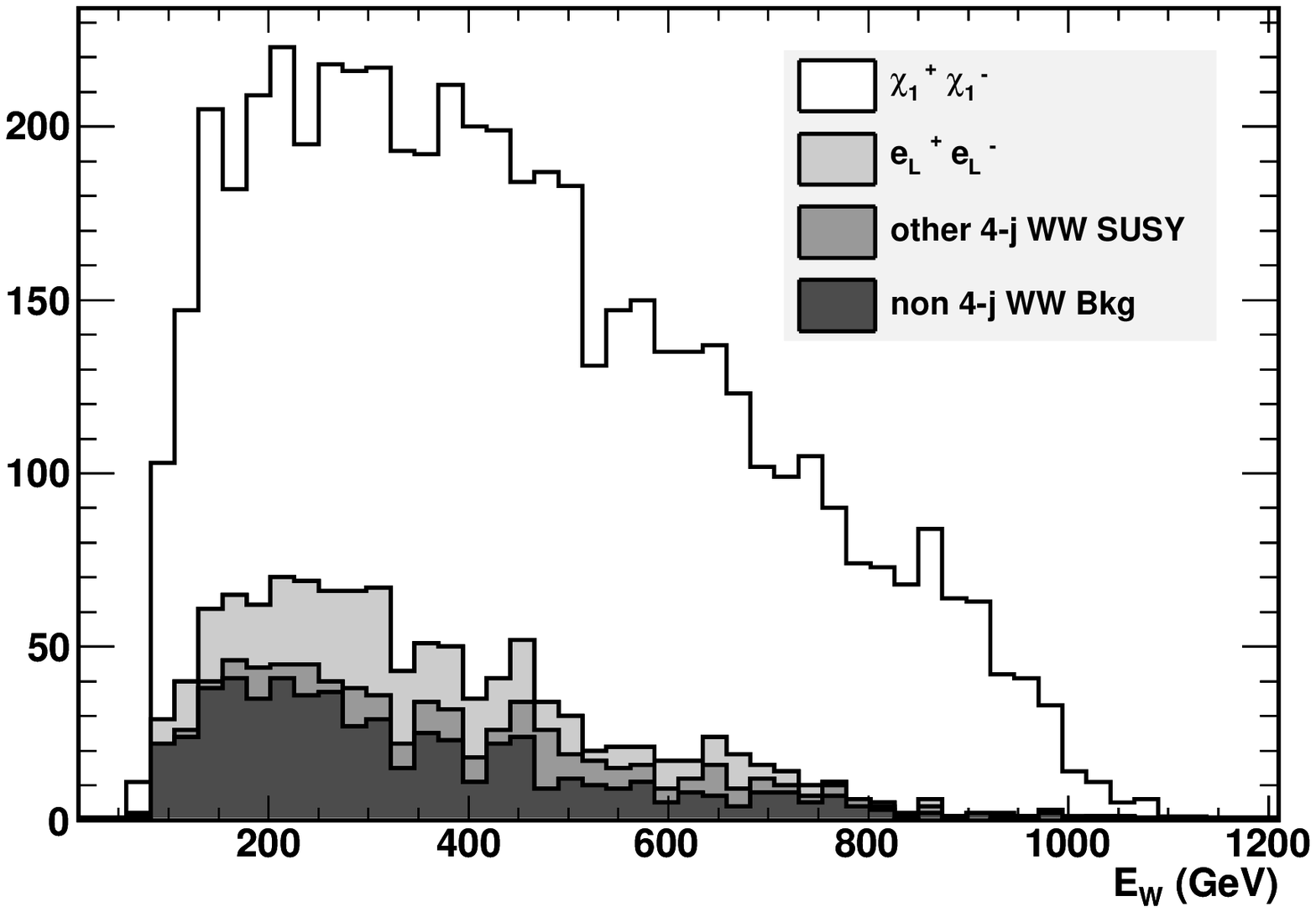} &
\includegraphics[width=7.0cm]{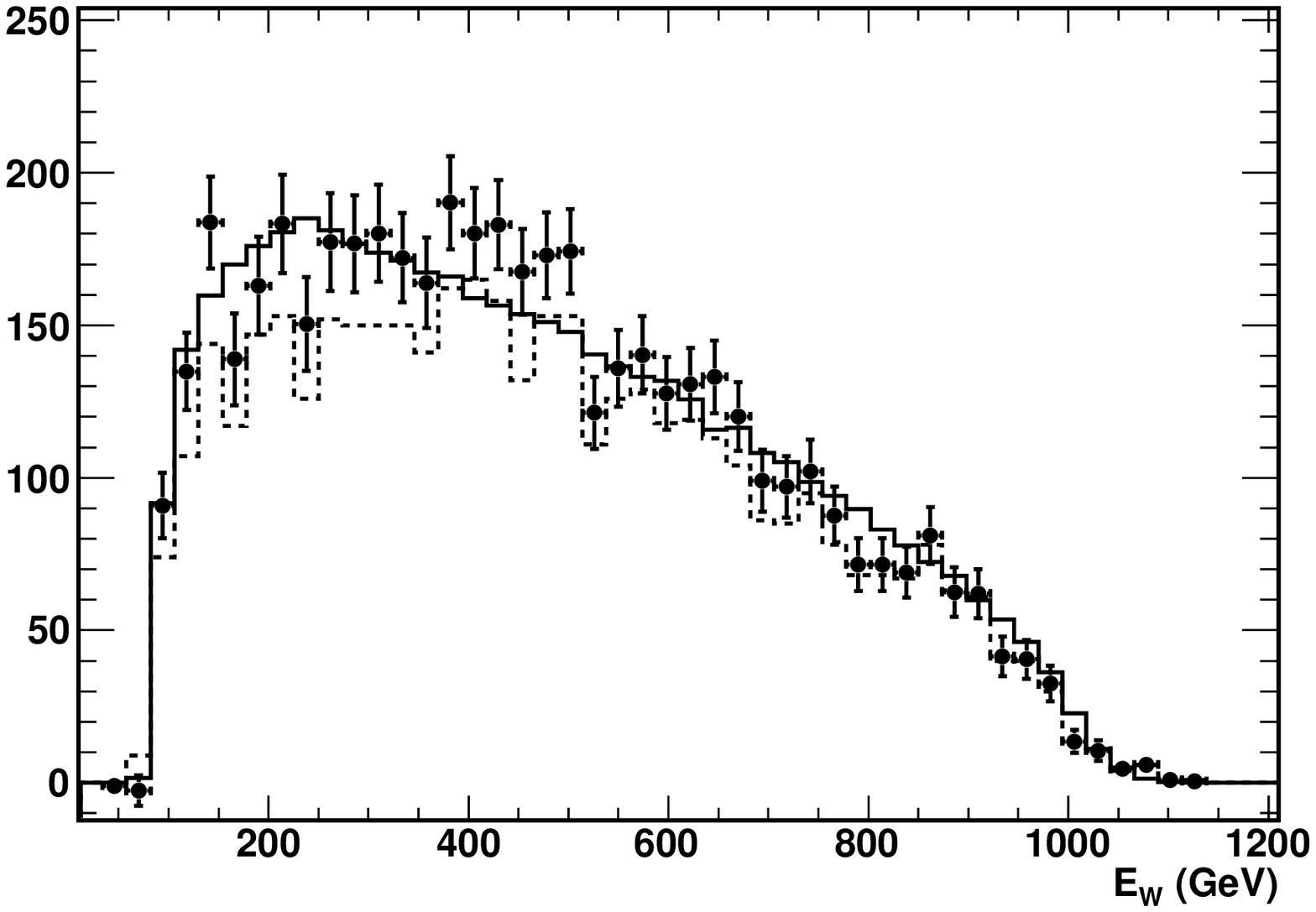} \\
\end{tabular}
\end{center}
\caption{Fits of fully simulated and reconstructed events in model~I: (left) $W^{\pm}$ energy spectrum of selected 
4-jet $WW$ candidate events energy spectrum and (right) background subtracted spectrum (points with error bars) 
with the result of the 3-par fit (continuous line).}
\label{fig:efitWDST}
\end{figure}
\begin{figure}[h!]
\begin{center}
\begin{tabular}{cc}
\includegraphics[width=7.0cm]{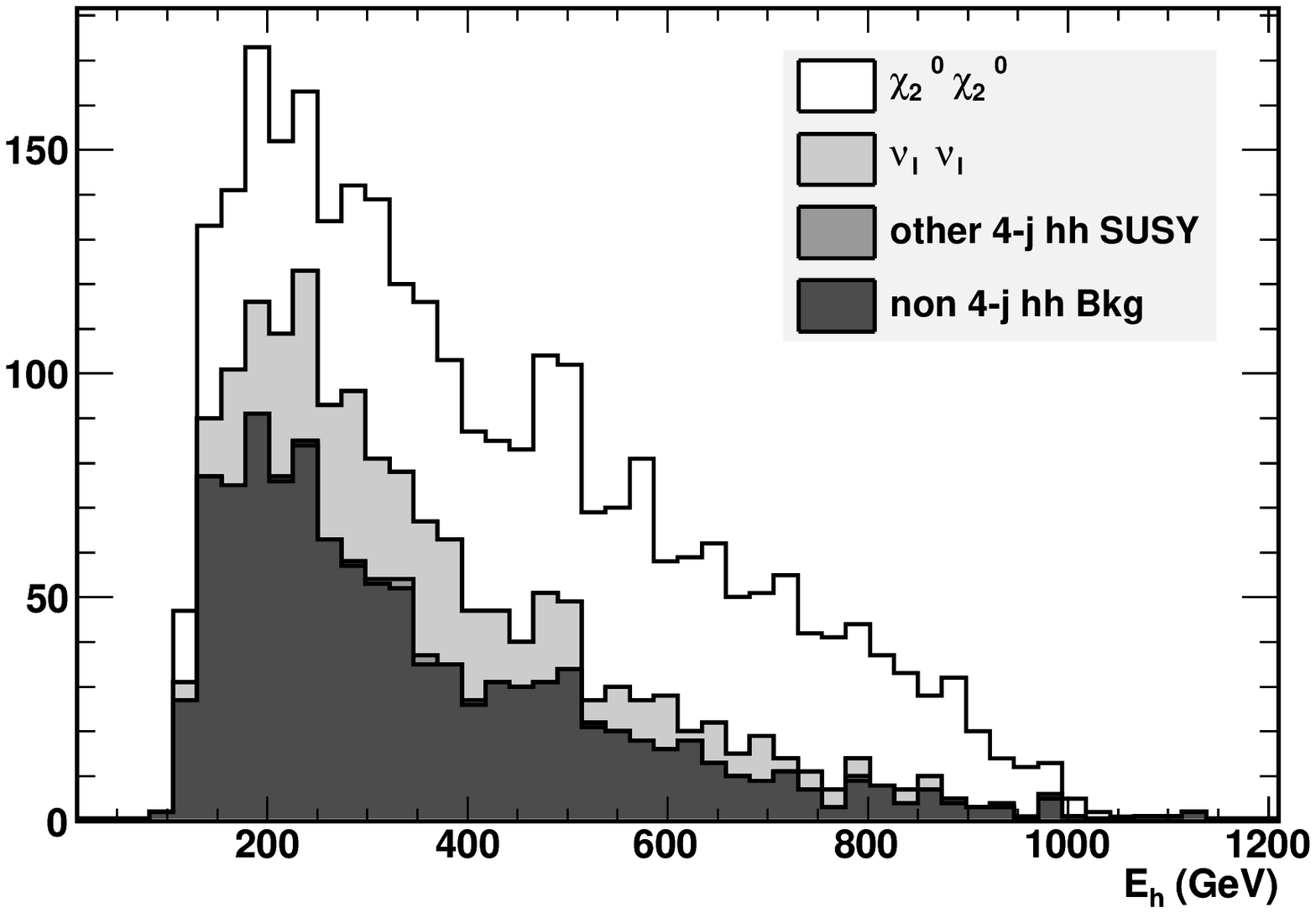} &
\includegraphics[width=7.0cm]{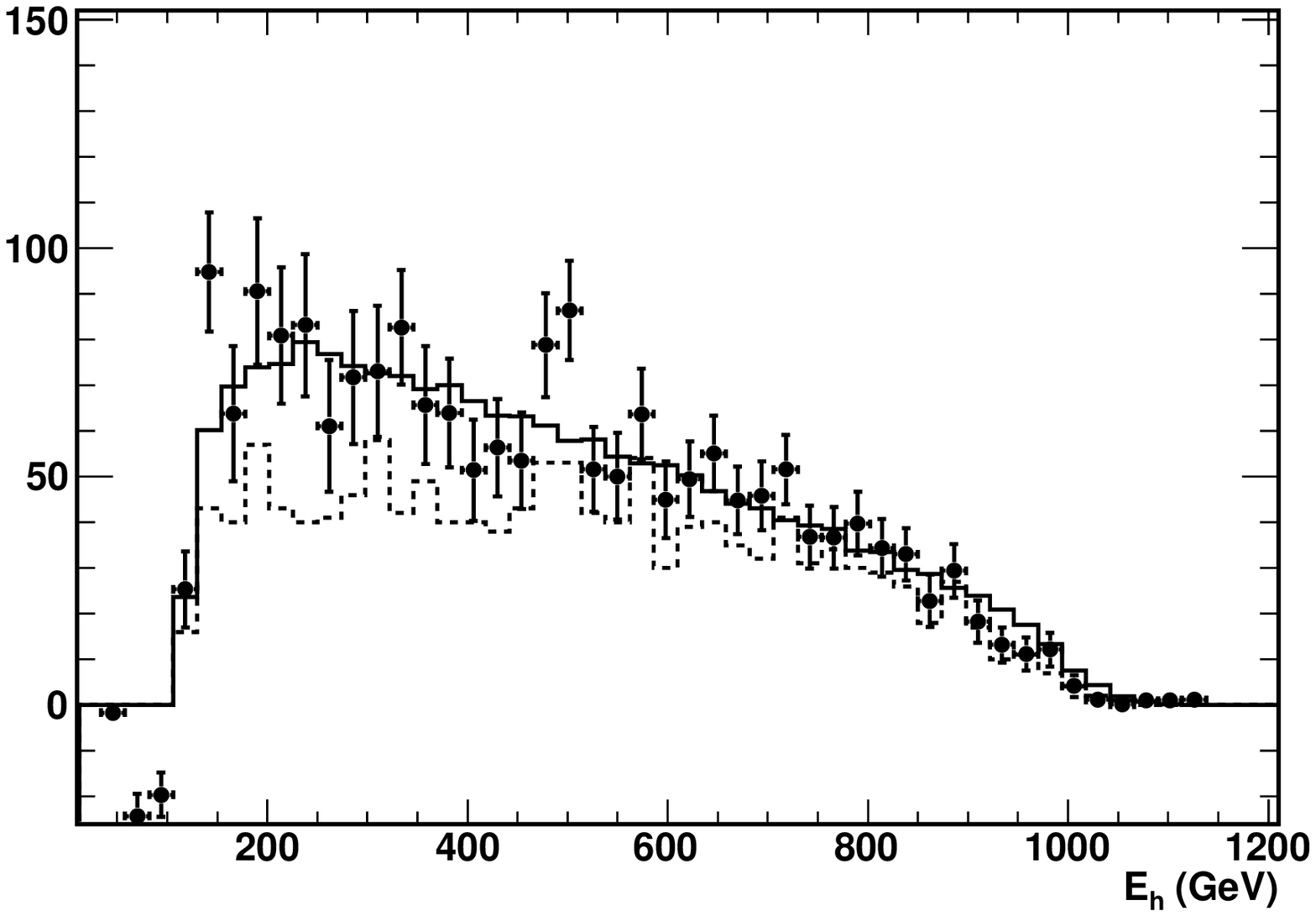} \\
\end{tabular}
\end{center}
\caption{Fits of fully simulated and reconstructed events in model~I: (left) $h^{0}$ energy spectrum of selected 
4-jet $hh$ candidate events energy spectrum and (right) background subtracted spectrum (points with error bars) 
with the result of the 3-par fit (continuous line).}
\label{fig:efitHDST}
\end{figure}
This efficiency is independent on the $W$ energy and the reconstruction and selection criteria do not 
introduce any significant bias to this distribution (see Figure~\ref{fig:effW}). 
The sample of selected 4-jet $WW$ candidate events has a purity of 86\% and consists of 78\% signal 
$\chi^+_1 \chi^-_1 \rightarrow W^+ \chi^0_1 W^- \chi^0_1$, 7.5\%  
$\tilde e_L^{+} \tilde e_{L}^{-} \to \chi_1^{+}  \chi_1^{-} \nu_e \bar \nu_e$ 
and 0.2\% of other 4-jet $WW$ SUSY processes (see the left panel of 
Figure~\ref{fig:mfitDST2}). The fraction of background fake-$WW$ events is obtained from the fit to the di-jet invariant 
mass distribution and the shape of their di-jet energy spectrum directly extracted from the reconstructed events, 
using the di-jet mass side-bands 40 $< E_{jj} <$ 60~GeV and 140 $< E_{jj} <$ 160~GeV, and subtracted.   

The multi-parameter fit to the $\chi^{\pm}_1$ and $\tilde e_L$  masses is repeated on the background-subtracted 
$W^{\pm}$ energy distribution of selected fully simulated and reconstructed events (see Figure~\ref{fig:efitWDST}).
The result is $M_{\chi^{\pm}_1}$ = (643 $\pm$ 14)~GeV and $M_{\tilde e^{\pm}_L}$ = (1100 $\pm$ 104)~GeV, 
where the statistical accuracies are consistent to those obtained on a smeared simulated spectrum of equal 
statistics, $M_{\chi^{\pm}_1}$ = (643 $\pm$ 12)~GeV and $M_{\tilde e^{\pm}_L}$ = (1100 $\pm$ 110)~GeV.
Finally, we select events with di-jets compatible with the $hh$ hypothesis. The total selection efficiency 
is 69\% for $\chi^0_2 \chi^0_2 \rightarrow h^0 \chi^0_1 h^0 \chi^0_1 \to b \bar b \chi^0_1 b \bar b \chi^0_1$ 
The multi-parameter fit to the $\chi^0_2$ and $\tilde \nu_{\ell}$  masses is repeated on the background-subtracted 
$h^0$ energy distribution of selected fully simulated and reconstructed events (see Figure~\ref{fig:efitHDST}).
The result is $M_{\chi^{0}_2}$ = (643 $\pm$ 26)~GeV and $M_{\tilde \nu_{\ell}}$ = (1097 $\pm$ 148)~GeV, 
where the statistical accuracies are consistent to those obtained on a smeared simulated spectrum of equal 
statistics, $M_{\chi^{0}_2}$ = (643 $\pm$ 21)~GeV and  $M_{\tilde \nu_{\ell}}$ = (1097 $\pm$ 123)~GeV. The 
degradation of the statistical accuracy in the analysis of the fully simulated and reconstructed data compared to
that on the smeared generator-level events, is likely due to the larger background from fake 4-jet $hh$ events, 
which is not included at generator level. In this study we do not consider jet flavour tagging for consistency with 
the simple procedure adopt for the generator level study. However, by applying $b$-tagging to the four jets, the
background, which does not contain two light Higgs bosons, can be largely reduced. 

In conclusion, the analysis of fully simulated and reconstructed SUSY events where 4-jet, $WW$ and $hh$ candidates 
are selected based on the reconstructed topology and di-jet invariant mass shows that reconstruction efficiencies 
are quite large and flat with the boson energy. Backgrounds from other final states can be reliably estimated and 
subtracted in a model-independent way. The statistical accuracies obtained on the extraction of gaugino and slepton 
masses from these data are found to be comparable to those from the smeared generator, once the reconstruction 
efficiencies are taken into account. 

\section{Mass Determination by Threshold Energy Scans}

An $e^+e^-$ linear collider  with tunable beam energy can determine the 
sparticle masses by performing energy scans of their pair production cross 
section near threshold. In principle, this method provides a better mass
accuracy, compared to the kinematic end-point method discussed above. 
Threshold energy scans put significant requirements on the 
machine performance and versatility. Not only the beam energy needs to be 
varied over a broad range, but, since the cross section at threshold is small, 
a large luminosity must be preserved in lower energy operation. Beamstrahlung 
effects are important at threshold, while SUSY background are reduced, at least for 
the lighter states. We study the processes 
\begin{itemize}
\item $e^+e^- \to \chi^+_1 \chi^-_1 \to W^+ \chi^0_1 W^- \chi^0_1$; $W \to q \bar q'$,
\item $e^+e^- \to \chi^0_2 \chi^0_2 \to h^0 \chi^0_1 h^0 \chi^0_1$; $h \to b \bar b$,
\item $e^+e^- \to \chi^+_2 \chi^+_2 \to W^+ \chi^0_2 W^- \chi^0_1 \to W^+ h^0 \chi^0_1 W^- \chi^0_1$; 
$h \to b \bar b$, $W \to q \bar q'$
\item $e^+e^- \to \chi^+_2 \chi^+_2 \to h^0 \chi^+_1 W^- \chi^0_1 \to h^0 W^+ \chi^0_1 W^- \chi^0_1$; 
$h \to b \bar b$, $W \to q \bar q'$
\end{itemize}
for model~I. For model~II we study
\begin{itemize}
\item $e^+e^- \to \chi^+_1 \chi^-_1 \to W^+ \chi^0_1 W^- \chi^0_1$; $W \to q \bar q'$
\end{itemize}
\begin{figure}
\begin{center}
\begin{tabular}{cc}
  \subfloat[$\chi^{\pm}_1$]{\includegraphics[width=7.0cm]{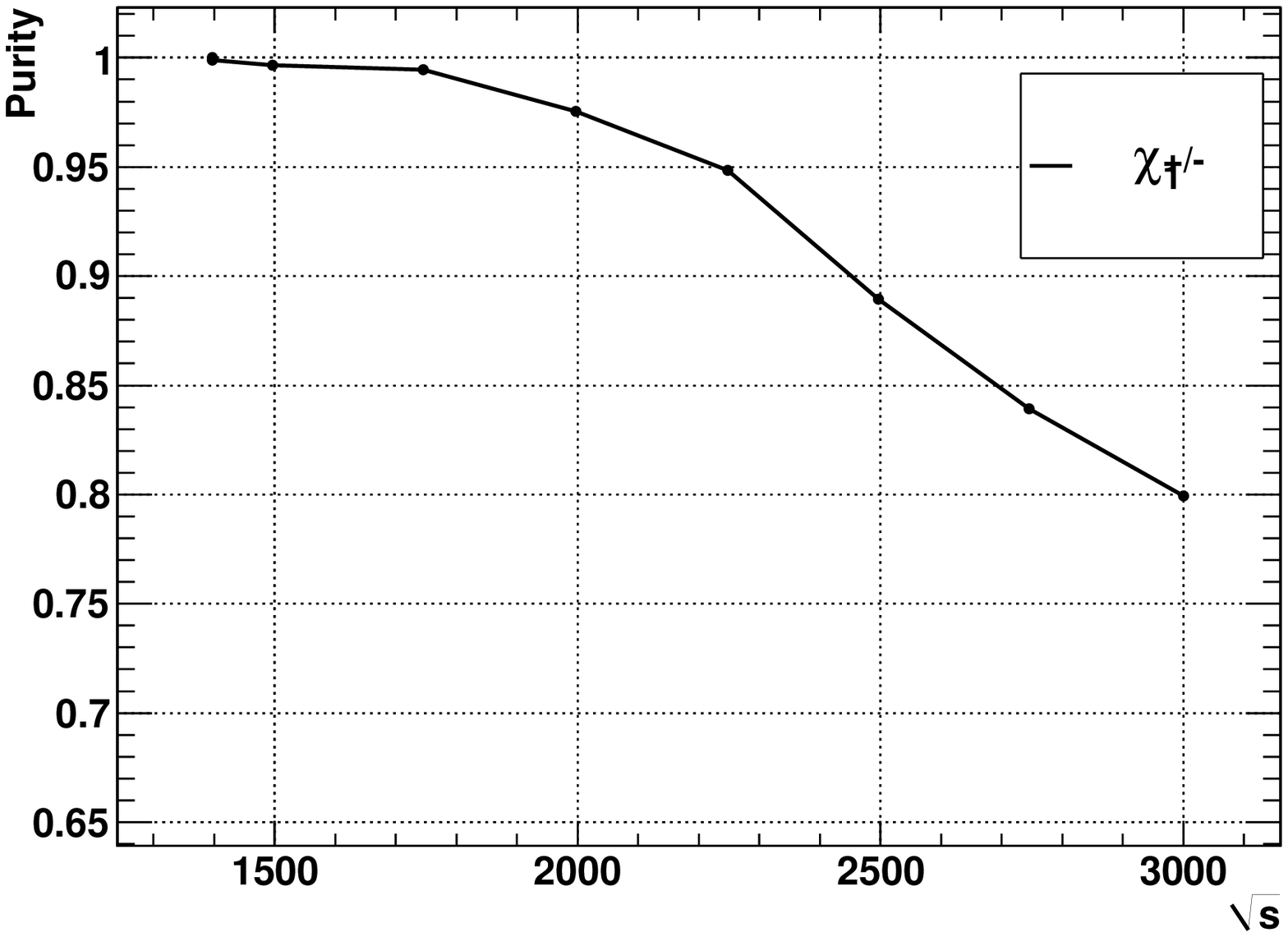}} &
\subfloat[$\chi^0_2$]{\includegraphics[width=7.0cm]{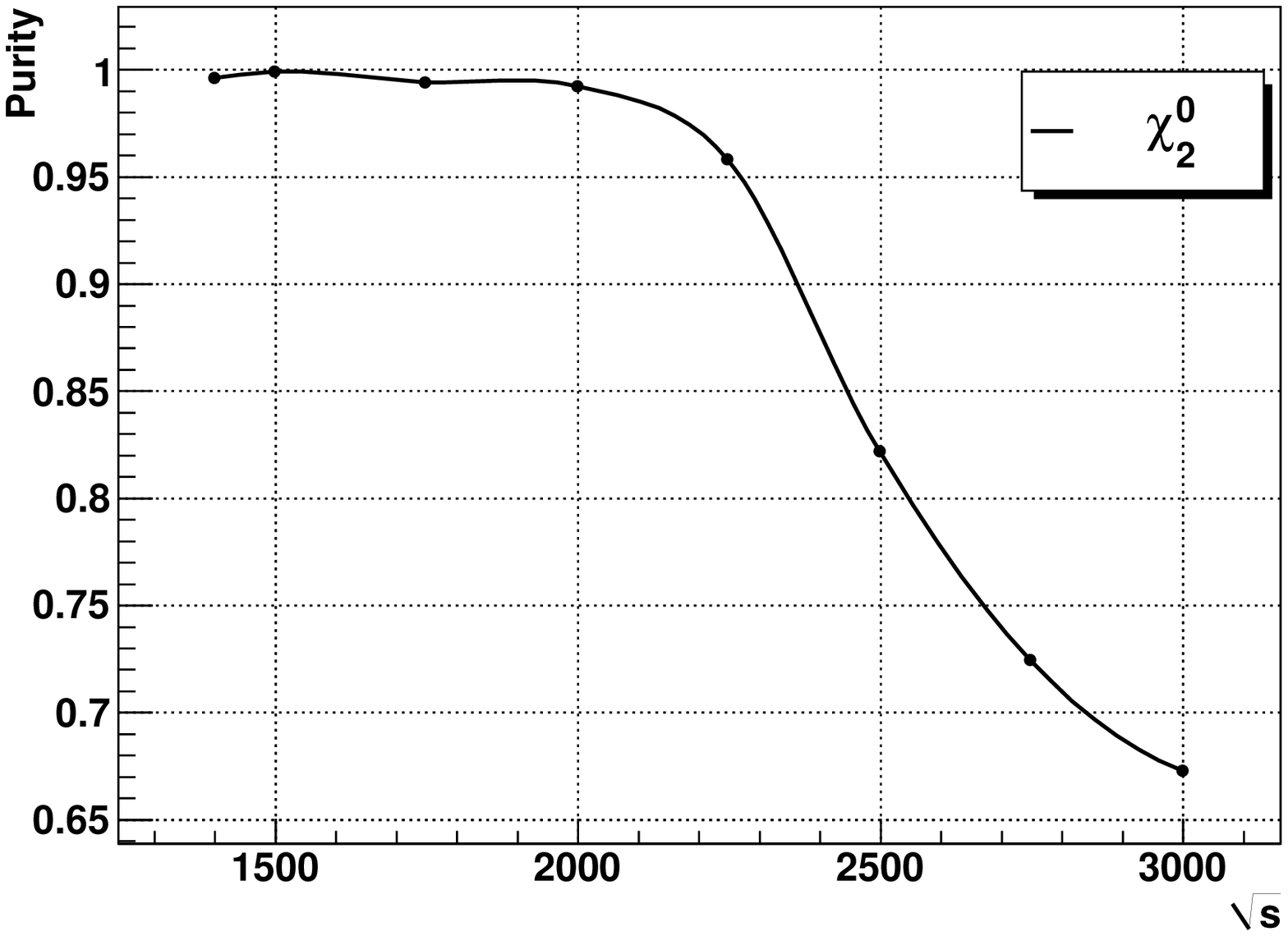}} \\
\end{tabular}
\end{center}
\caption{Purity for the $\chi^{+}_1 \chi^{-}_1$ and $\chi^0_2 \chi^0_2$ process in 4-jet final state 
as a function of the $\sqrt{s}$ energy.}
\label{fig:purity}
\end{figure}

\begin{figure}[h!]
\begin{center}
\begin{tabular}{cc}
\includegraphics[width=6.0cm]{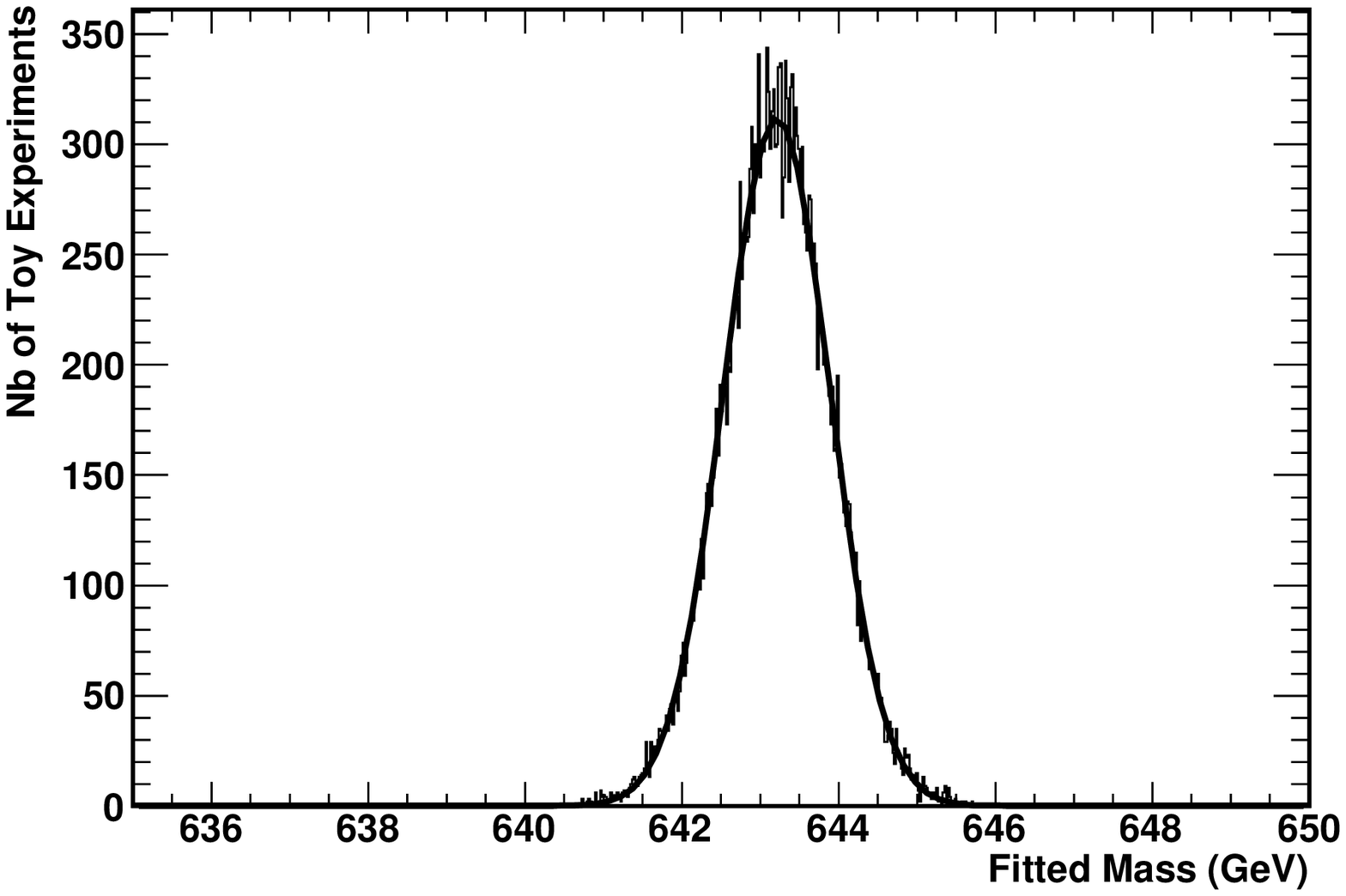}  & 
\includegraphics[width=6.0cm]{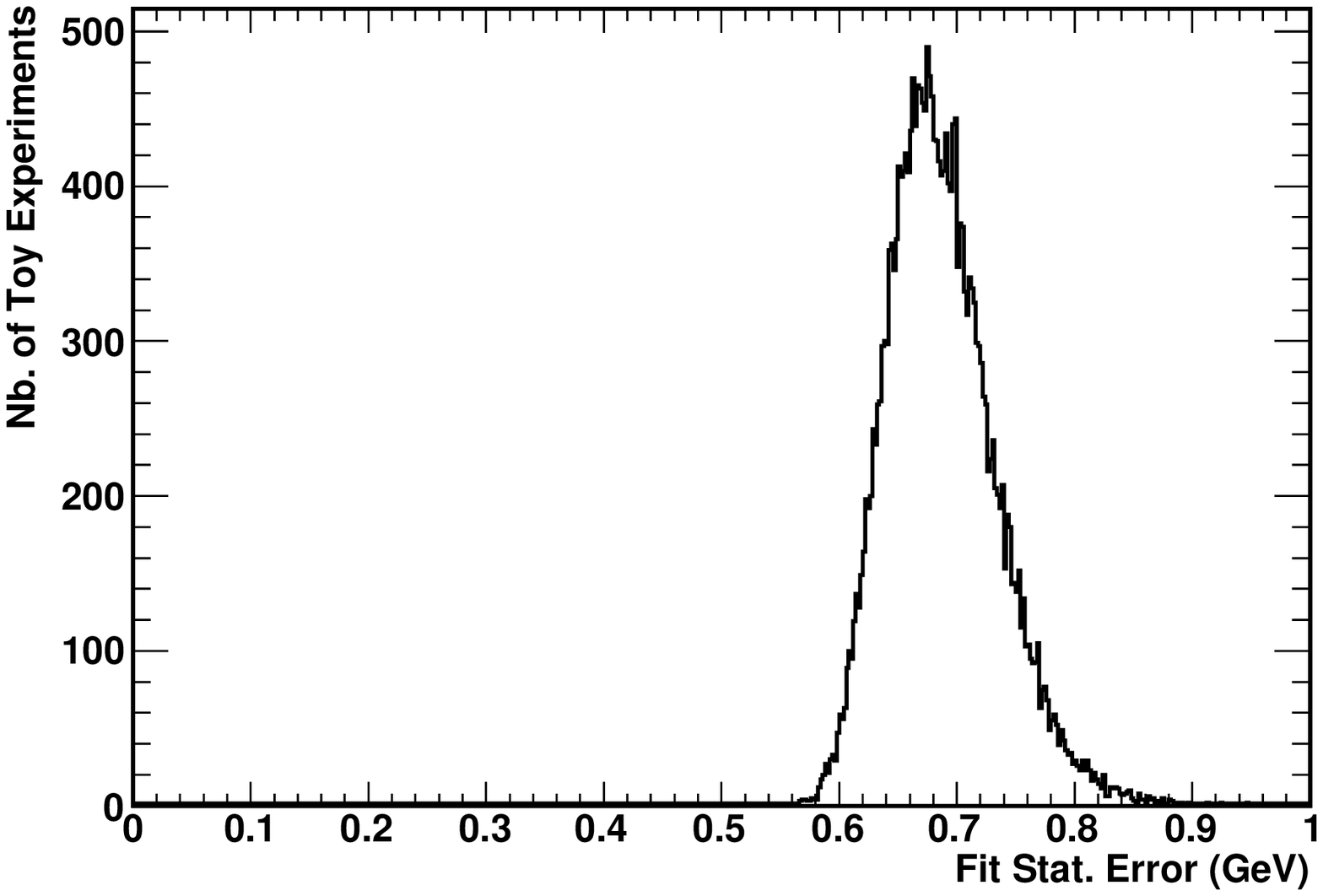}  \\
\end{tabular}
\end{center}
\caption{Toy test validation of the threshold scan fit results for $\chi^{\pm}_1$ for model~I. 
The corresponding $\chi^2$ fit result is (643.2$\pm$ 0.68)~GeV.}
\label{fig:val}
\end{figure}
\begin{figure}[h!]
\begin{center}
\includegraphics[width=8.0cm]{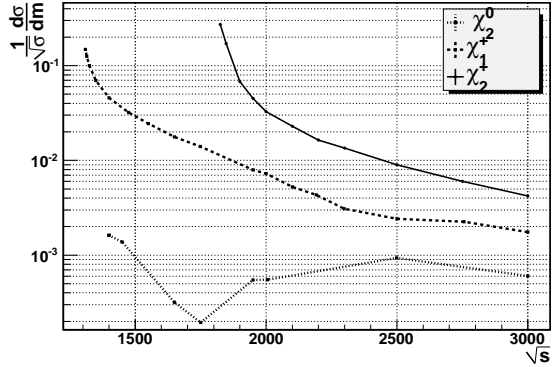}
\end{center}
\caption{Sensitivity $\frac{1}{\sqrt{\sigma}} \frac{d \sigma}{dm} $ to sparticle masses in the threshold 
scan as a function of $\sqrt{s}$ for model~I.}
\label{fig:sens}
\end{figure}
We assume a total integrated luminosity of 3~ab$^{-1}$, where the 2~ab$^{-1}$ taken at the maximum 
energy, as assumed above, are supplemented by 1~ab$^{-1}$ of statistics dedicated to the scan of 
sparticle pair production thresholds at lower energies.
At, or below, $\sqrt{s}$ = 2~TeV, the $\chi^{+}_1 \chi^-_1$ and $\chi^0_2 \chi^0_2$ pair production 
saturates the final states with 4-jet and $WW$ or $hh$, respectively, since slepton production is below 
threshold due to the larger $\tilde e_L$ and $\tilde \nu_{\ell}$ masses, as shown in 
Figure~\ref{fig:purity}.
A 1-par $\chi^2$ fit to the cross section values at the chosen operating energies is performed to extract 
the mass value.
The mass and its uncertainty are obtained by assuming a given number of cross section measurements
of the relevant pair production process at the $\sqrt{s}$ values for the scan points. The cross section 
is computed both at Born level and also adding ISR and beamstrahlung effects using {\tt Pythia}.
 The fit results are validated using toy tests. In these we repeat the fits by varying 
the cross section within its statistical uncertainty at each $\sqrt{s}$ value and we plot the result. 
We verify that the result is centred on the simulated mass and its width is consistent with the $\chi^2$ fit 
uncertainty (see Figure~\ref{fig:val}).
In order to define a suitable scan strategy, we first study the sensitivity to the sparticle masses as a 
function of $\sqrt{s}$. The sensitivity is defined as $1/\sqrt{\sigma} d \sigma/d m$, where $\sigma$ is 
the pair production cross section and $m$ the mass, as in ref.~\cite{Blair:2001cz}. 
\begin{table}[ht!]
\caption{Statistical accuracy on sparticle masses from energy scans under different assumptions for 
2~ab$^{-1}$ of integrated luminosity at 3~TeV and 1~ab$^{-1}$ at two energies near threshold.}
\begin{center}
\begin{tabular}{|l|c|cccccc|}
\hline
Particle        & Mass  & Born       & ISR        & ISR+BS    & ISR+BS     & w/ Pol    &  w/ Pol    \\
                & (GeV) &            &            &           &  +Bkg      & (+0.8/0)   &  (+0.8/-0.6) \\
\hline
Model~I         &       &           &             &           &             &           &            \\
$\chi^{\pm}_1$  & 643.2 & $\pm$~0.6 & $\pm$~0.6   & $\pm$~0.7 & $\pm$~0.7    & $\pm$~0.5  &  $\pm$~0.4  \\
$\chi^{0}_2$    & 643.1 & $\pm$~4.3 & $\pm$13.8   & $\pm$24.1 & $\pm$25.6    & $\pm$23.9  &  $\pm$18.1 \\
$\chi^{\pm}_2$  & 916.7 & $\pm$~0.8 & $\pm$~0.9   & $\pm$~1.3 & $\pm$~1.4    & $\pm$~1.1  &  $\pm$~0.9  \\
\hline
Model~II        &       &           &             &           &             &           &            \\
$\chi^{\pm}_1$  &1062.2 & $\pm$~6.2 & $\pm$~6.4   & $\pm$~6.9 &             & $\pm$~5.1  &  $\pm$~2.8  \\
\hline
\end{tabular}
\end{center}
\label{tab:scan}
\end{table}
We compute the cross section $\sigma$ at various $\sqrt{s}$ values for a set of closely spaced masses and 
obtain 
the derivative $ d \sigma/d m$ of the change of the cross section at each energy per unit of mass change. 
Results are shown in Figure~\ref{fig:sens}, which indicate that the maximum of the sensitivity to the mass is 
achieved near threshold.   
The number of scan points and the share of the statistics among them is optimised by studying the mass 
uncertainty obtained from the fit for different assumptions. We find that it is preferable to concentrate 
the luminosity in a small number of scan points. For example, the statistical accuracy on the mass of the 
$\chi^{\pm}_1$ in the model~I varies from $\pm$0.85~GeV, obtained for a four-point scan 
(1310$\le \sqrt{s} \le$1950~GeV), to $\pm$0.45~GeV, 
when the luminosity is split between just two points, one of which at the peak of the sensitivity 
($\sqrt{s}$=1350~GeV) and the second close to threshold ($\sqrt{s}$=1310~GeV). This confirms the findings 
of~\cite{Blair:2001cz} for lower sparticle masses and different luminosity spectrum. Finally, we consider 
the option of operating the collider with polarised beams. 
Results are summarised in Table~\ref{tab:scan}. In all cases, except the $\chi^{\pm}_2$, the mass 
accuracies obtained with a dedicated threshold scan improve on those resulting from the kinematic edge 
analysis at 3~TeV by factors of 2 or more. The use of polarised beam further improves these accuracies, 
effectively compensating for the loss of sensitivity due to ISR and BS.

\section{Conclusions}

The determination of chargino and neutralino masses in high-mass SUSY scenarios with two-body 
decays into $W^{\pm}$, $Z^0$ and $h^0$ bosons provides us with a mean to quantify the effect 
of radiation, by ISR and beamstrahlung, and parton energy resolution on the accuracy achievable 
in supersymmetric particle mass measurements at a multi-TeV $e^+e^-$ linear collider. In our
analysis both fits to boson energy spectra and threshold scans are considered for fully hadronic 
final states. Results from generator-level quantities are validated using fully simulated and 
reconstructed events in the $W^+W^- + E_{\mathrm{missing}}$ and $h^0h^0 + E_{\mathrm{missing}}$ 
final states.
Not accounting for reconstruction efficiencies, estimated to be $\simeq$60\% in four jet final 
states, the mass of charginos and neutralinos can be determined from the kinematic edges of 
the boson energy in inclusive SUSY event samples to a relative accuracy in the range 0.3\% to 1.0\% 
(0.6\% - 1.0\%) in absence of radiation and energy resolution effects to 0.8\% to 1.7\%  
(1.1\% - 2.0\%) accounting for ISR, BS and realistic energy resolution for the benchmark with 
particle masses in the range 600 - 900~GeV ($>$1000~GeV), respectively, with 2~ab$^{-1}$ of 
integrated luminosity at $\sqrt{s}$ = 3~TeV. The relative increase of the statistical uncertainty 
of the mass measurement is larger for the model~I which has the sparticles masses far way from pair 
the production thresholds. However, in absolute terms the larger production cross sections in this 
model yield better statistical accuracy in the mass determination. By adopting the criterion that 
the degradation to the mass measurement statistical accuracy from the parton energy resolution 
should not exceed that induced by ISR and BS, we derive the requirement of a relative energy 
resolution for jets, $\delta E/E \le$0.05.
If the accelerator can operate at energies below the nominal $\sqrt{s}$ (down to $\sqrt{s}$=1310~GeV 
for model~I and $\sqrt{s}$=2200~GeV for model~II) with comparable performance to collect about one 
third of the statistics at centre-of-mass energies close to the kinematic thresholds for sparticle 
pair production, the mass accuracies from these threshold scans improves by factors of 2 or more 
compared to those obtained from study of the kinematic edges at the maximum $\sqrt{s}$ energy. 
The availability of polarised beam in the scan further improves these accuracies, effectively 
compensating for the loss of sensitivity due to the effect of ISR and beamstrahlung.
%
%
\section{Acknowledgements}

We are grateful to the colleagues who contributed to this study. 
In particular to Jean-Jacques Blaising, Sabine Kraml and 
Abdelhak Djouadi for extensive discussion and their careful 
reading of the text. We are also thankful to by Dieter Schlatter 
for valuable suggestions on this note. 

%
%

\end{document}